\documentclass[a4paper,fleqn,10pt]{scrartcl}
\pdfoutput=1
\usepackage{authblk}

\usepackage{todonotes}
\presetkeys{todonotes}{inline}{}
\usepackage{amsmath}
\usepackage{amssymb}
\usepackage{commath}
\usepackage{array}
\usepackage{calc}
\usepackage{longtable}
\usepackage{multirow}
\usepackage{nicefrac}
\usepackage{pstricks}
\usepackage{graphicx}
\graphicspath{{figures/}}
\usepackage{xspace}
\usepackage{listings}
\usepackage{
  pgf,
  tikz}
\usetikzlibrary{
  patterns,
  shapes.multipart,
  arrows,
  trees,
  scopes,
  decorations.pathreplacing,
  decorations.pathmorphing,
  decorations.markings,
  decorations.text,
  positioning,
  calc
}
\usepackage{mciteplus}
\usepackage[a4paper,pdfborder={0 0 0}]{hyperref}
\usepackage[format=hang,labelfont=bf,hypcap=true]{caption}
\usepackage{subfig}
\usepackage{sectsty}
\usepackage{relsize}

\allsectionsfont{\sffamily}
\subsubsectionfont{\mdseries\itshape\large}

\let\spreprint\empty
\newcommand{\preprint}[1]{\def\spreprint{\protect#1}}
\let\sinstitute\empty

\makeatletter
\renewcommand{\maketitle}{\begingroup
  \null\thispagestyle{empty}%
    \ifx\spreprint\empty
      \vskip 5ex
    \else
      \flushright\large\spreprint\vskip 2ex
    \fi
    \vskip 5ex
    \flushleft
      {\sffamily\bfseries\huge\@title}\vskip 2ex
      \@author\vskip 2ex
      \ifx\sinstitute\empty
      \else
        {\small\sinstitute}
      \fi
    \vskip 5ex
  \endgroup
}
\makeatother
\renewenvironment{abstract}{\begin{center}
  {\large\sffamily\bfseries Abstract: }
  \begin{minipage}[t]{0.75\textwidth}
}{\end{minipage}\end{center}\vskip 10ex}

\lstdefinestyle{runcard}{frame=tb,
  language=Octave,
  aboveskip=3mm,
  belowskip=3mm,
  showstringspaces=false,
  columns=flexible,
  basicstyle={\small\ttfamily},
  numbers=left,
  numberstyle=\tiny\color{gray},
  keywordstyle=\color{blue},
  commentstyle=\color{gray},
  stringstyle=\color{mauve},
  breaklines=true,
  breakatwhitespace=true,
  tabsize=3
}

\numberwithin{equation}{section}

\newcommand{\done}{{\mathrm d}}

\newcommand{\bea}{\begin{align}}
\newcommand{\eea}{\end{align}}
\newcommand{\beq}{\begin{equation}}
\newcommand{\eeq}{\end{equation}}

\newcommand{\bs}{\begin{split}}
\newcommand{\es}{\end{split}}

\newcommand{\bi}{\begin{itemize}}
\newcommand{\ei}{\end{itemize}}
\newcommand{\bc}{\begin{center}}
\newcommand{\ec}{\end{center}}
\newcommand{\bac}{\begin{array}{c}}
\newcommand{\bacc}{\begin{array}{cc}}
\newcommand{\ea}{\end{array}}

\def\spa#1.#2{\langle#1\,#2\rangle}
\def\spb#1.#2{[#1\,#2]}

\newcommand{\kt}{\ensuremath{k_\mathrm{T}}\xspace}

\newcommand{\pt}{\ensuremath{p_\mathrm{T}}\xspace}
\newcommand{\ptzero}{\ensuremath{p_{\mathrm{T},0}}\xspace}

\newcommand{\pT}{\pt}

\DeclareRobustCommand{\PH}{{\ensuremath{h}}}
\DeclareRobustCommand{\PZ}{{\ensuremath{\mathrm{Z}}}}
\DeclareRobustCommand{\PW}{{\ensuremath{\mathrm{W}}}}
\DeclareRobustCommand{\PWpm}{{\ensuremath{\mathrm{W}^{\pm}}}}
\DeclareRobustCommand{\Pl}{{\ensuremath{\ell}}}

\newcommand{\alphaS}{\ensuremath{\alpha_\text{s}}\xspace}

\newcommand{\Nc}{\ensuremath{N_c}}

\newcommand{\Qcut}{\ensuremath{Q_\mathrm{cut}}}

\newcommand{\NNLO}{\ensuremath{\mathrm{NNLO}}\xspace}

\newcommand{\sla}[1]{\ensuremath{{#1\kern-0.45em/}}}

\newcommand{\QCD}{\text{QCD}}
\newcommand{\EW}{\text{EW}}

\newcommand{\EWapprox}{\ensuremath{\EW_\text{approx}}}

\newcommand{\QCDpEWapprox}{\ensuremath{\QCD+\EWapprox}}

\newcommand{\SecRef}[1]{Section~\ref{#1}\xspace}

\newcommand{\SecsRef}[2]{Sections~\ref{#1}--\ref{#2}\xspace}

\newcommand{\FigRef}[1]{Fig.~\ref{#1}\xspace}

\newcommand{\FigRefs}[2]{Figs.~\ref{#1} and~\ref{#2}\xspace}
\newcommand{\FigureRefs}[2]{Figures~\ref{#1} and~\ref{#2}\xspace}

\newcommand{\eg}{{\itshape e.g.}\xspace}
\newcommand{\ie}{{\itshape i.e.}\xspace}
\newcommand{\cf}{{\itshape cf.}\xspace}

\newcommand\babar{B\protect\scalebox{0.8}{A}B\protect\scalebox{0.8}{AR}\xspace}
\newcommand\lep{L\protect\scalebox{0.8}{EP}\xspace}
\newcommand\lepi{L\protect\scalebox{0.8}{EP}~1\xspace}
\newcommand\LEP{\lep}
\newcommand\LEPI{\lepi}

\newcommand\ALEPH{A\protect\scalebox{0.8}{LEPH}\xspace}

\newcommand\LHC{L\protect\scalebox{0.8}{HC}\xspace}
\newcommand\FCC{F\protect\scalebox{0.8}{CC}\xspace}
\newcommand\ATLAS{\atlas}
\newcommand\atlas{A\protect\scalebox{0.8}{TLAS}\xspace}
\newcommand\CMS{\cms}
\newcommand\cms{C\protect\scalebox{0.8}{MS}\xspace}

\DeclareRobustCommand{\plusplus}{\raisebox{0.2ex}{\smaller ++}}
\newcommand{\MCatNLO}{M\protect\scalebox{0.8}{C}@N\protect\scalebox{0.8}{LO}\xspace}

\newcommand{\LOPS}{L\scalebox{0.8}{O}P\scalebox{0.8}{S}\xspace}

\newcommand{\NLOPS}{N\scalebox{0.8}{LO}P\scalebox{0.8}{S}\xspace}

\newcommand{\UNNLOPS}{UN\ensuremath{^2}\scalebox{0.8}{LO}P\scalebox{0.8}{S}\xspace}
\newcommand{\LOOPsqPS}{L\scalebox{0.8}{OOP}$^2$P\scalebox{0.8}{S}\xspace}

\newcommand{\MEPSatNLO}{M\scalebox{0.8}{E}P\scalebox{0.8}{S}@N\protect\scalebox{0.8}{LO}\xspace}
\newcommand{\MEPSatLO}{M\scalebox{0.8}{E}P\scalebox{0.8}{S}@L\protect\scalebox{0.8}{O}\xspace}
\newcommand{\MEPSatLOOPsq}{M\scalebox{0.8}{E}P\scalebox{0.8}{S}@L\protect\scalebox{0.8}{OOP}$^2$\xspace}
\newcommand{\MINLO}{MIN\protect\scalebox{0.8}{LO}\xspace}
\newcommand{\Rambo}{R\protect\scalebox{0.8}{AMBO}\xspace}
\newcommand{\Vegas}{V\protect\scalebox{0.8}{EGAS}\xspace}

\newcommand{\Madgraph}{M\protect\scalebox{0.8}{AD}G\protect\scalebox{0.8}{RAPH}\xspace}

\newcommand{\UFO}{U\scalebox{0.8}{FO}\xspace}
\newcommand{\Herwig}{H\protect\scalebox{0.8}{ERWIG}\xspace}

\newcommand{\Pythia}{P\protect\scalebox{0.8}{YTHIA}\xspace}

\newcommand{\Caesar}{C\protect\scalebox{0.8}{AESAR}\xspace}

\newcommand{\Gosam}{G\protect\scalebox{0.8}{O}S\protect\scalebox{0.8}{AM}\xspace}

\newcommand{\BlackHat}{B\protect\scalebox{0.8}{LACK}H\protect\scalebox{0.8}{AT}\xspace}
\newcommand{\NJet}{NJ\protect\scalebox{0.8}{ET}\xspace}

\newcommand{\OpenLoops}{O\protect\scalebox{0.8}{PEN}L\protect\scalebox{0.8}{OOPS}\xspace}
\newcommand{\Recola}{R\protect\scalebox{0.8}{ECOLA}}

\newcommand{\Sherpa}{S\protect\scalebox{0.8}{HERPA}\xspace}
\newcommand{\Comix}{C\protect\scalebox{0.8}{OMIX}\xspace}

\newcommand{\Amegic}{A\protect\scalebox{0.8}{MEGIC}\xspace}
\newcommand{\CSS}{CSS\protect\scalebox{0.8}{HOWER}\xspace}
\newcommand{\Dire}{D\protect\scalebox{0.8}{IRE}\xspace}
\newcommand{\Ahadic}{A\protect\scalebox{0.8}{HADIC}\xspace}
\newcommand{\Phasic}{P\protect\scalebox{0.8}{HASIC}\xspace}
\newcommand{\Amisic}{A\protect\scalebox{0.8}{MISIC}\xspace}
\newcommand{\Hadrons}{H\protect\scalebox{0.8}{ADRONS}\xspace}

\newcommand{\Professor}{P\protect\scalebox{0.8}{ROFESSOR}\xspace}
\newcommand{\MCgrid}{MC\protect\scalebox{0.8}{GRID}\xspace}
\newcommand{\Rivet}{R\protect\scalebox{0.8}{IVET}\xspace}
\newcommand{\ApplGrid}{A\protect\scalebox{0.8}{PPL}G\protect\scalebox{0.8}{RID}\xspace}
\newcommand{\FastNLO}{F\protect\scalebox{0.8}{AST}NLO\xspace}
\newcommand{\NTUPLE}{N\protect\scalebox{0.8}{TUPLE}\xspace}

\newcommand{\HepMC}{H\protect\scalebox{0.8}{EP}MC\xspace}

\newcommand{\FeynRules}{F\protect\scalebox{0.8}{EYN}R\protect\scalebox{0.8}{ULES}\xspace}

\newcommand{\Fastjet}{F\protect\scalebox{0.8}{AST}J\protect\scalebox{0.8}{ET}\xspace}

\newcommand{\HepForge}{H\scalebox{0.8}{EP}F\scalebox{0.8}{ORGE}\xspace}
\newcommand{\LHAPDF}{L\protect\scalebox{0.8}{HAPDF}\xspace}

\newcommand{\cpp}{C\plusplus\xspace}

\newcommand{\ASCII}{A\scalebox{0.8}{SCII}\xspace}

\newcommand{\Sversion}{2.2.8\xspace}

\usepackage{placeins}
\usepackage{sectsty}
\allsectionsfont{\boldmath}

\hypersetup{
  pdfauthor={Enrico Bothmann, Gurpreet Singh Chahal, Stefan Hoeche, 
             Johannes Krause, Frank Krauss, Silvan Kuttimalai, 
             Sebastian Liebschner, Davide Napoletano, Marek Schoenherr,
             Holger Schulz, Steffen Schumann, Frank Siegert}, 
  pdftitle={Event Generation with Sherpa 2.2}
}

\preprint{FERMILAB-PUB-19-218-T\\SLAC-PUB-17433\\IPPP/19/42\\MCNET-19-11}

\author[1]{Enrico Bothmann}
\author[2]{Gurpreet Singh Chahal}
\author[3]{Stefan H{\"o}che}
\author[4]{Johannes Krause}
\author[2]{Frank Krauss}
\author[5]{Silvan Kuttimalai}
\author[4]{Sebastian Liebschner}
\author[6]{Davide Napoletano}
\author[2]{Marek Sch{\"o}nherr}
\author[7]{Holger Schulz}
\author[1]{Steffen Schumann}
\author[4]{Frank Siegert}

\affil[1]{Institut f{\"u}r Theoretische Physik,
  Georg-August-Universit{\"a}t G{\"o}ttingen, D-37077 G{\"o}ttingen, Germany}
\affil[2]{Institute for Particle Physics Phenomenology, Durham University,
  Durham DH1 3LE, UK}
\affil[3]{Fermi National Accelerator Laboratory, Batavia, IL, 60510-0500, USA}
\affil[4]{Institut f{\"u}r Kern- und Teilchenphysik, TU Dresden,
  D-01062 Dresden, Germany}
\affil[5]{SLAC National Accelerator Laboratory, Menlo Park, CA 94025, USA}
\affil[6]{IPhT, CEA Saclay, CNRS, Universit{\'e} Paris-Saclay,
  F-91191 Gif-sur-Yvette cedex, France}
\affil[7]{Department of Physics, University of Cincinnati, Cincinnati, OH 45219, USA}

\title{Event Generation with \Sherpa 2.2}

\begin{document}
\maketitle
\begin{abstract}
  \Sherpa is a general-purpose Monte Carlo event generator for the simulation of particle
collisions in high-energy collider experiments.  We summarise essential features and
improvements of the \Sherpa 2.2 release series, which is heavily used
for event generation in the analysis and interpretation of LHC Run 1 and Run 2 data.  We
highlight a decade of developments towards ever higher precision in the simulation of
particle-collision events.

\end{abstract}		
\clearpage
\tableofcontents

\section{Introduction}

Monte Carlo event generators are indispensable tools for the design, realisation,
analysis and interpretation of high-energy scattering experiments. In particular,
general-purpose generators such as \Pythia~\cite{Sjostrand:2014zea},
\Herwig~\cite{Bellm:2015jjp} and \Sherpa~\cite{Gleisberg:2008ta} are necessary to
address detailed aspects of the final states produced in individual scattering
events~\cite{Buckley:2011ms}. Typical experimental use cases comprise for example
the calibration of object-reconstruction algorithms, the evaluation of detector
acceptances, selection efficiencies, or the extrapolation of fiducial cross sections
to the full phase space. 

Furthermore, over the past decade, Monte Carlo event generators have been established
as a tool for precision predictions of scattering cross sections, differential
distributions and event topologies. Through the consistent inclusion of higher-order
perturbative corrections, in particular in QCD, but also in QED and in the electroweak
sector, they nowadays represent state-of-the-art theory calculations that make
precision analyses and data interpretation possible. Based on a high level of automation
they allow for both the realistic simulation of Standard Model production processes
and the description of almost arbitrary New Physics signals.  Monte Carlo
event generators form a vital cornerstone of collider-based particle physics, from
searches for new phenomena to actual Standard Model measurements.

\begin{figure}[th!]
  \centering
    \includegraphics[width=1.0\textwidth]{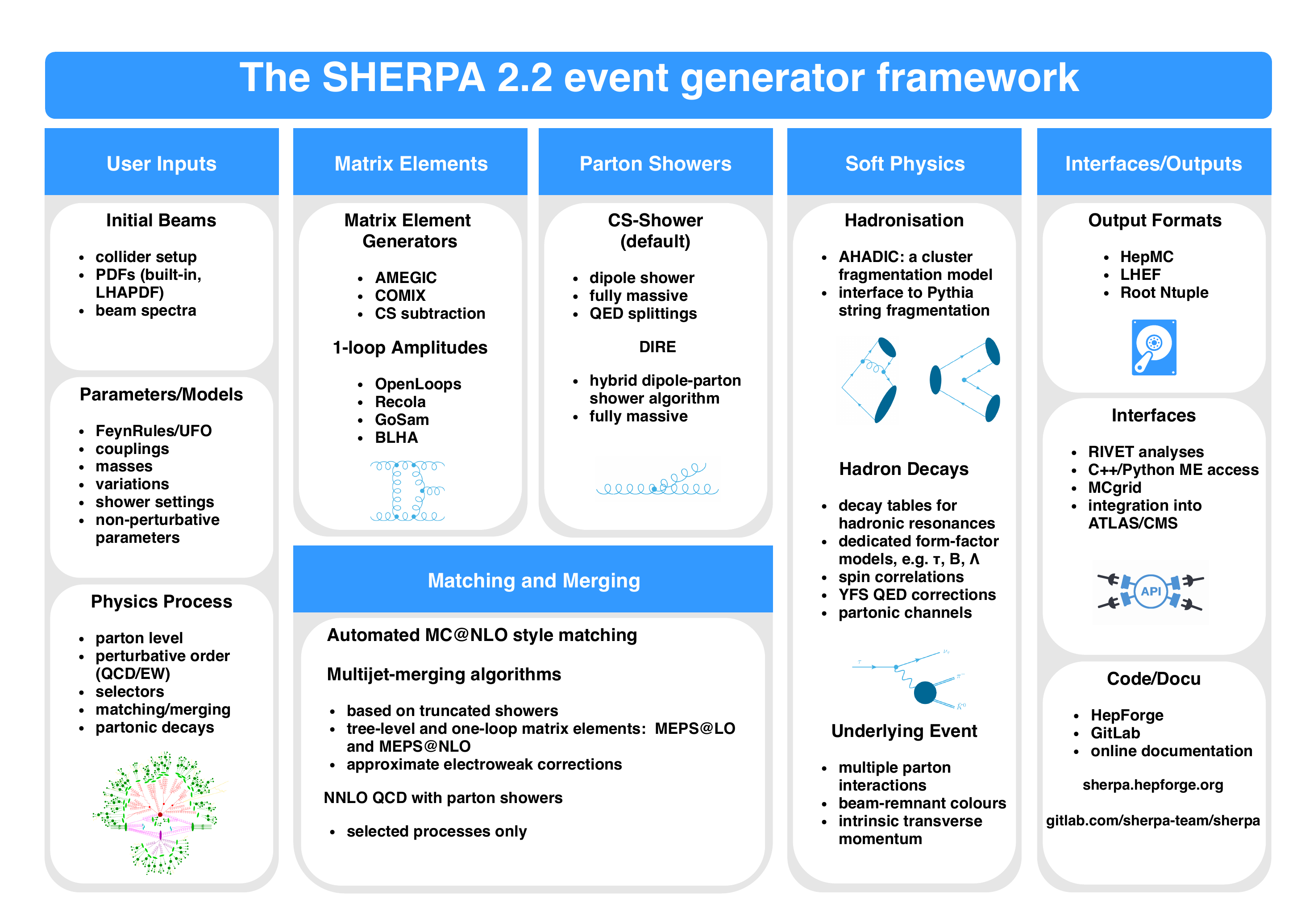}
    \caption{Overview of the \Sherpa 2.2 event generator framework.}%
    \label{fig:Overview}%
\end{figure}

The \Sherpa event generator framework, introduced about fifteen years ago
\cite{Gleisberg:2008ta,Gleisberg:2003xi}, is a general-purpose simulation
tool for particle collisions at high-energy colliders. It contains implementations
of all components needed for a factorised and probabilistic description of scattering
events at hadron-hadron, lepton-hadron and lepton-lepton colliders.

This paper summarises the current abilities and components of \Sherpa,
reflecting the legacy of the \Sherpa 2.2 series that was and is being used extensively
for the analysis of \LHC Run 1 and Run 2 data. A pictorial overview of the \Sherpa\
framework is given in Fig.~\ref{fig:Overview}. A generator
setup and the corresponding event generation is defined through a text file that
contains all non-default settings
needed to define the process of interest and to steer the event evolution. The latter
includes the setup of the initial beams, the physics model as well as parameters to
consider. \Sherpa\ features two built-in tree-level matrix element generators,
\Amegic~\cite{Krauss:2001iv} and \Comix~\cite{Gleisberg:2008fv,Hoche:2014kca}.
They are used for the simulation of parton-level events within the Standard Model and beyond,
and for the decay of heavy resonances such as $W$, $Z$, or Higgs bosons or top
quarks. Both include automated methods for efficient phase-space integration
and algorithms for the subtraction of infrared divergences in calculations at
next-to-leading order (NLO) in QCD~\cite{Catani:1996vz,Catani:2002hc,Gleisberg:2007md}
and the electroweak theory~\cite{Schonherr:2017qcj}. For the evaluation of
virtual corrections at NLO accuracy \Sherpa\ relies on interfaces to dedicated
one-loop providers, \eg \BlackHat \cite{Berger:2008sj},
\OpenLoops~\cite{Cascioli:2011va} and \Recola~\cite{Actis:2016mpe,Biedermann:2017yoi}.
The default parton-showering algorithm of the \Sherpa 2.2 series is the
\CSS~\cite{Schumann:2007mg}, based on Catani--Seymour dipole
factorisation~\cite{Catani:1996vz,Catani:2002hc,Nagy:2006kb}. 
As of version 2.2.0 \Sherpa\ also features an independent second shower
implementation, \Dire~\cite{Hoche:2015sya,Hoche:2017iem,Hoche:2017hno}. For the
matching of NLO QCD matrix elements with parton showers \Sherpa\ implements the MC@NLO
method~\cite{Frixione:2002ik,Hoeche:2011fd}. For NNLO QCD calculations the \UNNLOPS
method~\cite{Hoche:2014dla,Hoeche:2014aia} is used. The merging of multi-jet production
processes at leading order~\cite{Catani:2001cc,Krauss:2002up,Hoeche:2009rj} and
next-to-leading order~\cite{Hoeche:2010kg,Hoeche:2012yf} is based on truncated parton showers.
Multiple parton interactions are implemented via the Sj{\"o}strand--van-Zijl model~\cite{Sjostrand:1987su}.
 The hadronisation of partons into hadrons is modelled by
a cluster fragmentation model~\cite{Winter:2003tt}. Alternatively, in particular for
uncertainty estimations, an interface to the Lund fragmentation
model~\cite{Sjostrand:1982fn} of \Pythia~\cite{Sjostrand:2006za} is available.
\Sherpa\ provides a large library for the simulation of $\tau$-lepton and hadron decays,
including many form-factor models. Furthermore, a module for the simulation of QED final-state
radiation in particle decays~\cite{Schonherr:2008av}, which is accurate to first order in
$\alpha$ for many channels is built-in. To account for spin correlations in production and
subsequent decay processes the algorithm described in~\cite{Richardson:2001df}
is implemented. Events generated with \Sherpa\ can be cast into various output formats for
further processing, with the \HepMC~\cite{Dobbs:2001ck} format being the most commonly used.
In the specific case of parton-level events, at the leading and next-to-leading 
order in QCD, additional output formats are supported. They include Les Houches Event
Files~\cite{Alwall:2006yp}, \NTUPLE files for NLO QCD events~\cite{Bern:2013zja}
and cross-section interpolation grids produced via \MCgrid~\cite{DelDebbio:2013kxa,Bothmann:2015dba}
in the \ApplGrid~\cite{Carli:2010rw} and \FastNLO~\cite{Kluge:2006xs,Britzger:2012bs} formats.
To analyse events on-the-fly a runtime interface to the \Rivet package~\cite{Buckley:2010ar}
can be used conveniently. 

The \Sherpa\ Monte Carlo is publicly available from its \HepForge project page
{\url{sherpa.hepforge.org}}.  The actual code development and bug-tracking facilities
are hosted on {\url{gitlab.com/sherpa-team/sherpa}}. The current release version is
\Sherpa \Sversion. 

The paper is organised as follows. Sec.~\ref{sec:ingredients} will focus mainly on
highlighting and summarising the specific physics implementations and realisations in \Sherpa,
referring to more in-depth original literature where appropriate. This section also includes
a brief discussion on aspects related to the tuning of non-perturbative model parameters in \Sherpa.
In Sec.~\ref{sec:results} we present selected results obtained with recent versions of \Sherpa\
that shall illustrate typical use cases and highlight specific aspects of the simulation. We
present our conclusions and an outlook in Sec.~\ref{sec:conclusions}.

Please note, for more detailed and pedagogical reviews of general Monte Carlo event generation
techniques and their practical implementations we refer interested readers
to~\cite{Webber:1986mc,Buckley:2011ms,Hoche:2014rga}.


\section{Highlighting \Sherpa Components}
\label{sec:ingredients}
In the following we will briefly describe the central components of the \Sherpa\ framework.
We focus on the physics models and features available, providing references to the
original literature for more detailed theoretical derivations and discussions. 

The \Sherpa framework is written in \cpp in a highly modular structure, reflecting
the factorised ansatz to calculate the evolution of scattering events. The \Sherpa\
core module is responsible for steering the event generation process. It initialises the required
physics modules and iterates the steps of the simulation. The setup of each generator run, including
the specification of model parameters and all switches, is read from a simple
\ASCII file, called {\ttfamily{Run.dat}} per default. Parameters of a specific simulation aspect
are collated in blocks following a simple bracket syntax:
\begin{lstlisting}[style=runcard,numbers=none]
  (block_name){
    Parameter1 Value1;
    Parameter2 Value2;
    ...
  }(block_name)
\end{lstlisting}

Examples of blocks are {\ttfamily{(run)}}, where general settings are kept, while the specification
of the hard scattering process to be considered is compiled in {\ttfamily{(processes)}}. Settings
related to cuts on the hard scattering final state are given in {\ttfamily{(selectors)}}.
Specific run parameters will be highlighted along with the presentation of the physics models in
this section and the examples in Sec.~\ref{sec:results}. We organise
the discussion beginning with methods for the hard-process generation in Sec.~\ref{Sec::MEs},
followed by parton showers and the methods for matching and merging them with higher-order matrix
elements in Secs.~\ref{Sec::PS} and \ref{sec:matchmerge}, respectively. We present the
evaluation of perturbative uncertainties based on a reweighting method in
Sec.~\ref{sec:reweighting}. This is followed by a brief discussion of available beam spectra
and distribution functions in Sec.~\ref{sec:isr}. Section~\ref{sec:yfs} is devoted to the
discussion of higher-order QED and electroweak corrections in the decays of unstable particles.
Sec.~\ref{sec:ue} presents our treatment of beam remnants and the underlying event, while
Sec.~\ref{sec:hadronisation} describes the \Sherpa\ cluster hadronisation model. We close by
presenting our methods for $\tau$-lepton and hadron decays in Sec.~\ref{sec::HadronDecays}.

\subsection{Hard-Scattering Matrix Elements}
\label{Sec::MEs}

The simulation of individual events starts from a partonic hard-scattering
configuration, with momenta distributed according to the corresponding
squared QFT transition matrix element. Sampling those partonic events
allows one to determine the total production rate and differential distributions
of the final-state objects to a given perturbative fixed-order accuracy,
\eg at tree-level or at next-to-leading order in the strong or
electroweak coupling. Given the plethora of processes that users might
want to study -- both within the Standard Model and various theories for
New Physics -- a high level of automation is mandatory for the construction
and evaluation of matrix elements. 

In \Sherpa a large variety of fixed-order calculations are available,
ranging from the explicit implementation of some simple $2\to 2$
squared amplitudes at leading order (LO) and next-to leading order
(NLO), over automated matrix-element generators (MEGs) for tree-level
processes with large multiplicities of external particles, to
interfaces to external matrix-element implementations at tree- and one-loop
level. The respective MEG to be used in a simulation run is specified via:

\begin{lstlisting}[style=runcard,numbers=none]
  (run){
    ...
    ME_SIGNAL_GENERATOR Internal Amegic Comix BlackHat OpenLoops Recola ...;
    ...
  }(run)

  (processes){
    ...
    Loop_Generator Internal BlackHat OpenLoops Recola ...;
    RS_ME_Generator Amegic Comix;
    ...
  }(processes)
\end{lstlisting}

{\ttfamily{ME\_SIGNAL\_GENERATOR}} defines the global choice for the matrix-element
provider(s) to be used throughout the run. When specifying
several values they are consecutively asked to provide the requested
matrix element. By specifying {\ttfamily{Loop\_Generator}}
and/or {\ttfamily{RS\_ME\_Generator}} the generators for the loop amplitudes
and the subtracted real-emission terms may be chosen separately.

\paragraph{Built-in Matrix Element Generators}
\Sherpa includes two fully automated MEGs,
\Amegic~\cite{Krauss:2001iv} and \Comix~\cite{Gleisberg:2008fv},
for the calculation of fixed-order total and differential
cross sections and decay widths for multi-particle production and
decay processes at tree level.  Both MEGs are capable of simulating
complicated final states as chains of subsequent decays in the
narrow-width approximation, including a proper treatment of all
effects due to spin and colour correlations. \Comix allows for external
particles with spin-0, 1/2, and 1, while \Amegic also supports
external spin-2 particles~\cite{Gleisberg:2003ue}. In both MEGs
Majorana fermions are treated using the formalism presented
in~\cite{Denner:1992vza,*Denner:1992me}. Squared amplitudes
in both \Amegic and \Comix can be projected on arbitrary orders in
the contributing couplings. This permits, among others, the computation
of pure QCD contributions to the cross section or to exclusively select
interference terms, see \eg~\cite{Coradeschi:2015tna}.

To give an example, the definition of the tree-level
partonic processes for hadronic electron-positron-pair production in
association with two final-state partons reads: 
\begin{lstlisting}[style=runcard,numbers=none]
  (processes){
    % use light-jet container 93
    Process 93 93 -> 11 -11 93 93;
    % constrain orders in strong (1st) and ew (2nd) coupling
    Order (2,2);
    End process;
  }(processes)
\end{lstlisting}

Note, particles are referred to using their PDG Monte Carlo
number~\cite{Tanabashi:2018oca}. In addition, \Sherpa\ permits the 
utilisation of both predefined and user-specific particle containers.
In the above example, the predefined container {\ttfamily{93}} comprises 
all massless QCD partons, \ie gluons and massless quarks. The coupling
orders are counted at the squared matrix element level.

The factorisation and renormalisation scales used in the evaluation
of the hard process can be specified through
\begin{lstlisting}[style=runcard,numbers=none]
  (run){
    ...
    SCALES <scale-setter>{<fac-scale-definition>}{<ren-scale-definition>};
    ...
  }(run)
\end{lstlisting}

Possible scale setters for fixed-order calculations include {\ttfamily{VAR}}
and {\ttfamily{FASTJET}}. The first allows the use of simple user-defined
functions of the final-state momenta, the latter invokes jet finding via
\Fastjet~\cite{Cacciari:2011ma}. In both cases particle/jet
momenta are accessible through {\ttfamily{p[<i>]}}, where {\ttfamily{i=0,1}} labels
the initial-state momenta and final-state particles or $p_T$-ordered jets use {\ttfamily{i>1}}.
Examples to set both the factorisation and the renormalisation scales to either
the invariant mass of the two jets or their scalar sum read:

\begin{lstlisting}[style=runcard,numbers=none]
% VAR scale setter
SCALES VAR{Abs2(p[4]+p[5])}{Abs2(p[4]+p[5])};

% FASTJET scale setter
SCALES FASTJET[A:antikt,PT:30.,R:0.4,M:0]{H_T2}{H_T2};
\end{lstlisting}

To regularise the phase space, cuts on the final-state leptons and partons
need to be applied. A possible event selection may read (again using
the \Fastjet\ package for jet finding):
\begin{lstlisting}[style=runcard,numbers=none]
  (selector){
    % window cut on di-lepton invariant mass
    Mass 11 -11 80. 100.;
    % transverse momentum cut pT>15 GeV on leptons
    PT  11 15. E_CMS;
    PT -11 15. E_CMS;
    % require at least 2 anti-kt jets with R=0.4 and pT>30 GeV 
    FastjetFinder antikt 2 30. 0 0.4;
  }(selector)
\end{lstlisting}

\Amegic effectively performs a colour decomposition of the full
amplitude, leading to gauge-invariant subsets of amplitudes for
each colour structure. 
These terms are each composed of Feynman diagrams expressed as helicity
building blocks based on~\cite{Kleiss:1985yh,*Ballestrero:1992dv}.
In this amplitude construction process, common sub-amplitudes are identified
and algebraically factored out~\cite{Gleisberg:2003bi}, thereby dramatically
reducing evaluation times later on. The resulting expressions are written
out as \cpp source code, compiled and linked into dynamic libraries. During
the event-generation phase, these libraries are automatically located and loaded to the
main code.

\Comix implements the colour-dressed Berends--Giele recursive
relations~\cite{Duhr:2006iq}, a tree-level equivalent of the
Dyson--Schwinger equations~\cite{Dyson:1949ha,*Schwinger:1951ex,*Schwinger:1951hq},
to construct off-shell currents that are fused into amplitudes.
Information about the valid current and vertex assignments in the process
is written to disk in the form of text files such that subsequent runs of
the generator requesting the same process can commence faster. \Comix uses the
colour-flow representation~\cite{tHooft:1973jz,*Maltoni:2002mq} and colour
sampling to compute cross sections including QCD particles. This explicit
computation of colour-ordered amplitudes turns out to be advantageous in the context
of matrix-element parton-shower matching and merging.

\paragraph{BSM Simulations}
An interface to \FeynRules~\cite{Alloul:2013bka,Christensen:2008py,Christensen:2009jx}, \ie
the \UFO model definition files~\cite{Degrande:2011ua}, allows the user to consider
a wide range of models.  In \Amegic, however, only vertices with up
to four external particles are supported, imposing some limits on its
abilities, while in \Comix this number is limited only by computing power,
allowing calculations in more complicated theories. With the physics-model
information encoded in the standard \UFO format~\cite{Degrande:2011ua},
\Sherpa creates and links complete \cpp source code necessary to compute
arbitrary scattering processes, employing an automatic generator for
Lorentz~\cite{Hoche:2014kca} and colour~\cite{Krauss:2016ely} structures
which represent the elementary vertices of the theory. The generators
\Amegic and \Comix have been extensively benchmarked, internally and against other codes, for example
in the SM~\cite{Gleisberg:2003bi} and the MSSM~\cite{Hagiwara:2005wg}\footnote{
  We would like to note that since version \Sherpa-2.0 the
  realisation~\cite{Gleisberg:2003ue} of the ADD model of large extra dimensions
  is no longer supported. }.

\paragraph{Phase-Space Integration}
\Sherpa uses various methods to efficiently integrate multi-particle
phase spaces, implemented in its \Phasic module. These can be classified
as importance-sampling techniques, where phase-space points are generated using
suitable approximations for the desired target distribution,
that is given by the squared matrix element. For this purpose a set of
phase-space maps (called channels) is automatically constructed by the MEGs according
to the propagator and vertex structures of contributing Feynman diagrams 
or current topologies.
The full set of contributing integration channels is combined into a
multi-channel integrator that features an automatic optimisation of the
individual channel weights~\cite{Kleiss:1994qy}.

In \Amegic this leads to the construction of typically one channel per
diagram~\cite{Krauss:2001iv}. Within \Comix, the method is recast into a
recursive algorithm, reducing the factorial growth in the number of channels
to an exponential one~\cite{Gleisberg:2008fv}. Both \Amegic and \Comix further
optimise the integration over propagator masses and polar angles in decays,
using a re-mapping of random numbers base on~\Vegas~\cite{Lepage:1980dq,Ohl:1998jn}
for each channel.

\paragraph{Resonance Decays}
Intermediate unstable resonances, as they frequently appear in
extensions of the Standard Model, can produce high-multiplicity
final states through cascade decays.  In \Sherpa there are two ways of
treating such effects. The first is to select solely $s$-channel 
diagrams/current topologies of the requested intermediate resonances, 
thereby automatically taking into account finite-width and 
spin-correlation effects while possibly violating gauge invariance of
the overall amplitude. An example for the production and decay of
top quarks in electron-positron annihilation reads
\begin{lstlisting}[style=runcard,numbers=none]
  (processes){
    % enforce intermediate top-quarks
    Process 11 -11 -> 6[a] -6[b];
    % decays t -> bW
    Decay 6[a] -> 5 24[c];
    Decay -6[b] -> -5 -24[d];
    % decays W+ -> mu+ nu, W- -> qqb'
    Decay 24[c] -> -13 14;
    Decay -24[d] -> 94 94;
    End process;
  }(processes)
\end{lstlisting}

Alternatively, employing a strict narrow--width-type factorisation of 
production and decay, resonances can be produced as external particles
and then decayed through separate decay matrix elements.
By default, a posteriori the decay kinematics is adjusted to a Breit--Wigner
distribution using the resonance's width.
Spin correlations are retained through the algorithm worked out
in~\cite{Collins:1987cp,Knowles:1987cu,Knowles:1988vs,Richardson:2001df}.
For the latter case, \Sherpa automatically constructs the decay
tables and computes the partial widths and branching ratios at tree level.
It is possible for users to overwrite any of the automatically generated
branching ratios, and to enable or disable any subset of decay channels.
This can be useful, for example, to include NLO K-factors or to better
match known (and measured) branching ratios. The setup corresponding to
top-quark production and decay from above in the factorised approach reads:
\begin{lstlisting}[style=runcard,numbers=none]
  (run){
    % enable hard decays
    HARD_DECAYS On;
    % enforce decay W+ -> mu+ nu
    HDH_STATUS[24,-13,14] 2;
    % switch off decays W- -> l- nu
    HDH_STATUS[-24,-12,11] 0;
    HDH_STATUS[-24,-14,13] 0;
  }(run)

  (processes){
    % produce final-state tops, decay through hard decay module
    Process 11 -11 -> 6 -6;
    End process;
  }(processes)
\end{lstlisting}

\paragraph{NLO Calculations and One-Loop Providers}
The inclusion of NLO QCD corrections to a given scattering process has become
a de-facto standard in today's event generators, including their matching to 
parton showers. As both virtual and real-emission corrections are separately
infrared divergent, a cancellation procedure is required. In \Sherpa
this has been realised for the first time through the automation of the
Catani--Seymour dipole subtraction formalism~\cite{Catani:1996vz,Catani:2002hc}
in~\cite{Gleisberg:2007md}. Renormalised QCD virtual corrections are obtained 
either through dedicated interfaces from programs and libraries like
\BlackHat~\cite{Berger:2008sj,Berger:2009ep,Berger:2010vm,Berger:2010zx},
\Madgraph~\cite{Hirschi:2011pa,Alwall:2014hca},
\OpenLoops~\cite{Cascioli:2011va} and
\Recola~\cite{Actis:2012qn}, or through the generic Binoth Les-Houches Accord
interface~\cite{Binoth:2010xt,Alioli:2013nda} from codes like 
\Gosam~\cite{Cullen:2011xs,Cullen:2014yla} or \NJet~\cite{Badger:2012pg}. 
An example process declaration including the evaluation of NLO QCD corrections 
in a fixed-order computation reads:
\begin{lstlisting}[style=runcard,numbers=none]
  (processes){
    Process 93 93 -> 11 -11 93 93;
    % use asterisk wild card for strong coupling 
    Order (*,2);
    % evaluate NLO QCD corrections in fixed order scheme
    NLO_QCD_Mode Fixed_Order;
    % include Born (B), Virtual (V), Integrated Subtraction (I), Real (R) and Subtraction terms (S)
    NLO_QCD_Part BVIRS;
    Loop_Generator <One-Loop Provider>;
    End process;
  }(processes)
\end{lstlisting}

\noindent
Examples of NLO QCD calculations performed with \Sherpa include:
\begin{itemize}
\item vector-boson production with up to five jets at NLO QCD~\cite{Bern:2013gka,Anger:2017nkq},
\item Higgs-boson production in association with up to three jets, taking into account
  finite-mass corrections~\cite{Greiner:2016awe},
\item top-quark pair production with up to three jets~\cite{Hoche:2016elu},
\item diphoton plus up to three jets production~\cite{Badger:2013ava,Bern:2014vza},
\item up to five-jet production at the \LHC~\cite{Bern:2011ep,Badger:2012pf,Badger:2013yda}\,.
\end{itemize}

The generalisation of the subtraction formalism to electroweak corrections 
has been implemented in~\cite{Schonherr:2017qcj} and renormalised electroweak one-loop 
corrections can at present be obtained from \Gosam, \OpenLoops and \Recola. 
They are, however, not yet available in \Sherpa-2.2. Processes that have
already been evaluated at full EW one-loop order with a development version of
\Sherpa\ include:
\begin{itemize}
\item three-jet production at the \LHC~\cite{Reyer:2019obz},
\item four-lepton production~\cite{Kallweit:2017khh,Biedermann:2017yoi,Bendavid:2018nar},
\item $t\bar{t}h$ production~\cite{Biedermann:2017yoi},
\item $W$ ($Z$) production with up to three (two) jets~\cite{Kallweit:2014xda,Biedermann:2017yoi},
\item $\gamma\gamma W$ and $\gamma\gamma Z$ production~\cite{Greiner:2017mft},
\item and $\gamma\gamma j$ production~\cite{Chiesa:2017gqx}.
\end{itemize}

\paragraph{NNLO QCD Calculations}
For a few phenomenologically highly relevant processes \Sherpa allows for the
evaluation at NNLO QCD precision, using the $q_\mathrm{T}$-slicing 
method based on the ideas of \cite{Catani:2007vq,Catani:2009sm}.
QCD NNLO cross sections can be computed in \Sherpa for neutral and charged
current Drell--Yan processes~\cite{Hoeche:2014aia,Alioli:2016fum} and for
Higgs-boson production~\cite{Hoche:2014dla}.\footnote{
  A development version of \Sherpa has also been used to compute NNLO cross sections
  for di-boson production at hadron colliders~\cite{Dawson:2016ysj}.}
Note, the NNLO facilities are not distributed with 
the public code releases, but can be obtained in the form of plugins
from \url{http://www.slac.stanford.edu/~shoeche/pub/nnlo/}.

\paragraph{External Matrix Elements}
\Sherpa provides a generic interface to external amplitude generators, that
can be used in particular to compute cross sections for loop-induced processes,
like \eg $gg\to W^+W^-$ or $gg\to HH$. For matrix elements provided by
\OpenLoops~\cite{Cascioli:2011va}, the interface is fully automated and
can be used as a blueprint to access other external MEGs as well. 

\noindent
For the phase-space integration of externally provided matrix elements, a set
of process-specific phase-space generators is available in \Sherpa. If they
need to be extended, a phase-space generator for a process with similar
characteristics can be generated with \Amegic and then used as a plug-in.
Alternatively, phase space can be sampled uniformly, using the \Sherpa
implementation of the \Rambo algorithm~\cite{Kleiss:1985gy}.


\clearpage
\subsection{Parton Showers}
\label{Sec::PS}

QCD parton showers form an indispensable part of any multi-purpose
event generator. They account for the successive emission of QCD or QED quanta
off the initial- and final-state partons of the hard process.
In doing so, showers relate a few-parton hard-scattering
configuration at momentum scale $Q^2_{\text{hard}}$ to a set of partons with
typical inter-parton separation scales down to $Q^2_0\approx 1\,\text{GeV}^2$. This
solves the evolution of arbitrary hard-scattering processes from high
to low scales, where ultimately a non-perturbative hadronisation process sets
in, transforming the final-state partons into primary hadrons. 

Formally, parton showers provide approximate numerical solutions for the
all-orders resummation of large kinematical logarithms. A statement on
the logarithmic accuracy for an arbitrary observable evaluated with a shower
algorithm cannot easily be made. However, in recent years investigations on
the correspondence of parton showers to resummation approaches have been
fruitful, see for instance~\cite{Hoeche:2017jsi,Dasgupta:2018nvj,Bewick:2019rbu}. 

Furthermore, the need to match parton showers to higher-order matrix elements,
in particular multi-leg tree-level or one-loop matrix elements, has served as a
development paradigm. This raises issues about matching the exact singularity and
colour structure of QCD matrix elements, preserving their fixed-order accuracy,
without compromising on the resummation property of the parton shower. This has
for instance led to the formulation of shower algorithms based on NLO QCD
infrared subtraction schemes.

\Sherpa comprises two different parton showers, based on different construction
paradigms, and implementing different ways to fill the phase space for multiple
emissions of secondary particles.

\paragraph{\CSS}
The default shower of the \Sherpa-2 series is based on Catani--Seymour dipole
factorisation~\cite{Catani:1996vz,Catani:2002hc}, first
proposed in~\cite{Nagy:2006kb}. The technique was implemented in
\Sherpa~\cite{Schumann:2007mg} and at the same time in~\cite{Dinsdale:2007mf},
building on a set of generic operators for particle emission off a dipole in
unintegrated and spin-averaged form in the large-$N_c$ limit.  Each dipole
contains a splitting parton and a colour-connected spectator parton. The shower
evolves through sequential splittings of such dipoles.
In the \Sherpa implementation, all QCD splittings within the Standard
Model and the MSSM, as well as the emission and splittings of photons are
incorporated, evolving QCD and QED quanta on an equal footing~\cite{Hoeche:2009xc}.
Note that the \CSS fully supports finite-mass effects. This is important in particular
for the production and evolution of $b$-quarks~\cite{Schumann:2007mg,Hoeche:2009xc},
thus allowing for systematic studies of $b$-quark associated/initiated processes
in the four- and the five-flavour scheme~\cite{Krauss:2016orf,Krauss:2017wmx}. 
With this, the \CSS, in essence, implements a fully-differential general-mass 
variable flavour-number scheme (GM-VFNS), generating massive quark thresholds 
through momentum conservation.
Furthermore, general electroweak splittings are implemented in
\CSS~\cite{Krauss:2014yaa}. However, as the chirality of
fermions can currently only be treated in an approximated form, these splittings
are disabled by default.

\noindent
In the dipole picture of the \CSS, soft-gluon emissions are mapped onto
two dipoles, which consist of the same partons, but with the roles of spectator
and emitter interchanged. The splittings are ordered by their associated
transverse momenta. For final-state splitters, this is the transverse
momentum between the two daughters, whereas for initial-state splitters the
transverse momentum is taken with respect to the emitting beam particle. In
contrast to the original formulation~\cite{Schumann:2007mg}, where the
kinematics of the Catani--Seymour formalism is used, in the current default configuration
recoil from the emission is either compensated by the spectator if the emitter
is a final-state parton, or otherwise distributed equally among all final-state
particles. This modified recoil scheme was first proposed in~\cite{Platzer:2009jq} and
refined to include massive partons in~\cite{Hoeche:2009xc}. It is crucial
to obtain reliable predictions for Deep Inelastic Scattering (DIS) processes~\cite{Hoeche:2009xc,Carli:2009cg}.

\noindent
The above choices were made with the matching and merging of the shower with
hard matrix elements in mind (these techniques are described in
Sec.~\ref{sec:matchmerge}). Building the splitting kernels on top of the
subtraction formalism used to calculate NLO matrix elements allows one to write the
\MCatNLO formalism in the most simple form. Using the transverse momentum as the
ordering variable removes the need to veto splittings with scales that are larger
than the scale set by the hard process.  And finally, local energy-momentum conservation
enables the translation of a multi-leg matrix element into a history of parton-shower
emissions, which is needed for attaching showers to multi-parton amplitudes~\cite{Hoeche:2009rj}.

\paragraph{\Dire}

The second parton shower implemented in \Sherpa  is \Dire~\cite{Hoche:2015sya},
it presents a hybrid between the colour-dipole picture~\cite{Gustafson:1986db} and
standard collinear parton evolution. Similar to the \CSS, it is based on
Catani--Seymour dipole subtraction~\cite{Catani:1996vz,Catani:2002hc}, but 
uses the inverse of the soft eikonal as evolution variable.
The soft-enhanced part of the splitting functions is defined by a
partial fraction of the soft eikonal of the colour dipole~\cite{Catani:1996vz},
giving the correct soft-anomalous dimension at one-loop order.
The collinear remainder of the splitting kernels is determined by the constraint
that they reproduce the known collinear anomalous dimensions, while respecting
flavour and momentum sum rules.

\noindent
The resulting splitting functions can be negative, leading to negative emission
probabilities which necessitate the weighted Sudakov veto algorithm, introduced
in~\cite{Hoeche:2009xc,Platzer:2011dq,Lonnblad:2012hz}. The negative prefactor
is then moved to an analytic event weight. In the same way, \Dire can also deal
with negative values of PDFs without resorting to an unphysical emission cut-off.
The event-weight variance imposed by this approach is typically small.

\noindent
\Dire uses the same recoil strategy as \CSS, and as for \CSS, massive partons are
supported, with the additional construction principle that the evolution variable
of \Dire is still mapped to the soft-enhanced term of the full matrix element.
A unique feature of \Dire is that it has also been implemented in
\Pythia~\cite{Sjostrand:2014zea}, allowing extensive cross validation
between the two generators thus enabling stringent consistency checks of
event samples produced for experimental analyses.

\noindent
Within the framework of \Dire, it has been shown that triple-collinear and
double-soft NLO corrections to the splitting functions can consistently be
included in a parton shower~\cite{Hoche:2017iem,Dulat:2018vuy}. A complete
treatment of higher-order corrections will be available in a future version
of \Sherpa.


\subsection{Matching and Merging}\label{sec:matchmerge}

Having discussed the methods used for calculating hard-scattering matrix elements and
parton showering raises the question how to combine these two complementary approaches,
while preserving their respective strengths. Consider a well-defined inclusive
$n$-jet type observable. A tree-level calculation at ${\cal{O}}(\alphaS^n)$
will typically provide the lowest-order prediction. Subsequent emissions from a parton
shower provide a (leading) logarithmic approximation for the higher jet rates,
preserving the leading-order $n$-jet rate. In contrast, an exact NLO QCD calculation, \ie
including the virtual and real corrections, yields an NLO accurate prediction for the
$n$-jet cross section, while the $(n+1)$-jet rate is approximated to leading order.

{\em Matching} matrix elements and parton showers resolves the double-counting of the NLO
corrections in the matrix-element calculation with the first parton-shower emission.
Multijet {\em merging}, on the other hand, allows for the combination of final states of increasing
matrix-element parton multiplicity, evolved by a parton shower, into an inclusive description.
This enables prediction for higher jet rates at NLO or LO accuracy, depending on the order of the
underlying matrix-element calculation, up to a certain maximum matrix-element parton
multiplicity. Yet higher jet numbers are accounted for by the parton shower off the
highest-multiplicity matrix element. 

Multijet merging, first introduced in~\cite{Catani:2001cc,Krauss:2002up}, has been
one of the cornerstones of \Sherpa since its inception.  Promoting the idea underlying
multijet merging to the inclusion of higher-order matrix elements builds on the exact
matching of these matrix elements to the subsequent parton showering, delivering
precise simulations in their own right. In this section we describe the methods for
matching and merging in \Sherpa, including the incorporation of NNLO QCD corrections
for a few processes and means to account for approximate NLO electroweak contributions 
relevant in particular in high--momentum-transfer regions. 

\paragraph{Matching of NLO Matrix Elements and Parton Showers}
For the matching of next-to leading order matrix elements, \Sherpa uses a variant of
the \MCatNLO method~\cite{Frixione:2002ik}. Its basic idea is the realisation that parton showers
organise their radiation pattern, and in particular the first emission, by identifying
and factorising the singular soft and collinear limits of the emission matrix elements.
In parton showers the notion of a resolution parameter in the emission phase
space of the secondary quanta regularises the singularities, leading to the appearance of
logarithms in the cut-off parameter. In NLO calculations, however, these singular terms
must be identified and subtracted from the real-emission matrix elements.
This enables the decomposition of the parton-level calculation into two parts with well-defined,
finite cross sections: an infrared-subtracted real-emission contribution, where the
subtraction is identified as the first parton-shower emission off an underlying Born
configuration, and a part consisting of the original Born-level calculation supplemented
with the virtual correction and the integrated infrared subtraction terms, both of which
share the same Born kinematics.  Parton showers are attached to both parts, with
starting conditions reflecting the respective kinematics. In \Sherpa, this idea has been
recast in a form that maximises the benefit of using identical kernels for infrared
subtraction and parton showering~\cite{Hoeche:2011fd}.

\clearpage
\noindent
In the past decade, the \MCatNLO matching in \Sherpa has been continuously developed and
refined. Specific aspects and applications have been discussed in a series of dedicated
publications, including pure jet production at the \LHC~\cite{Hoche:2012wh}, the hadronic
production of electroweak gauge bosons and up to three jets~\cite{Hoeche:2012ft},
$t\bar{t}b\bar{b}$ production~\cite{Cascioli:2013era}, $s$- and $t$-channel single-top
production~\cite{Bothmann:2017jfv} or Higgs-boson pair production~\cite{Jones:2017giv}.
The \MCatNLO approach is nowadays routinely used in Standard Model simulations with
\Sherpa. Furthermore, it forms the basis for all {\em merging} approaches involving
NLO QCD matrix elements. To give an example, the process definition of an \MCatNLO
matched simulation of Drell--Yan lepton-pair production in association with two
jets reads:
\begin{lstlisting}[style=runcard,numbers=none]
  (run){
    ...
    SHOWER_GENERATOR CSS
    ...
  }(run)
  (processes){
    Process 93 93 -> 11 -11 93 93;
    Order (*,2);
    % evaluate NLO QCD corrections in MC@NLO scheme
    NLO_QCD_Mode MC@NLO;
    Loop_Generator Recola;
    End process;
  }(processes)
\end{lstlisting}
Note, selector definitions similar to the ones stated in Sec.~\ref{Sec::MEs} apply
here as well. 

\paragraph{NNLO Matrix Elements and Parton Showers: First Steps} 

Using the \UNNLOPS method proposed in~\cite{Hoche:2014dla}, and relying on 
$q_T$-slicing~\cite{Catani:2007vq,Catani:2009sm} to regulate the additional 
infrared singularities, it is possible to also include NNLO-correct
matrix elements for the production of colour-singlets at hadron colliders into a parton shower
framework. In \Sherpa this has been achieved for two processes, Drell--Yan and Higgs-boson
production~\cite{Hoche:2014dla,Hoeche:2014aia}, thereby providing an important alternative
to the \MINLO-based implementations of~\cite{Hamilton:2013fea,Karlberg:2014qua}. 
More recently, the application to hadronic final state
production in Deep Inelastic Scattering~\cite{Hoche:2018gti} has been discussed. 
However, here the projection-to-Born method~\cite{Cacciari:2015jma}, rather 
than the $q_T$-slicing technique has been used to regulate the additional 
infrared singularities appearing at NNLO. 
Note, the NNLO+PS facilities are not distributed with the public code releases.
The Drell--Yan generator can be obtained from
\url{http://www.slac.stanford.edu/~shoeche/pub/nnlo/}.

\paragraph{Multijet Merging at LO and NLO}
The multijet-merging approach uses the notion of jets -- usually defined through a
$k_T$-type measure -- to classify emissions as either jet production or jet evolution,
and to additively combine towers of exact matrix elements with increasing jet
multiplicities into one inclusive sample. Denoting the separation scale of 
both regimes as $Q_{\text{cut}}$,
emissions with $Q\ge Q_{\text{cut}}$ get accounted for by exact matrix elements, while
radiation with $Q<Q_{\text{cut}}$ is described by the parton shower instead. In turn,
hard jet-emission configurations will follow the fixed-order matrix-element kinematics, while
the inner-jet evolution and the production of additional softer jets is in the realm
of the parton shower's emission kernels. 
The resummation of emission-scale hierarchies is provided by the parton shower 
in both regimes.

The classification into two disjoint, complementary regimes avoids the explicit
double-counting of emissions, while the logarithmic accuracy of the parton 
shower is recovered by both the matrix elements and the parton shower's emission 
kernels having the same infrared limits (at leading \Nc) and using the 
parton shower's resummation in both regions. Originally these ideas have been proposed
for the combination of tree-level matrix elements in~\cite{Catani:2001cc,Krauss:2002up},
and have been implemented, in variations, in all event
generators~\cite{Lonnblad:2001iq,Mangano:2001xp,Alwall:2007fs,Lavesson:2007uu,
  Hoeche:2009rj,Hamilton:2009ne,Hamilton:2010wh,Hoeche:2010kg,Lonnblad:2011xx,
  Lonnblad:2012ng}. A dedicated comparison can, for example, be found in~\cite{Alwall:2007fs}.

\clearpage
The \Sherpa merging algorithm for tree-level matrix elements, called \MEPSatLO, has been
detailed in~\cite{Hoeche:2009rj}. It relies on a truncated parton shower, \ie the
shower explicitly generating the Sudakov form factor for lines between reconstructed
matrix--element-type emissions. Broadly speaking the algorithm proceeds as follows: 

\begin{itemize}
\item initial cross sections for the multijet matrix elements to be considered are
  evaluated,
\item according to the total cross section a specific jet multiplicity is picked, then
  a flavour channel and an event kinematics are generated,
\item for the given flavour assignment and kinematics a clustering algorithm
  is applied that inverts the parton shower algorithm until a unique
  core process and subsequent emission scales in the full matrix-element 
  configuration are identified,
\item a scale choice is made for the strong-coupling factors, comprising the respective 
  contributions for both the identified core process and the individual emission 
  scales, identical to those used in the parton shower,
\item the truncated parton shower starts from the core configuration,
  reconstructing the identified matrix-element emissions when the shower
  evolution parameter crosses their pre-determined scales,
  and the event is vetoed when the
  parton shower produces an emission above the resolution scale $Q_{\text{cut}}$, 
  implementing the Sudakov factor of the parton-shower resummation in the 
  matrix-element region.
\end{itemize}

This procedure allows one to add event configurations exclusive in the emission scale down
to $Q_{\text{cut}}$ into an inclusive sample, thereby cancelling the dependence on the
separation parameter to the logarithmic accuracy of the parton shower. Note, the sample
of highest matrix-element multiplicity has to be exclusive down to the lowest matrix-element
emission scale only, \ie $Q^{\text{ME}}_{\text{last}}\ge Q_{\text{cut}}$.

The well-established LO approach has been promoted to include matrix elements at
NLO accuracy in QCD, called \MEPSatNLO, and implemented in \Sherpa in 
\cite{Hoeche:2012yf,Gehrmann:2012yg,Hoeche:2014rya}. 
It combines \MCatNLO matched samples of increasing jet multiplicities, 
separated by the resolution parameter $Q_{\text{cut}}$ into an inclusive sample. 
In general, the approach follows the outline above, only the usual care with 
overlapping descriptions through NLO matrix elements and parton showers 
is taken, and any overlaps are carefully removed to fully maintain 
the respective accuracies throughout. 
Other formulations and approximations thereof have been 
presented in \cite{Lavesson:2008ah,Frederix:2012ps,Lonnblad:2012ix,Platzer:2012bs}. 

\paragraph{NLO EW Corrections in Matching and Merging}
In \cite{Kallweit:2015dum,Gutschow:2018tuk}, an approach to incorporate 
approximate electroweak and subleading mixed QCD-EW corrections into the above 
described \MEPSatNLO QCD method was introduced, dubbed \MEPSatNLO \QCDpEWapprox. 
There, the Born-like input cross section into the \MCatNLO matched samples 
of the multijet-merged calculation are supplemented with exact NLO EW 
renormalised virtual corrections as well as approximated NLO EW real-emission
corrections integrated over their real-emission phase space. 
This approximation is tailored to reproduce the exact NLO EW corrections 
in regions with large momentum transfers where they are dominated by 
virtual weak-boson exchanges and renormalisation corrections. 
The integrated-out real-photon emission part of the electroweak correction, 
which are of prime importance for leptonic final states, are recovered 
in a full event simulation by including a soft-photon resummation, 
\cf Sec.\ \ref{sec:yfs}.
Subleading tree-level contributions may be added where relevant. 

The \MEPSatNLO approach defines the current standard in simulating QCD-associated
Standard Model production processes with \Sherpa. Examples of validation and application
of the method include:
\begin{itemize}
\item $V$+jets production with up to two jets described at NLO QCD and
  approximate NLO EW \cite{Kallweit:2015dum},
\item $h$+jets production with up to three jets described at NLO QCD and
  5 jets at LO \cite{Hoeche:2014lxa,Badger:2016bpw},
\item four-lepton production~\cite{Cascioli:2013gfa},
\item triple vector boson production \cite{Hoeche:2014rya},
\item Higgs production in association with a gauge boson \cite{Hoeche:2014rya},
\item application to loop-induced production processes~\cite{Goncalves:2015mfa,Cascioli:2013gfa},
\item and top-quark pair production in association with up to three jets~\cite{Hoeche:2014qda}.
\end{itemize}

In Sec.~\ref{sec:results} we illustrate results for a variety of processes based on the
matching and merging of matrix-element elements and parton showers and compare
them with actual data from the \LHC. 

\noindent
We close this section with an example for the process setup of Drell--Yan production in
association with QCD jets, based on NLO QCD matrix elements for up to two jets:
\clearpage
\begin{lstlisting}[style=runcard,numbers=none]
  (processes){
    % process definition: Drell-Yan + 0,1,2 jets @ NLO QCD
    Process 93 93 -> 11 -11 93{2};
    Order (*,2);
    % merging scale parameter corresponding to Qcut=30 GeV
    CKKW sqr(30/E_CMS);
    NLO_QCD_Mode MC@NLO;
    RS_ME_Generator Comix;
    Loop_Generator OpenLoops;
    % include approx. EW corr. and first subleading tree-level 
    Associated_Contributions EW|LO1;
    End process;
  }(processes)
\end{lstlisting}


\subsection{Internal Reweighting}\label{sec:reweighting}

The advancements of state-of-the-art QCD calculations as described in this
publication led to a considerable growth in computational cost per event, a
limiting factor in current and future applications of event generators.
One place where this cost can be addressed relatively easily is in studies targeting
theory uncertainties for QCD input parameter and scale choices. Traditionally, this
involved re-running the whole event-generation chain with different PDFs, values for
the strong coupling $\alphaS$, or with varied choices for the renormalisation and
factorisation scales $\mu_{R,F}$. Nowadays this is achieved by appropriately reweighting
the default prediction, significantly reducing the computational costs. Furthermore,
\Sherpa\ allows for a reweighting of the nominal NLO QCD calculation to include the
associated approximate NLO EW corrections and subleading tree-level contributions.

\paragraph{Implementation}
The parameter-reweighting techniques available in \Sherpa have been described
in~\cite{Bothmann:2016nao}. Like in other generators, \cf~\cite{Mrenna:2016sih,Bellm:2016voq},
they are calculated \emph{on-the-fly} and cover scale variations, different PDF
choices and modified values for coupling constants. They can furthermore include the effects
these choices have on the parton shower, without rerunning it. The shower-emission
reweighting uses the generalised Sudakov Veto Algorithm presented in~\cite{Hoeche:2009xc}.  
Relative weights that emerge from different choices of these inputs are provided either
in the \HepMC event output~\cite{Dobbs:2001ck} or directly passed through the internal
interface to the \Rivet analysis framework~\cite{Buckley:2010ar}. Especially when the
events are stored on disk, this also reduces the necessary disk space by potentially large
factors, replacing full events for each variation by single numbers.

Reweighting in \Sherpa can be applied to fixed-order calculations, both at LO
and NLO using the \NTUPLE decomposition~\cite{Bern:2013zja}.
When applied to matched or merged calculations, both the \CSS and \Dire parton
showers are supported. A lower bound on the parton-shower evolution scale can be set to omit
the reweighting of very soft emissions. This allows for a trade-off between speed
and accuracy. 

An extensive example that invokes 7-point scale variations, variations over
several PDF sets (including all their error replica/eigenvector PDFs, and sets
with varied $\alpha_S(m_Z)$), and adding electroweak corrections as separate
variations, is given by the following snippet:
\begin{lstlisting}[style=runcard,numbers=none]
  (run){
    % pairs of factors multiplying default squared scales muR,muF   
    SCALE_VARIATIONS 0.25,0.25 0.25,1. 1.,0.25 1.,1. 1.,4. 4.,1. 4.,4.;
    % event variation for given (error) sets of PDFs
    PDF_VARIATIONS CT14nlo[all] MMHT2014nlo68cl[all] NNPDF30_nlo_as_0118[all] \
                     NNPDF30_nlo_as_0115 NNPDF30_nlo_as0121;
    % reweight nominal QCD to QCD+EW and QCD+EW+subLO
    ASSOCIATED_CONTRIBUTIONS_VARIATIONS EW EW|LO1;
    % enable consistent variations of parton-shower splittings
    CSS_REWEIGHT 1;
    % reweight the alpha_s that multiplies the splitting probability
    REWEIGHT_SPLITTING_ALPHAS_SCALES 1;
    % reweight the PDF ratios for initial-state splittings
    REWEIGHT_SPLITTING_PDF_SCALES 1;
  }(run)
\end{lstlisting}
Note that although the production will be considerably faster compared to
producing separate event samples for each variation, the inclusion of hundreds
of variations (as in the example above) can still slow down the production
significantly, especially when enabling the parton-shower reweighting.


\subsection{Initial State Radiation and PDFs}\label{sec:isr}

As a multi-purpose generator, \Sherpa can be used to simulate collisions for
various different collider setups, \eg $pp$, $e^+e^-$, $ep$ or $\gamma\gamma$, or,
more exotically, $\mu^+\mu^-$. This requires in particular the proper
modelling of beam spectra and (partonic) substructures. 

\paragraph{Beam Particles}
\Sherpa allows for a two-step definition of particles entering a hard
interaction: \texttt{BEAM} particles are specified, which may be subjected to a
spectrum, modifying their energy, or, possibly get converted to other particles,
that are refered to as \texttt{BUNCH}.
For the latter, two examples are available in \Sherpa, namely
\begin{itemize}
\item Laser Backscattering, where initial beam leptons are ``converted'' into
  bunch photons through Compton
  scattering~\cite{Badelek:2001xb,Zarnecki:2002qr,Archibald:2008zzb}; and
\item equivalent photons in the Weizs\"acker--Williams approximation,
  where the beam particles act as quasi classical sources of collinear
  photon fluxes \cite{vonWeizsacker:1934sx,Williams:1934ad,Budnev:1974de}.
\end{itemize}

\noindent
By default initial beams are considered monochromatic and will directly enter
the second stage, where their potential substructure is resolved.

\paragraph{Encoding Partonic Structure: Available PDFs}
The emerging beam particles, that initiate the hard scattering, may feature a
partonic structure, described by a parton distribution function (PDF). This in
particular applies to protons, photons, or leptons, whose constituents then
form the initial states of the matrix-element calculations described in
Sec.~\ref{Sec::MEs}.

\noindent
For these beam particles \Sherpa provides built-in PDFs that are shipped with
the code, namely
\begin{itemize}
\item various proton PDFs, in particular the default set NNPDF~3.0~\NNLO~\cite{Ball:2014uwa}, 
\item the GRV leading-order photon PDF set~\cite{Gluck:1991ee,Gluck:1991jc},
\item and an analytic QED lepton structure function in different approximations~\cite{Nicrosini:1986sm,Berends:1987ab,Blumlein:2011mi,Blumlein:2019srk}.
\end{itemize}

\noindent
In addition, \Sherpa can be built with an interface to the \LHAPDF
library~\cite{Whalley:2005nh,Buckley:2014ana}, allowing the user ample choice
in particular of proton PDFs, including their respective error and variational
sets. 

\noindent
An example beam setting, assuming proton-proton collisions at $\sqrt{s}=13$ TeV using
the MMHT 2014 NLO PDF set~\cite{Harland-Lang:2014zoa} via \LHAPDF reads:
\begin{lstlisting}[style=runcard,numbers=none]
  (run){
    BEAM_1 2212; BEAM_ENERGY_1 6500.;
    BEAM_2 2212; BEAM_ENERGY_2 6500.;

    PDF_LIBRARY LHAPDFSherpa;
    PDF_SET MMHT2014nlo68cl;
    % use the PDF implementation of running aS
    USE_PDF_ALPHAS 1;
  }(run)
\end{lstlisting}

In setups which include PDFs, \Sherpa automatically uses a consistent value of
$\alpha_S$ and order of its running throughout the event generation. When
using \LHAPDF, it is optionally possible to use the actual implementation of
the running within the given PDF library.


\clearpage
\subsection{Higher-Order QED and EW Corrections to Decays}
\label{sec:yfs}

Higher-order QED and electroweak corrections can be computed in \Sherpa
using the soft-photon resummation of Yennie, Frautschi and Suura
(YFS)~\cite{Yennie:1961ad}, which exploits the universal structure of
soft real and virtual photon emissions to construct an all-order
approximation while all mass effects are retained.  
The implementation in \Sherpa~\cite{Schonherr:2008av,Krauss:2018djz} 
focusses on higher-order corrections to particle decays, both for 
elementary particle decays (e.g.\ $\PWpm$, $\PZ$, $\PH$, $\tau^\pm$) as well 
as for hadron decays. 

\paragraph{Implementation}
The soft-photon resummed higher-order QED corrections in \Sherpa are applied 
to decay processes that involve colourless particles only, while those that 
involve coloured particles -- quarks and gluons -- are subjected to a regular
parton shower.  By default, exact first-order QED corrections are applied to $\PZ\to\Pl\Pl$, 
$\PW\to\Pl\nu$, $\PH\to\Pl\Pl$ and $\tau\to\Pl\nu_\Pl\nu_\tau$ and some hadron
decays~\cite{Schonherr:2008av,Badger:2016bpw,Alioli:2016fum,Bernlochner:2010fc}.
In all other cases, the eikonal approximation underlying the all-orders
resummation is corrected in the hard collinear emission regime through
subtracted Catani--Seymour dipole splitting functions~\cite{Schonherr:2008av}. 

\paragraph{Treatment of Resonances}
To meaningfully dress the complex final state of a hard scattering process 
with QED radiation it is mandatory to preserve its internal resonance 
structures.  In \Sherpa this is achieved through universal resonance identification 
described in~\cite{Kallweit:2017khh}.  It identifies all possible resonances by
first scanning the final state of a scattering process for possible 
recombinations into resonant states present in the employed physics model. Then, all
possible combinations are ordered by the difference of invariant mass of
the decay products and the mass of the resonance, scaled by its width:
$\Delta=|m_\text{kin}^\text{inv}-m_\text{res}|/\Gamma_\text{res}$. Resonances are
identified in ascending order of $\Delta$, and configurations with $\Delta>\Delta_\text{res}$
are classified as a non-resonant production of the respective final state, where the
arbitrary parameter $\Delta_\text{res}$ is set to 10 by default.  The kinematics of the
radiation off the thus identified resonant decay is subject to the condition that
the invariant mass of the system is maintained.  Non-resonantly produced final states
are corrected for QED effects using the universal YFS exponential coupled with universal
collinear-emission correction factors.

The main switches steering the YFS corrections are given by
\begin{lstlisting}[style=runcard,numbers=none]
  (run){
    % apply QED corrections to hard scattering - On/Off
    ME_QED On;
    % threshold \Delta_res to differentiate resonant and non-resonant regions
    ME_QED_CLUSTERING_THRESHOLD 10.;
    % general YFS switch: 0 - Off, 1 - soft photons only, 2 - soft and hard photons
    YFS_MODE 2;
    % apply exact first order QED matrix element corrections: 0 - Off, 1 - On
    YFS_USE_ME 1;
  }(run)
\end{lstlisting}


\subsection{Underlying Event and Beam Remnants}\label{sec:ue}

The inner structure and finite size of incident hadrons in collisions,
\eg at the \LHC, allow for effects beyond the hard process and secondary radiation.
These are collectively called the underlying event (UE). In particular, partons inside the
hadron may have some non-perturbative transverse momentum, and the break-up
of the hadrons will produce further colour charges that will have an impact on the
hadronisation of the partons. Furthermore, and maybe most prominently, it is
possible to have more than one parton--parton interaction per hadron--hadron
scatter. Such multiple parton interactions (MPIs) alter the overall particle
yield in collisions, and they influence observables such as jet rates and
jet shapes. The parameters introduced in the models addressing the underlying
event are subject to tuning and need to be determined by comparing generator
predictions to actual collider data, cf.~Sec.~\ref{sec:tuning}.

\paragraph{Modelling Multiple Parton Interactions}
The first model successfully simulating MPIs as the dominant effect in the UE was
proposed by Sj\"ostrand and van der Zijl in~\cite{Sjostrand:1987su},
and it is also the MPI model implemented in \Sherpa. It is based on partonic
$2\to 2$ QCD scatters and the observation that their cross section exceeds the
total hadronic cross section even for moderate transverse momenta above
$\sim$2--5 GeV. This is interpreted as having more than one parton--parton scatter per
hadronic collision.  The scatters are ordered by their transverse momentum,
acting as an ``evolution parameter'' for the UE, which dresses the primary
interaction with secondary scatters, through an expression similar to the
Sudakov form factor in the parton shower:
\begin{equation}\label{eq:ue_sudakov}
 P_{\text{no}}(p_{\perp,\mathrm{min}}) = \exp\left(-\frac{1}{\xi\sigma_{\mathrm{ND}}}\,\int\limits_{p^2_{\perp,\mathrm{min}}}
    \done p_\perp^2\frac{\done\hat\sigma}{\done p_\perp^2}\right)\,,
\end{equation}
where $\sigma_{\mathrm{ND}}$ is the non-diffractive hadron--hadron cross section.
Furthermore, $\hat\sigma$ denotes the parton-level $2\to 2$ scattering cross section,
including parton distribution functions, where the potential singular structure of
the differential cross section, introduced by the $t$-channel singularity in the
scattering amplitude at small momentum transfers and the divergent behaviour of
the strong coupling at small scales, is tamed by supplementing the transverse
momentum with a regulator $\ptzero$, \ie $\pt^2\to \pt^2+\ptzero^2$. The evolution
terminates when the transverse momentum of the secondary scatters falls below a cut-off
value $p_{\perp,\mathrm{min}}$, usually of the order of a few GeV. $\xi$ is a dimensionless
parameter, allowing to rescale the non-diffractive cross section.  

In their paper, Sj\"ostrand and van der Zijl also extended their model to describe
Minimum Bias events; this, however, is not realised in \Sherpa.

\paragraph{Implementation}
The Sj\"ostrand--van-der-Zijl model~\cite{Sjostrand:1987su}
has been implemented in \Sherpa by precalculating and tabulating the
partonic $2\to 2$ scattering cross sections, using the results for the
Sudakov-like factor driving the evolution of the MPIs in the transverse-momentum
scale. These tables are either calculated and stored or read in during the
initialisation phase of the run. \Sherpa uses all partonic channels in MPIs,
including processes with photons in the final state, and it supplements the
scatters with a parton shower that starts at the transverse momentum of the scatter.
The \Sherpa implementation also features an impact-parameter dependence, given by
the matter-density profile $\rho(r)$ of the incident hadrons. Available options
are a simple Gaussian, an exponential, and, the default, a double Gaussian profile
supporting a more compact matter core of radius $r_2$, containing the fraction
$f_{\text{mat}}$ of the hadronic matter, surrounded by a larger sphere of radius $r_1$:
\begin{equation}
  \rho(r) \propto (1-f_{\text{mat}}) \frac{1}{r^3_1}\exp\left(-\frac{r^2}{r^2_1}\right)+f_{\text{mat}} \frac{1}{r^3_2}\exp\left(-\frac{r^2}{r^2_2}\right)\,.
\end{equation}
The corresponding profile parameters $f_{\text{mat}}$, $r_1$ and $r_2$, as well as the
cut-off scale $p_{\perp,\mathrm{min}}$, the regulator $\ptzero$ and $\xi$ are subject of tuning
to reference data. The \Sherpa\ module for the underlying event is called \Amisic.

\paragraph{Intrinsic Transverse Momentum}
Partons inside hadrons are assigned a transverse momentum $\kt$ of the order of
up to a few $\Lambda_{\mathrm{QCD}}$. This is most visible for the case of Drell--Yan
production of lepton pairs at small transverse momenta.  There is a finite probability of
the parton shower ending with no emissions down to its cut-off scale of about $1$ GeV,
which would lead to a visible peak at zero combined transverse momentum of the lepton pair.
Instead the intrinsic $\kt$ washes out this unwanted and unphysical feature, and
marginally shifts the overall distribution.  In \Sherpa, the intrinsic $\kt$ of
partons in a beam hadron is chosen flavour- and $x$-independently according to a
Gaussian distribution, parametrised by a mean value and a width:
\begin{equation}\label{eq:ue_intrinsic_kt}
  \mathcal{P}(k_\perp) \propto \exp\left(-\frac{(k_\perp-\langle k_\perp\rangle)^2}{\sigma^2}\right)\,.
\end{equation}
It is applied to all partons stemming
from the hadron break-up: the initiators of the parton shower at the cut-off scale
for both the signal process and the MPIs as well as for all other partons that
are added to guarantee flavour sum rules.

\paragraph{Beam Remnants}
One subtlety in the modelling of the MPIs is the treatment of flavours and colours.
For the former, flavour sum rules must be respected, which may necessitate to
add extra quarks during the breakup of the hadron in the collision. Similarly,
starting from a colour-neutral hadron it is clear that the colours of the
partons must compensate each other, which offers some freedom in the colour
assignments. In \Sherpa this freedom is used to assign the colours such that the
total length of the colour connections in momentum space, parametrised through
the Lund measure~\cite{Andersson:1985qr,Sjostrand:1987su}, is minimal.

\noindent
For collisions that are not initiated by hadrons, the treatment of beam remnants is
significantly less involved; in the case of initial-state radiation off leptons or
similarly simple configurations, the beam remnant will be collinear to the incident beam
but with reduced energy.


\subsection{Hadronisation}\label{sec:hadronisation}

There are currently two successful approaches included in event generators
to describe the transition from the quanta of perturbative QCD, the quarks
and gluons, to the observable hadrons, namely the Lund string
model~\cite{Sjostrand:1982fn,Andersson:1983jt} used in
\Pythia~\cite{Sjostrand:2006za,Sjostrand:2014zea}
and cluster fragmentation models~\cite{Webber:1983if}, such as the ones implemented
in \Herwig~\cite{Gieseke:2003hm} and \Sherpa~\cite{Winter:2003tt}.

\paragraph{Underlying Principles}
In both models, the parton configurations coming from the parton showers,
underlying event and beam remnants are cast into the form of colour-connected
singlets, which will decay non-perturbatively. These decays proceed by
``popping'' flavour/anti-flavour pairs and inserting them into the singlet
structure, which in turn decays into more singlets with reduced masses.
The only flavours being allowed to be produced in this way are the light
$u$, $d$, and $s$ quarks and, possibly, diquarks made from them. The latter
are hypothetical bound states of two quarks or two anti-quarks, forming a
colour sextet or anti-sextet, which in the large-$N_c$ limit is re-interpreted
as a colour anti-triplet or triplet. The diquarks also carry the baryonic
quantum numbers -- in this picture baryons are bound states of a
quark and a diquark.

\noindent
The Lund string and the cluster fragmentation models differ in the logic in which the
non-perturbative flavour production proceeds. In the string model the singlets
form coloured lines (strings) of the type $qgg\dots gg\bar{q}$, which decay from
their ends into a hadron and a ``shorter'' string. The flavour necessary to
form the hadron is compensated by the anti-flavour of the new string end.
In contrast, in the cluster model, gluons decay non-perturbatively to
form colour-neutral quark/anti-quark or quark/diquark clusters. The clusters
are interpreted as massive hadron resonances and undergo binary decays, until
clusters are formed that are light enough to be hadrons.

\paragraph{Cluster Fragmentation in \Ahadic}
In \Sherpa, the cluster model is implemented in the module \Ahadic.  It starts
with non-perturbative gluon decays at the end of the perturbative phase which result
in the production of quark/anti-quark and of diquark/anti-diquark pairs.
In what follows the term ``quark'' is used such that it also includes diquarks.
Their selection is driven by the phase space available for them, defined by
their constituent masses, and by further flavour-specific suppression weights.

The splitting kinematics is realised in a dipole frame, where the gluons
remain massless and the necessary recoil is provided by the spectator object.
The splitting kinematics of the gluon decays is
defined by two parameters $y$ and~$z$, with distributions given by
\begin{eqnarray}
  \mathcal{P}(y) &\propto& y^\eta\,
  \exp\left(-\frac{\left(y\hat{s}-\frac{y}{(1-y)} m^2_{\mathrm{spect}}\right)-
      4m_{\mathrm{min}}^2}{4p^2_{\perp,0}}\right)\,,\\
  \mathcal{P}(z) &\propto& z^2+(1-z)^2\,,
\end{eqnarray}
where $\hat{s}$ is the mass of the splitter--spectator system, $m_{\mathrm{spect}}$ is the
mass of the spectator, $m_{\mathrm{min}}$ is a minimal mass of the resultant quark--anti-quark
system, given by their constitutent masses. The parameters $\eta$ and $p^2_{\perp,0}$ are
subject to tuning, cf. Sec.~\ref{sec:tuning}.

The $y$-dependent term in the exponential denotes the invariant mass squared of the
quark--anti-quark system, $m^2_{qq}$, which defines the allowed mass range of their flavours,
\begin{equation}
  m_{qq}^2 \equiv y\hat{s}-\frac{y}{1-y}m^2_{\mathrm{spect}}\,.
\end{equation}
In a centre-of-mass system, where the original gluon and spectator have momenta $p_A$ and $p_B$
oriented along the positive and negative $z$-axis, the momenta of the two quarks and the spectator after
splitting are then given by
\begin{eqnarray}
  \begin{array}{lcrcrcr}
    p_q^\mu       &=& (1-z)\left(1-\beta\right)\cdot p_A^\mu &+&
                     zy\cdot p_B^\mu \hphantom{\,,}&+&\hphantom{\,,} \vec{k}_\perp\,,\\
    p_{\bar{q}}^\mu &=& z\left(1-\beta\right)\cdot p_A^\mu &+&
                     (1-z)y\cdot p_B^\mu \hphantom{\,,}&-&\hphantom{\,,} \vec{k}_\perp\,,\\
    p_{\mathrm{spect}}^\mu &=& \beta\cdot p_A^\mu &+&
                     (1-y)\cdot p_B^\mu\,,&&
  \end{array}
\end{eqnarray}
where the terms involving $\beta = \frac{m^2_{\mathrm{spect}}}{\hat{s}(1-y)}$
ensure that the spectator is on-shell.
The transverse momentum $\vec{k}_\perp$ is distributed isotropically in the transverse plane.
Its absolute value is given by
\begin{equation}\label{eq:had_kt}
  k_\perp^2 = z(1-z)m_{qq}^2-m_q^2\,.
\end{equation}

After the gluon decays, \Ahadic proceeds with the formation of colour-neutral clusters,
by combining colour-connected quarks and anti-quarks. Depending on their mass, these clusters
either decay into hadrons or into further clusters.
For both types of decays, flavour pairs have to be
created again, using the same suppression weights as for gluon splittings.

For cluster decays
into two clusters, a similar kinematics is built, with a new, non-perturbatively ``popped''
quark--anti-quark pair.  Its mass is given by
\begin{equation}
  m_{qq}^2 = xy\hat{s}\,,
\end{equation}
where $x$ and $y$ are the momentum fraction taken away from
the splitter and the spectator partons within the decaying cluster, respectively.
Their distributions are again given by
\begin{equation}
  \mathcal{P}(x) = x^{\eta_x} P(m_{qq}^2) \quad \text{and} \quad
  \mathcal{P}(y) = y^{\eta_y} P(m_{qq}^2)\,,
\end{equation}
where
\begin{equation}
  P(m^2_{qq}) \propto \exp\left(-\frac{m_{qq}^2-4m^2_{\mathrm{min}}}{4p_{\perp,0}^2}\right)\,.
\end{equation}
The exponents $\eta_{x,y}$ are determined depending on whether the quark associated to it is
leading, \ie has been produced perturbatively, or not, and whether it is the ``splitter'' or the
``spectator'', cf. Sec.~\ref{sec:tuning}.

To fix the kinematics of the new quark pair, again an additional energy-splitting variable $z$
is uniformly selected, such that the four-vectors of splitter, spectator, and new
quarks are given by
\begin{eqnarray}
  \begin{array}{lcrcrcr}
    p_q^\mu             &=& zx\cdot  p_A^\mu  &+&  (1-z)y\cdot p_B^\mu \hphantom{\,,} &+& \hphantom{\,,} k_\perp^\mu\,,\\
    p_{\bar q}^\mu       &=& (1-z)x\cdot  p_A^\mu  &+&  zy\cdot p_B^\mu \hphantom{\,,} &-& \hphantom{\,,} k_\perp^\mu \,,\\
    p_{\mathrm{spect}}^\mu &=& \alpha(1-x)\cdot  p_A^\mu &+& (1-\beta)(1-y)\cdot  p_B^\mu\,,\\
    p_{\mathrm{split}}^\mu &=& (1-\alpha)(1-x)\cdot  p_A^\mu &+& \beta(1-y)\cdot  p_B^\mu\,,&&
  \end{array}
\end{eqnarray}
where $\alpha$ and $\beta$ are determined by the on-shell constraints of splitter and
spectator, $p^2 = m^2$, and the squared transverse momentum is given by Eq.~(\ref{eq:had_kt}).

Clusters that are too light will decay into two hadrons; this is determined by comparing the
cluster mass $M_c$ with a critical mass $M_{\mathrm{crit}}$ determined by a combination of the
masses of the lightest and heaviest hadron pairs, $M_-$ and $M_+$ that could emerge in the cluster
decay,
\begin{equation}
  M_{\mathrm{crit}} = M_-(1-\kappa) + M_+\kappa\,,
\end{equation}
with an off-set parameter $\kappa$. If the cluster made of quarks $q_1\bar{q}_2$ is lighter than $M_c$ it will decay into two hadrons;
the relative probabilities of an individual decay channel $C\to h_1h_2$ is determined by a product
of the ``popping'' probability of the necessary additional $q\bar{q}$ pair, $\mathcal{P}_q$, the
flavour component of the wave function of the two hadrons, $|\psi_{1,2}|^2$, their meson or baryon
multiplet weights, $\mathcal{P}_\text{multi}$, the decay phase-space weight,
and a mass-dependent factor with parameter $\chi$,
\begin{equation}
  \mathcal{P}(C\to h_1h_2) =
  \mathcal{P}_q\,|\psi_1(q_1\bar{q})|^2|\psi_2(q\bar{q}_2)|^2\,
  \mathcal{P}_{\mathrm {multi}}\,\sqrt{(M_c^2-m_1^2-m_2^2)^2-4m_1^2m_2^2}{8\pi M_c^2}\,
  \left(\frac{(m_1+m_2)^2}{M_c^2}\right)^\chi\,.
\end{equation}

Default values for the hadronisation parameters used to model gluon splittings, quark--anti-quark
pair creation and cluster decays in \Sherpa are compiled in App.~\ref{app:tune}.

\paragraph{Interface to Lund String Fragmentation}
In addition to its native cluster model implementation, \Sherpa also provides a
link to the Lund string fragmentation model implemented in \Pythia 6.4~\cite{Sjostrand:2006za}.
The parameters of this model can be directly set through the run cards steering \Sherpa.


\subsection{Hadron Decays}
\label{sec::HadronDecays}

Primary hadrons formed during the hadronisation stage are often unstable
and will decay further into secondary hadrons. The same is also true for the
$\tau$ lepton, which is unstable and predominantly decays into hadrons.
Since decay products are often unstable themselves, a cascade of decays
emerges.

\paragraph{Organisation of Decay Chains}
Hadron decays and their cascades in \Sherpa are handled by its \Hadrons
module in a recursive approach, based on individual $1\to n$ decays,
first simulated assuming the incident hadron is on-shell. Spin correlations
across the propagator of the decaying particle can be taken into account by
the algorithm introduced in~\cite{Richardson:2001df}.  Off-shell kinematics
is imposed a posteriori with a relativistic Breit--Wigner distribution,
through the application of a reverse Rambo algorithm~\cite{Kleiss:1985gy}
which shifts the momenta to their new mass shells while preserving momentum
conservation in the decay cascade. Since decaying particles have a finite
lifetime they will travel in space before they decay and the resulting vertex
offset is included in the simulation.

\paragraph{Decay Widths and Kinematics}
Due to the plethora of observed hadron decay channels and the limited
theoretical framework to predict them precisely, the decay tables are based
on measured branching ratios~\cite{Olive:2016xmw}. For some particles the
branching ratios of observed decays do not add up to unity. In such cases,
the branching ratios are rescaled within their known uncertainties
to add up to one. It is also possible that the known decay modes are not
sufficient -- this is particularly true for heavy mesons and baryons.
For them the known decay tables are amended with partonic decays of
one of the constituent quarks with subsequent parton showering and
hadronisation. This could of course lead to an exclusive final state already
present in the decay table -- in such a case the resulting hadronic final
state is vetoed and the procedure repeated, until a legitimate final state
is produced.

\noindent
The kinematics of each decay step is generated according to generic matrix
elements representing the spin structure of the involved particles.
Furthermore, in many cases and in particular for weak decays involving
hadrons, a wide variety of form-factor models are implemented, thus
parametrising the weak decay of quarks in a bound state beyond the generic
spin matrix elements.
State-of-the-art decay tables and form-factor implementations are provided
\eg for decays of the $\tau$, $B^0$, $B^\pm$, $B_s$, $B_c^\pm$, $D^0$,
$D^\pm$, $D_s$, $\Lambda_b$, $\Lambda_c^\pm$.


\subsection{Tuning non-perturbative model parameters}\label{sec:tuning}

The non-perturbative models used to address the parton-to-hadron transition,
the intrinsic motion of partons bound in composite initial states and the underlying
event involve a number of parameters, that are not determined by first principles.
Rather, they need to be adjusted through an iterative comparison of corresponding \Sherpa
predictions with experimental data. This tuning procedure is achieved in three consecutive
steps, implicitly assuming that the respective phases of the event generation factorise
sufficiently. Input to these models are well-defined perturbative matrix-element
calculations with parton showers attached, that, in turn, evolve the hard-process particles
into a parton ensemble with minimal inter-parton separations of order the parton-shower
cut-off scale, independent of the hard-process momentum transfer. These perturbative
calculations are specified by a set of input parameters, that also affect the subsequent
non-perturbative evolution. Most importantly the strong coupling $\alpha_s$ and the
parton density functions. Per default, in \Sherpa, the value and the running of $\alpha_s$
is set in accordance with the PDFs employed. The standard PDF set of \Sherpa is
NNPDF~3.0~\NNLO~\cite{Ball:2014uwa}, with $\alphaS(M_Z)=0.118$ and two-loop QCD running. 
For leptonic initial states we assume $\alphaS(M_Z) = 0.118$ and use a two-loop running
as well. 

\paragraph{Tuning of the cluster hadronisation}

In a first step, the parameters of the cluster hadronisation are tuned, usually with
respect to data from \LEPI, such as 
\begin{itemize}
\item the mean and distribution of the charged-particle multiplicity;
\item the yields of individual hadron species, in particular charged and neutral pions
  and kaons, protons, lambdas, heavy mesons and baryons;
\item the distribution of charged particles in phase space with respect to the thrust
  axis, \ie their rapidities and transverse momenta inside and outside the event
  plane;
\item the fragmentation function of $B$ hadrons; and
\item event shapes, and especially thrust, thrust major and minor as well as
  differential jet rates.
\end{itemize}
Given that the \Sherpa cluster fragmentation model features about 20 parameters to
describe the non-perturbative splitting of final-state gluons, the formation of mesonic and
baryonic clusters, their subsequent decays and ultimately the creation of primary hadrons,
cf. Sec.~\ref{sec:hadronisation}, their tuning procedure is largely automated. It relies
on the use of the \Professor tuning tool~\cite{Buckley:2009bj} to optimise the description
of the reference data. The resulting main parameters of the \Sherpa cluster-hadronisation
model are compiled in App.~\ref{app:tune}. Assuming universality of the hadronisation model
these parameters are kept fixed also for other collision energies and beam particles. 

\paragraph{Tuning of the intrinsic transverse momentum}

Assuming an incoming proton beam two parameters determine the intrinsic transverse-momentum
distribution of its constituents, \ie the mean and width of the hypothesised Gaussian distribution,
cf. Eq.~(\ref{eq:ue_intrinsic_kt}). These are adjusted by studying the transverse-momentum
distribution of Drell--Yan lepton pairs in proton--proton collisions at $\sqrt{s}\equiv E_{\text{ref}}=7$
TeV. We thereby assume identical parameter values for the two proton beams. The
current default tune results $\langle k_\perp\rangle = 1.1$ GeV and $\sigma=0.85$ GeV. The
determined width parameter is scaled to other centre-of-mass energies according to
 \begin{equation}
    \sigma(E_{\text{cms}}) =  \sigma(E_{\text{ref}})\left(\frac{E_{\text{cms}}}{E_\text{ref}}\right)^{0.55}\,.
\end{equation}
%


\paragraph{Tuning of the underlying event}

With the parameters of the cluster hadronisation and the intrinsic transverse-momentum adjusted,
the tuning of the underlying-event model remains. To this end the relevant quantities of the
impact-parameter dependent multiple-parton-interaction model, cf.~Sec.~\ref{sec:ue}, are adjusted. 
We use the \Professor tuning tool for this task and employ dedicated measurements from Tevatron and
LHC as reference data. The obtained default set of model parameters is again summarised in
App.~\ref{app:tune}. Note, the quoted values for $p_{T,\text{min}}$ and $p_{T,0}$ are for a reference
collision energy of $\sqrt{s}\equiv E_{\text{ref}}=1.8$ TeV. They get evolved to the actual collider energy $E_{\text{cms}}$
according to 
 \begin{equation}
    p_{T,i}(E_{\text{cms}}) =  p_{T,i}(E_\text{ref})\left(\frac{E_{\text{cms}}}{E_\text{ref}}\right)^\alpha\,,
\end{equation}
with the power $\alpha$ set to $0.244$.



\clearpage
\section{Highlighting \Sherpa\ Applications}
\label{sec:results}
In this section we present selected results obtained with releases of the \Sherpa-2 series.
The purpose mainly lies in giving illustrative applications of the calculational methods
and physics models introduced in Sec.~\ref{sec:ingredients}. While presenting these examples,
we will highlight specific features and aspects of the simulation chain. Where available, we
directly compare to experimental data, gauging the quality of our predictions. In fact, in
many cases \Sherpa\ has been used in the actual analysis of the data providing
state-of-the-art signal and background samples, being vital for the proper interpretation
of the measurements.

\SecsRef{sec:res-mm:zjets}{sec:res-mm:diphoton} focus on aspects of the combination of
QCD matrix elements and parton showers when applied to processes such as jet-associated
vector- and Higgs-boson production, top-quark single and pair production, or vector-boson
pair creation, including channels with photons. Sec.~\ref{sec:res-mm:bsm} is devoted to
simulations of physics beyond the Standard Model, while Sec.~\ref{sec:full_results} focuses
on non-perturbative aspects of the simulation: the hadronisation and hadron decays, as well
as the underlying event in proton-proton collisions. If not stated explicitly, in all
applications presented here the default \Sherpa tune, \ie the respective set of
non-perturbative model parameters listed in App.~\ref{app:tune}, has been used.

\subsection{\texorpdfstring{{$Z(\to\ell\ell)$}}{Z->ll} production in association with jets}
\label{sec:res-mm:zjets}
We begin the discussion with the most prominent testbed for calculational schemes combining
QCD matrix elements with parton showers, namely the production of a massive vector boson in
association with jets. These processes feature significant production rates at the \LHC and
probe a wide range of kinematic configurations, from almost exclusive vector-boson production
to signatures featuring very hard jets and a gauge boson at a rather small transverse
momentum. While studies of these scenarios are interesting on their own, they reflect specific
situations where $V$+jets production has to be considered as an important Standard Model
background in New Physics searches. Accordingly, a realistic simulation needs to address
not only the jet-production rates, but also their distributions in the bulk and the tails
of various observables.  Furthermore, kinematic correlations between the final-state objects
need to be modelled correctly. 

\begin{figure}[t!]%
  \centering
  \subfloat[inclusive jet cross sections]{\label{fig:Znjet}
    \includegraphics[width=0.47\textwidth]{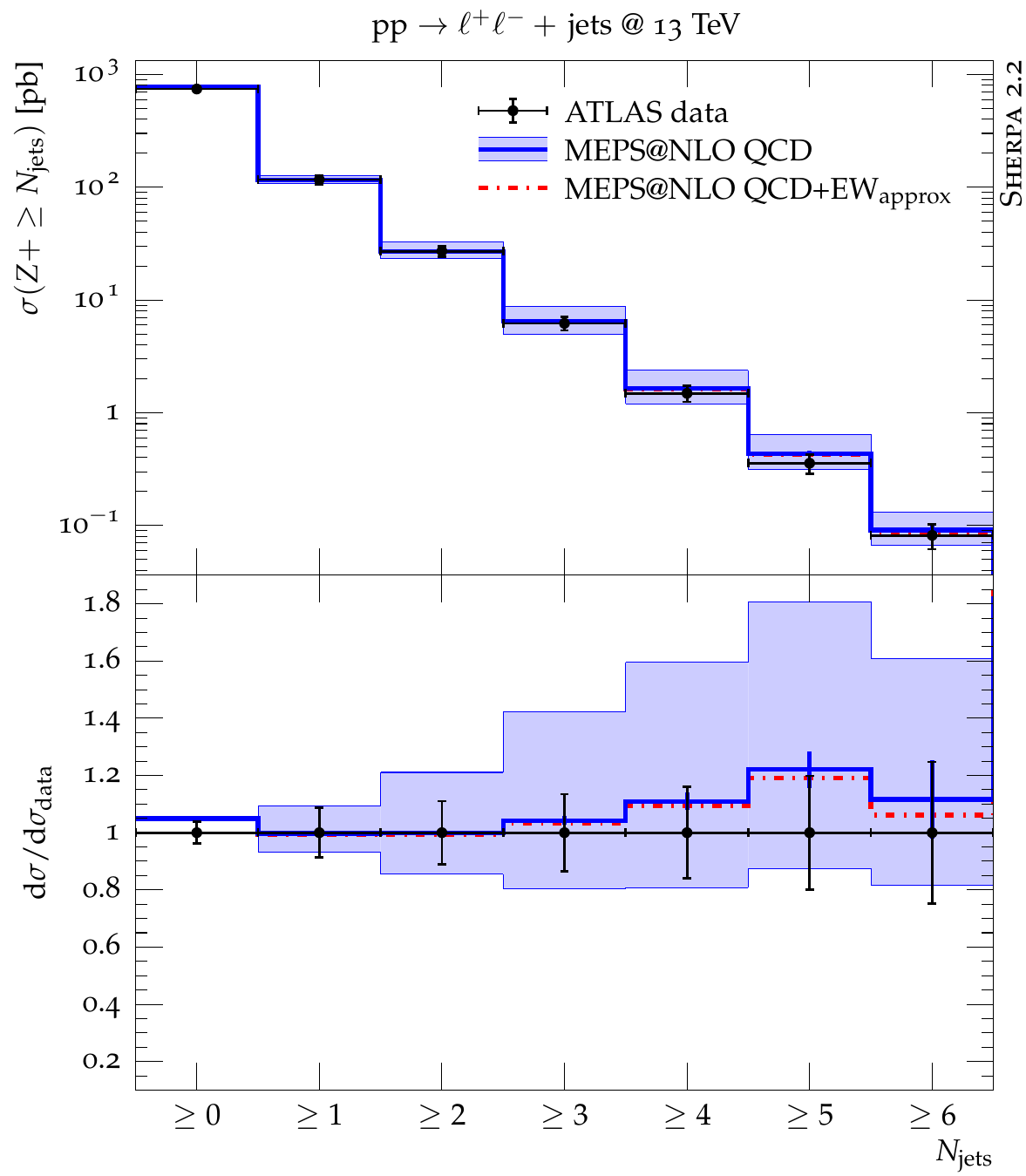}}
  \hfill
  \subfloat[azimuthal correlation of the two leading jets]{\label{fig:Zdphi}
    \includegraphics[width=0.47\textwidth]{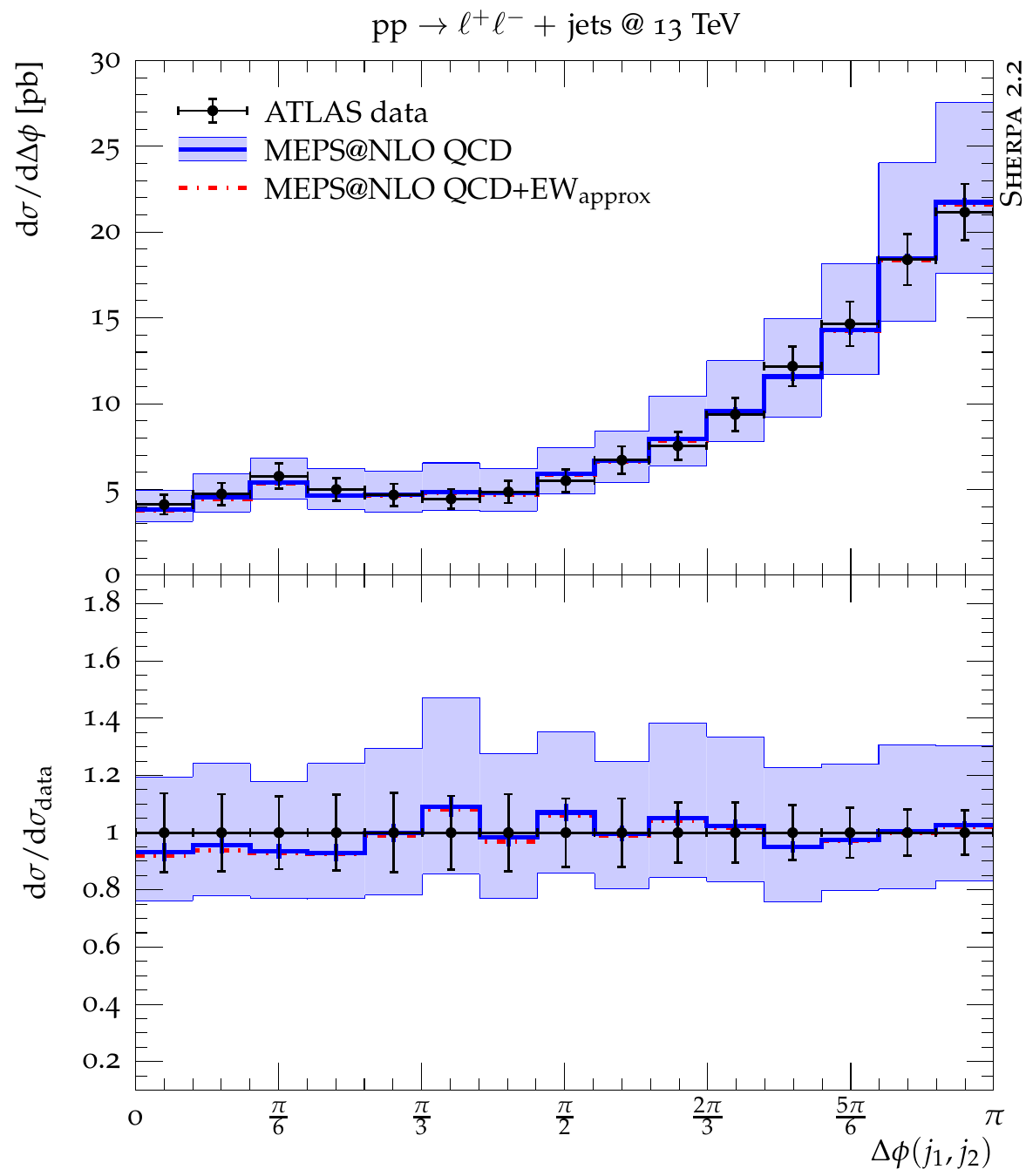}}
  \\
  \subfloat[scalar sum of the dilepton transverse mass and the jet transverse momenta]{\label{fig:Zht}
    \includegraphics[width=0.47\textwidth]{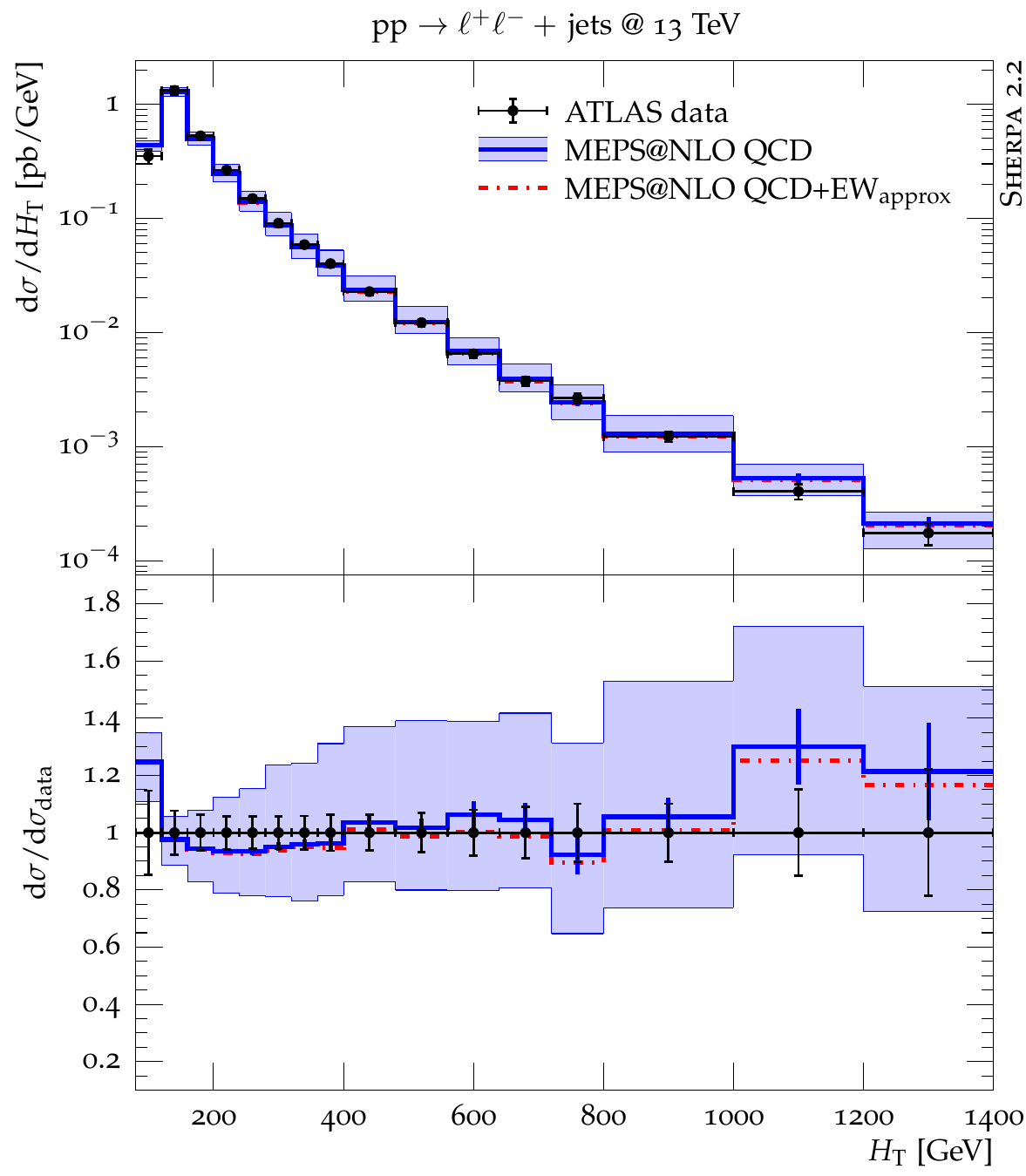}}
  \hfill
  \subfloat[invariant mass of the two leading jets]{\label{fig:Zmjj}
    \includegraphics[width=0.47\textwidth]{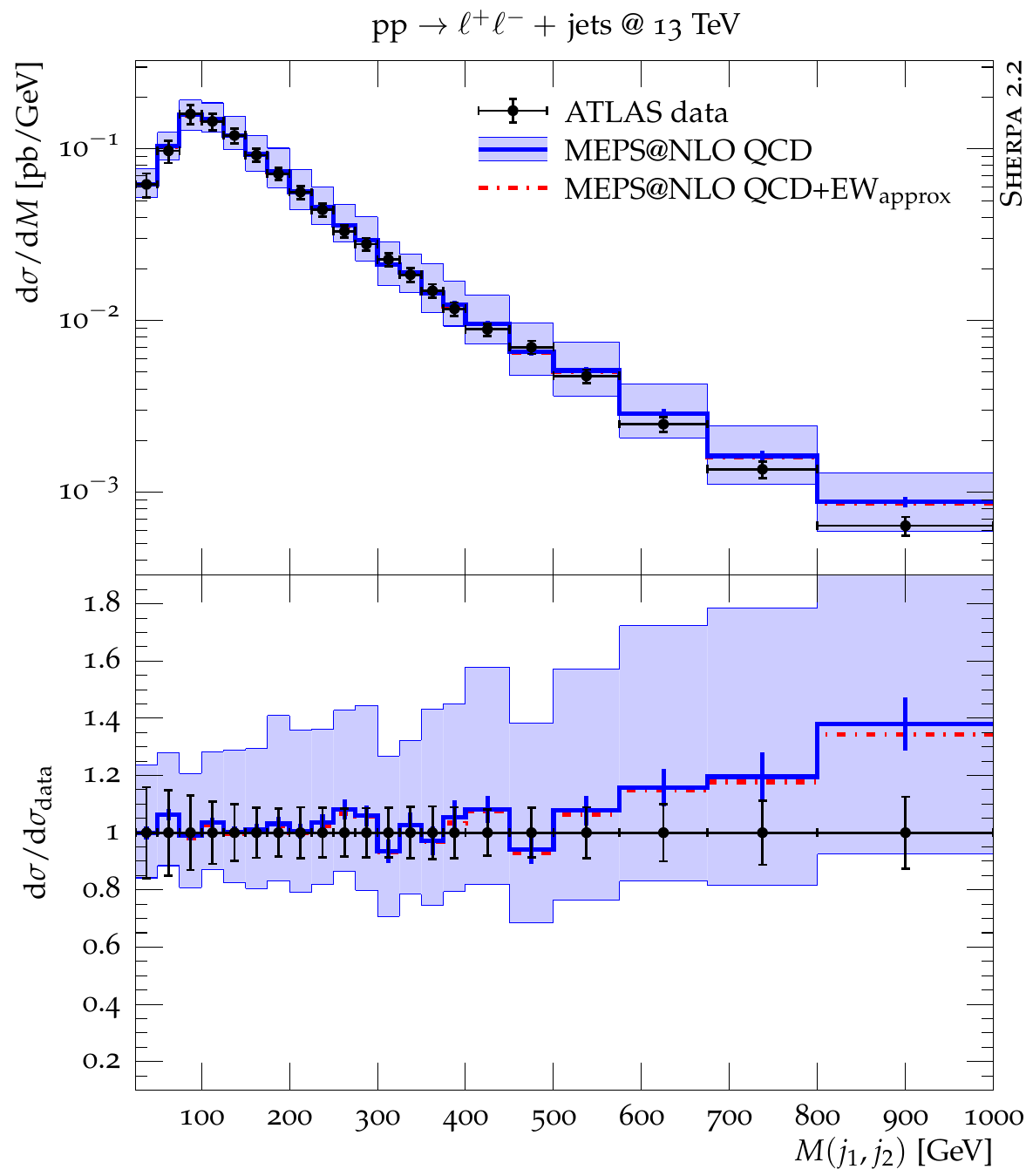}}
  \caption{Results for various observables in $Z$+jets production at the \LHC.
    The uncertainty bands for the \Sherpa\ predictions correspond to the 
    envelope over a 7-point scale variation, whereas their error bars 
    indicate the Monte-Carlo error. In addition, the effect
    of adding approximate electro-weak corrections to the nominal predictions
    is shown.
  }
  \label{fig:zjets}
\end{figure}

In \FigRef{fig:zjets} we present a few results for $Z$+jets production in
proton-proton collisions at $\sqrt{s}=13\;\textrm{TeV}$ with off-shell decays
for $Z\to\ell\ell$ and compare the \Sherpa\ predictions with data from \ATLAS~\cite{Aaboud:2017hbk}.
We refer the reader to \cite{Aaboud:2017hbk} which describes the event-selection criteria used.
\FigRefs{fig:Znjet}{fig:Zdphi} show the distribution for the number of
jets, $N_\text{jets}$, and the azimuthal correlation between the two leading
jets, \ie $\Delta \phi(j_1,j_2)$.%
\footnote{Note that a jet is called ``leading'' if it is
the one with the highest transverse momentum, and that the jets $j_i$
($i=1,\dots,N_\text{jets}$) are ordered descending in their transverse
momentum.}
While the $N_\text{jets}$ distribution probes multijet production
rates, $\Delta \phi(j_1,j_2)$ is sensitive to kinematic correlations between jet momenta.
\FigureRefs{fig:Zht}{fig:Zmjj} show the scalar sum of jet and lepton transverse momenta, commonly
referred to as $H_\mathrm{T}$,
and the invariant mass of the pair of leading jets
$M(j_1, j_2)$. $H_\mathrm{T}$ is sensitive to the $p_\mathrm{T}$ spectra of the leading jets. The tail
of the distribution probes higher multiplicities, and can therefore not be described by the parton shower
alone.  The invariant mass distribution is sensitive to non-perturbative effects at
small values and forms an important background for New-Physics searches at large values.

The \Sherpa\ prediction in these plots is obtained from multijet merging, applying the \MEPSatNLO method described
in Section~\ref{sec:matchmerge}. In practice, we consider matrix elements for the production of
an electron anti-electron pair with zero, one and two jets computed at NLO accuracy in the strong coupling, matched to
the parton shower with the \MCatNLO prescription, while the $Z$+3- and $Z$+4-jets calculations
are included at LO only. The merging cut parameter is set to $\Qcut=20\,\text{GeV}$.
The parton showered events are hadronised by the cluster fragmentation
and the underlying event is simulated through the \Amisic module.
QED corrections are enabled for the leptonic decay of the
intermediate $\gamma^*/Z$, \cf\ Sec.~\ref{sec:yfs}. The scale-variation
band shown in \FigRefs{fig:Znjet}{fig:Zmjj} is obtained through the ``on-the-fly'' reweighting
described in \SecRef{sec:reweighting} for a 7-point scale variation with factors of \nicefrac{1}{2}
and 2, including the scale dependence of both the fixed-order and the parton-shower calculation
in a consistent way. We further include approximate NLO EW corrections.
These, however, have negligible impact for the considered observables.

In the \Sherpa\ calculation up to four jets might be seeded by hard matrix-element partons,
jet multiplicities beyond that originate from the parton shower. We observe that the $N_\text{jets}$
distribution is well modelled by \Sherpa\ even up to six jets. The azimuthal correlation between the
two hardest jets is well described by the \MEPSatNLO method, in contrast to calculations where
the first and/or second jet originate from a spin-averaged parton-shower emission. The
$H_\mathrm{T}$ distribution, shown in \FigRef{fig:Zht}, is well described by \Sherpa\
as is the invariant mass of the two leading jets, depicted in \FigRef{fig:Zmjj}.

\subsection{\texorpdfstring{{$W(\to\ell\nu)$}}{W->ln} production in association with jets}
\label{sec:res-mm:wjets}
We proceed with the inclusive production of a leptonically decaying $W$ boson.
Besides the importance of incorporating higher-order QCD matrix elements in the
simulation, we illustrate the impact of electroweak one-loop corrections. In \FigRef{fig:W}
we present the gauge-boson transverse momentum distribution in proton-proton collisions at
$\sqrt{s}=13\;\textrm{TeV}$ evaluated in various approximations. The $W$ boson is reconstructed
from the charged lepton and the missing transverse momentum, where only modest acceptance cuts
are applied.

The standard \MEPSatNLO QCD prediction is contrasted with its LO variant \MEPSatLO.
In both calculations matrix elements with up to two jets, at NLO QCD and LO QCD accuracy,
respectively, have been matched to the parton shower and merged into an inclusive sample.
For the merging criterion we chose $\Qcut=20\,\text{GeV}$. Further details on the calculational
setups can be found in Ref.~\cite{Kallweit:2015dum}. Comparing the blue and green
uncertainty bands, it is apparent that the prediction based on exact NLO QCD matrix elements
features a significantly reduced theoretical uncertainty.

Including approximate NLO EW corrections in \MEPSatNLO \QCDpEWapprox, \cf Sec.~\ref{sec:matchmerge}, 
has an important impact for $W$ production at large transverse momenta, exhibiting the familiar
structure of the well-known EW Sudakov suppression. The corresponding one-loop virtual amplitudes for
up to $W+2j$ production have been obtained from \OpenLoops. Subleading mixed QCD-EW tree-level
contributions are provided by \Comix. Their impact is very marginal on this observable, however,
this is different for the leading-jet transverse-momentum distribution, \cf~\cite{Kallweit:2015dum}.

\subsection{\texorpdfstring{{$gg\to h$}}{gg->h} production in association with jets}
\label{sec:res-mm:hjets}
Higgs-boson production processes form a centre piece of the \LHC physics program.
This is in particular true for the case of the gluon-fusion channel, as it features the
largest cross section. It is commonly described in the Higgs Effective Field Theory
(HEFT) approach. In the complete Standard Model it constitutes a loop-induced process, with
top and bottom quarks propagating. With \Sherpa\ both approaches can be employed.

We present results based on inclusive Higgs production as well as Higgs production
in association with one jet at NLO accuracy in the strong coupling, while Higgs production
in association with two and three jets is described at LO accuracy, merged using the standard
\MEPSatNLO method, \cf~\cite{Hoeche:2014lxa}. While the HEFT computation proceeds
straight-forwardly, the full Standard Model computation reweights each component
of the NLO calculation with its loop-induced counter-part \cite{Buschmann:2014sia}. 
Only the virtual corrections, which are structurally of two-loop origin including
different and dynamic mass scales, and have only been calculated
recently~\cite{Jones:2018hbb}, are approximated by factorising the NLO correction in the
effective theory and the mass corrections at LO. This approximation has been shown to
well reproduce the shape of the full NLO results for the Higgs-boson $p_{\mathrm{T}}$ distribution~\cite{Jones:2018hbb}.

\begin{figure}%
  \centering
  \parbox{0.47\textwidth}{%
    \includegraphics[width=0.47\textwidth]{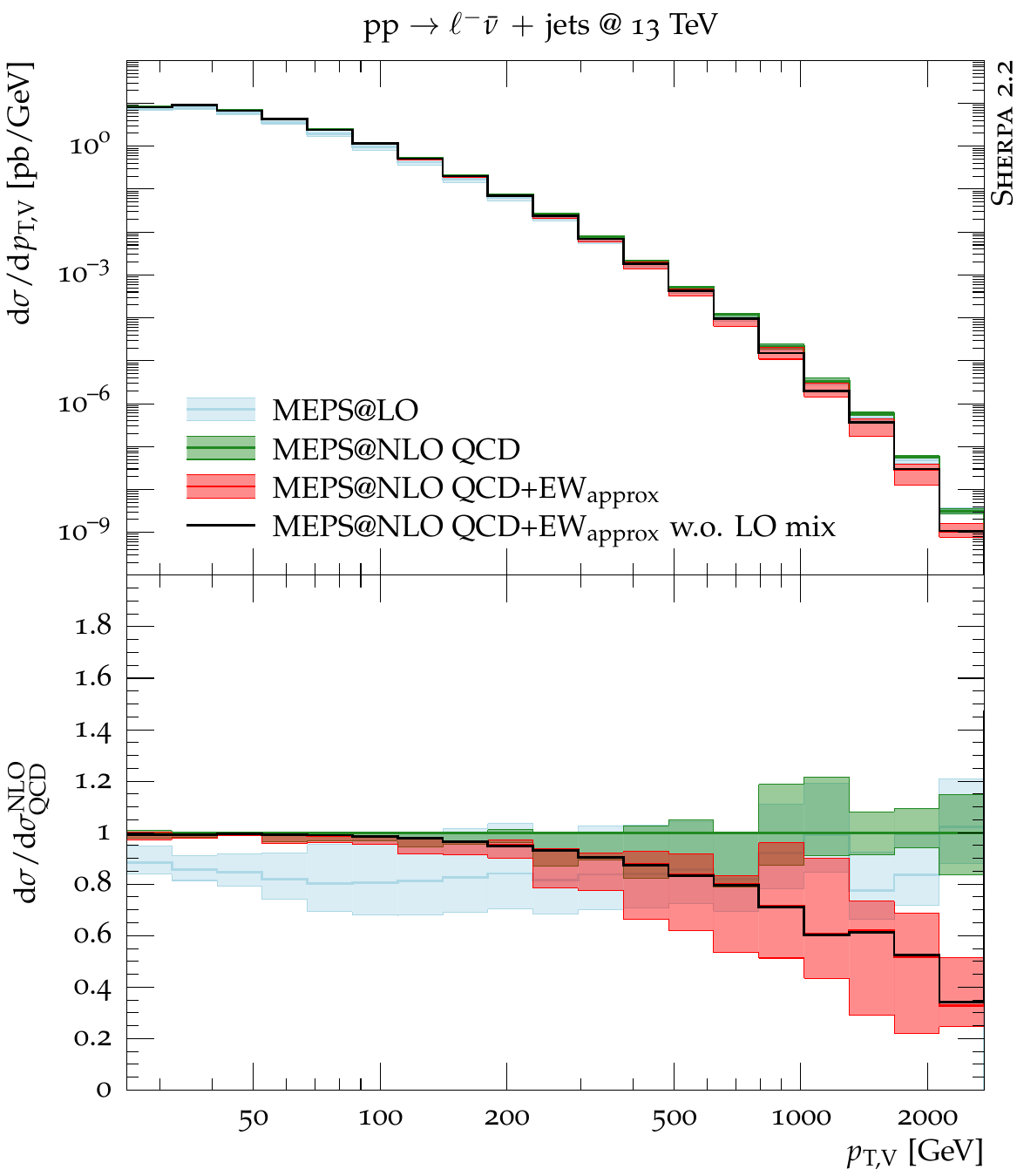}
    \caption{Predictions for the $W$-boson transverse momentum distribution at
      the \LHC.\\}%
    \label{fig:W}}%
  \hfill
  \parbox{0.47\textwidth}{%
    \includegraphics[width=0.47\textwidth]{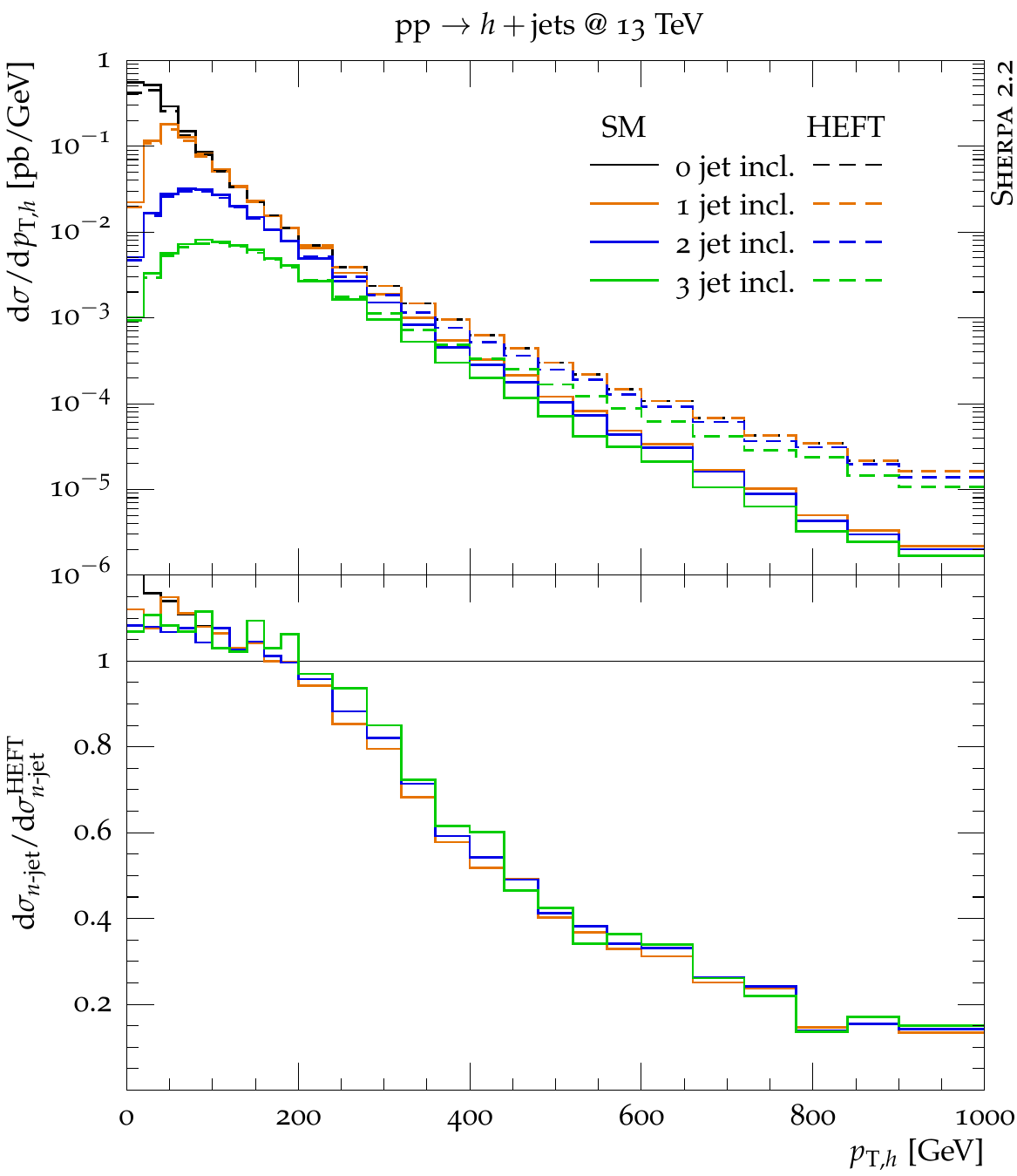}
    \caption{Predictions for the Higgs-boson transverse momentum distribution
      in gluon-fusion production at the \LHC.}%
    \label{fig:h}}%
\end{figure}

\FigRef{fig:h} details the 0-, 1-, 2- and 3-jet inclusive distributions of the transverse momentum
of the Higgs boson in gluon-fusion production. It is interesting to see, that the quark-mass corrections
introduced through the top-quark running in the loop in the exact Standard Model calculation, are
independent of the number of jets that are accompanying the Higgs boson. This mass suppression reaches
up to $-60\%$ at transverse momenta of around $500\;\textrm{GeV}$ and increases further towards higher 
$p_\mathrm{T}$. 
The finite mass corrections in dependence on the jet multiplicity 
have been further investigated in \cite{Greiner:2016awe}. 
With the exception of the scalar sum of transverse momenta, $H_\mathrm{T}$, 
no appreciable dependence was found among the observables investigated, 
both in inclusive Higgs boson production and in Higgs boson production 
through gluon fusion in VBF kinematics.

For completeness we list the following additional flags which instruct \Sherpa to perform the outlined
reweighting procedure using the appropriate loop-induced processes from the \OpenLoops library:

\begin{lstlisting}[style=runcard,numbers=none]
(run){
  % finite top mass effects
  KFACTOR GGH;
  OL_IGNORE_MODEL 1;
  OL_PARAMETERS preset 2 allowed_libs pph2,pphj2,pphjj2 psp_tolerance 1.0e-7;
}(run);
\end{lstlisting}

\FloatBarrier
\subsection{\texorpdfstring{{$t\bar{t}$}}{ttbar} production in association with jets}
\label{sec:res-ttbar}
The production of a top-quark pair in proton-proton collisions is particularly challenging
due to the non-negligible mass and finite life-time of the colour-charged tops.

Fig.~\ref{fig:ttbar} shows the visible energy ($H_\mathrm{T}$) distribution
in top-quark pair production as predicted by the NLO multijet merging
in \Sherpa. This calculation involves NLO fixed-order input predictions
with up to two light jets in addition to the top-quark pair. The top-quark
decays are calculated at leading order including spin correlations based on
the $t\bar{t}+$jets Born matrix elements using spin-density matrices, \cf
Sec.~\ref{Sec::MEs}. The one-loop matrix elements were obtained from \OpenLoops.
The \MCatNLO\ matching for heavy quarks applied in the simulation is based
on the massive Catani--Seymour dipole subtraction~\cite{Catani:2002hc}
and was originally constructed in~\cite{Hoeche:2013mua}. Further details on the
calculational setup can be found in~\cite{Hoeche:2014qda}.

Besides the \MEPSatNLO result we present the corresponding \MEPSatLO prediction.
Note the excellent agreement between the two predictions, after the
leading-order result has been multiplied by a global $K$-factor of $1.65$.
The first ratio panel in \FigRef{fig:ttbar} shows clearly, that, beyond this global
$K$-factor, the main effect of the higher-order corrections is a drastic reduction
of the scale uncertainty, which in this case has been determined by varying
the renormalisation and factorisation scales, but not the resummation scale.
The second lower panel shows the individual contributions of $t\bar{t}$ (solid),
$t\bar{t}j$ (dash-dotted) and $t\bar{t}jj$ (dotted) final states to the
overall result. At low $H_\mathrm{T}$ all components contribute to the overall result,
while at high $H_\mathrm{T}$ the prediction is given almost entirely by the
$t\bar{t}jj$ component.

\subsection{Single-top quark production}
\label{sec:res-mm:singletop}
In Ref.~\cite{Bothmann:2017jfv} a dedicated \Sherpa\ study of single-top quark production
in hadronic collisions has been presented which is challenging due to the various production
modes and their differing characteristics in how the final-state phase space is populated.
Our study includes the consistent evaluation in the four- and five-flavour PDF schemes
and process-definition ambiguities when considering higher-order corrections, where
a separation from top-quark pair production has to be defined. With \Sherpa\ single-top
quark production in the $s$, $t$ and $tW$ channels can be simulated using the \MCatNLO\
implementation. 

\begin{figure}%
  \centering
  \parbox{0.47\textwidth}{%
    \includegraphics[width=0.47\textwidth]{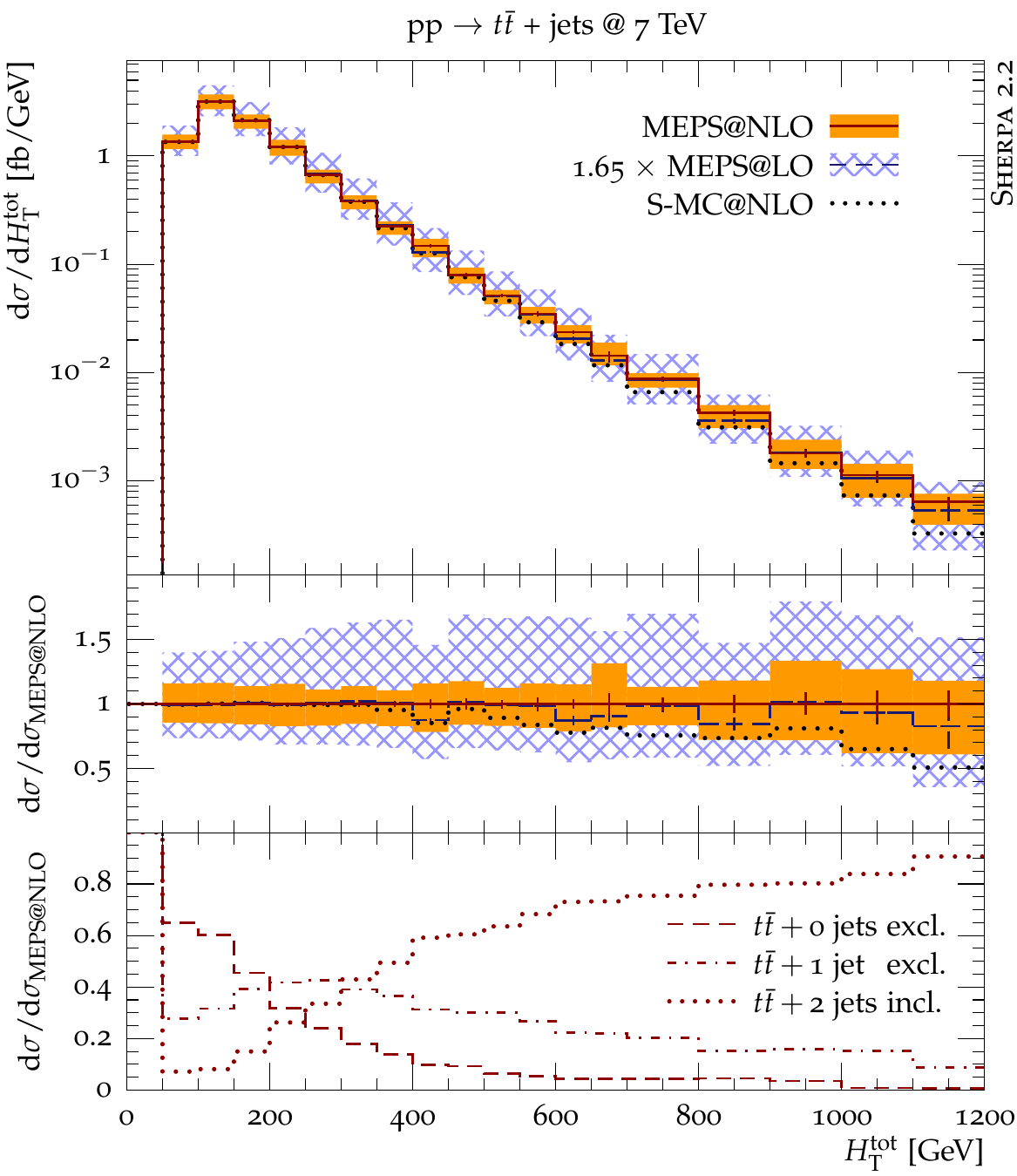}
    \caption{Predictions for the $H_\mathrm{T}$ distribution in top-quark pair
      production at the \LHC.
      \\\textcolor{white}{.}\\\textcolor{white}{.}\\\textcolor{white}{.}}%
    \label{fig:ttbar}}%
  \hfill
  \parbox{0.47\textwidth}{%
    \includegraphics[width=0.47\textwidth]{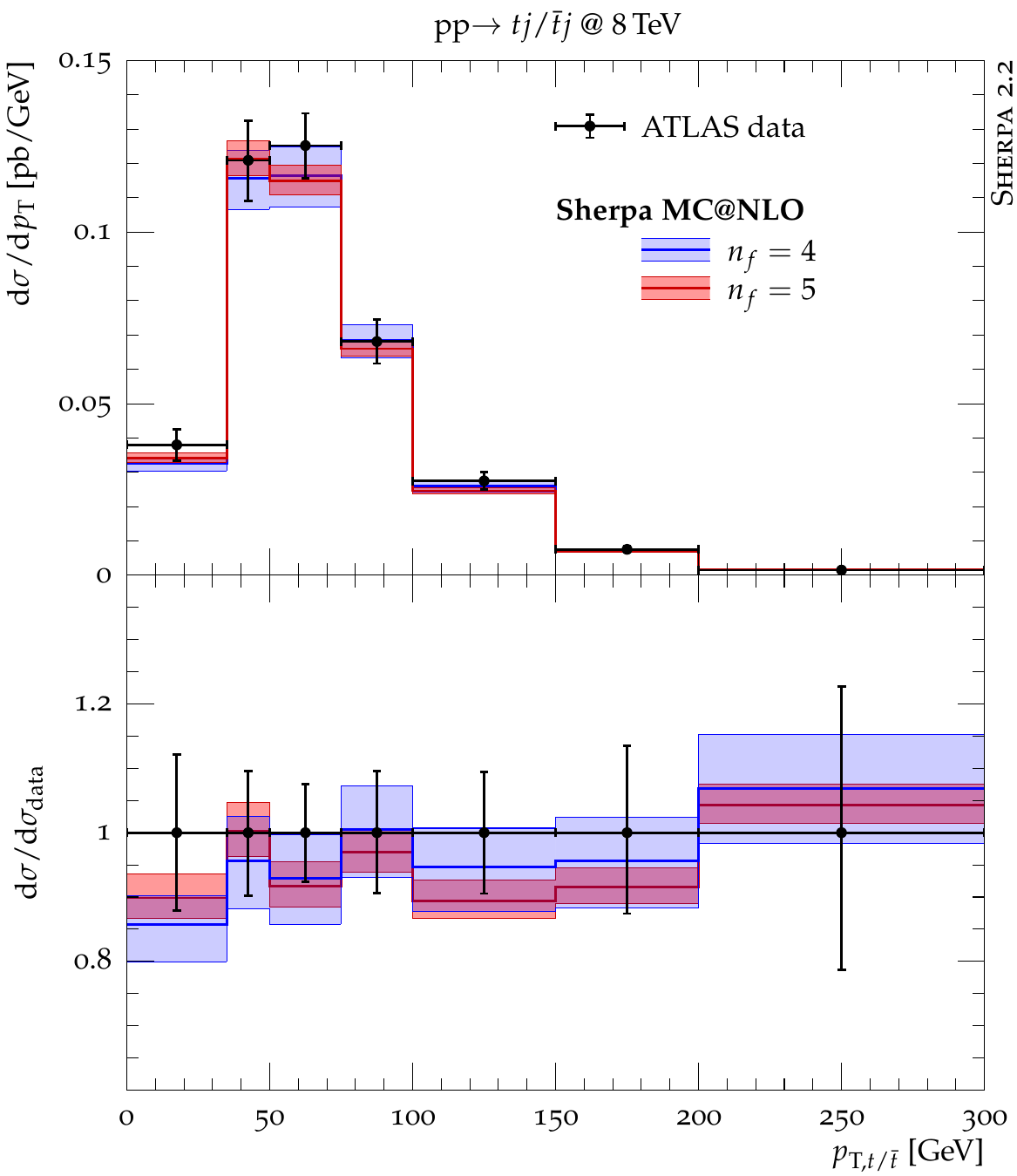}
    \caption{Results for the $p_\mathrm{T}$ distribution of reconstructed
      top-quarks in $t$-channel single-top production at the \LHC. The four-
      and five-flavour \MCatNLO\ are compared with data
      from~\cite{Aaboud:2017pdi}.}%
    \label{fig:singletop}}%
\end{figure}

In \FigRef{fig:singletop}, we compare \MCatNLO\ results for the reconstructed top-quark
transverse-momentum distribution in the $t$-channel production mode in the four- and five-flavour
scheme with ATLAS data taken at $\sqrt{s}=8$\,TeV~\cite{Aaboud:2017pdi}.
The bands correspond to the theory error convention used in~\cite{Aaboud:2017pdi}. That is,
the statistical, the strong coupling, the PDF and the (dominant)
7-point scale uncertainties, all added in quadrature.  The \Sherpa\ predictions and experimental
data agree within their respective uncertainties. For further details on the calculation and additional results,
see~\cite{Bothmann:2017jfv}. The minimal settings to generate $t$-channel
single-top production events at \MCatNLO\ with \Sherpa\ are:

\begin{lstlisting}[style=runcard,numbers=none]
  (run){
    % single-top specific scale definition
    CORE_SCALE SingleTop;
    % enable decays of produced top-quarks
    HARD_DECAYS On;
    ...
  }(run)
  (processes){
    Process 93 93 -> 6 93;
    Order (*,2); NLO_QCD_Mode MC@NLO;
    % require t-channel propagator
    Min_N_TChannels 1; 
    ...
    End process;
  }(processes)
\end{lstlisting}

\clearpage
\subsection{Diboson production in association with jets}
\label{sec:res-mm:diboson}
Another important class of benchmark processes at hadron colliders is
diboson production. This includes the pair production of massive gauge
bosons, \ie $W$ and $Z$, but also photon pairs or mixed $W\gamma$, $Z\gamma$
final states. These channels provide precision tests of the electroweak
sector, including triple- and even quartic gauge couplings. They form
irreducible backgrounds for Higgs-boson production with the Higgs decaying
into gauge bosons, or searches for New Physics. Besides the accurate
modelling of the diboson final states, a realistic description of
the associated QCD activity is vital, as it often provides the only handle
to separate signal from irreducible backgrounds. All the channels mentioned
above have loop-induced contributions such as $gg\to W^+W^-$, that are
phenomenologically important and require refinements in the techniques
for combining matrix elements with parton showers. 

\begin{figure}%
  \centering
  \parbox{0.47\textwidth}{%
    \includegraphics[width=0.48\textwidth]{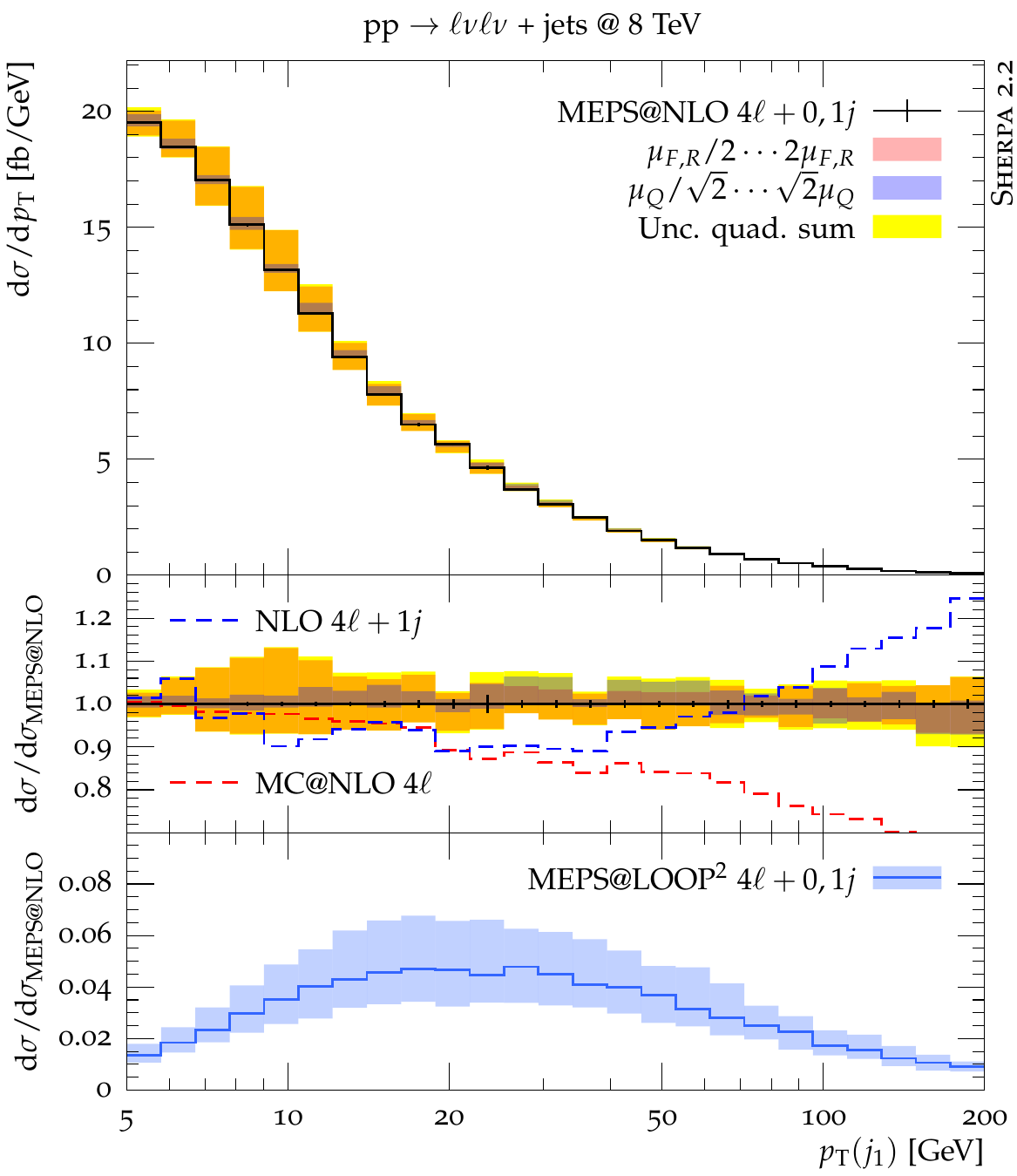}
    \caption{Prediction for the leading jet transverse-momentum distribution in
      $pp\to \ell\nu\ell\nu$ production in association with jets at the \LHC.}%
    \label{fig:vv}}%
  \hfill
  \parbox{0.47\textwidth}{%
    \includegraphics[width=0.48\textwidth]{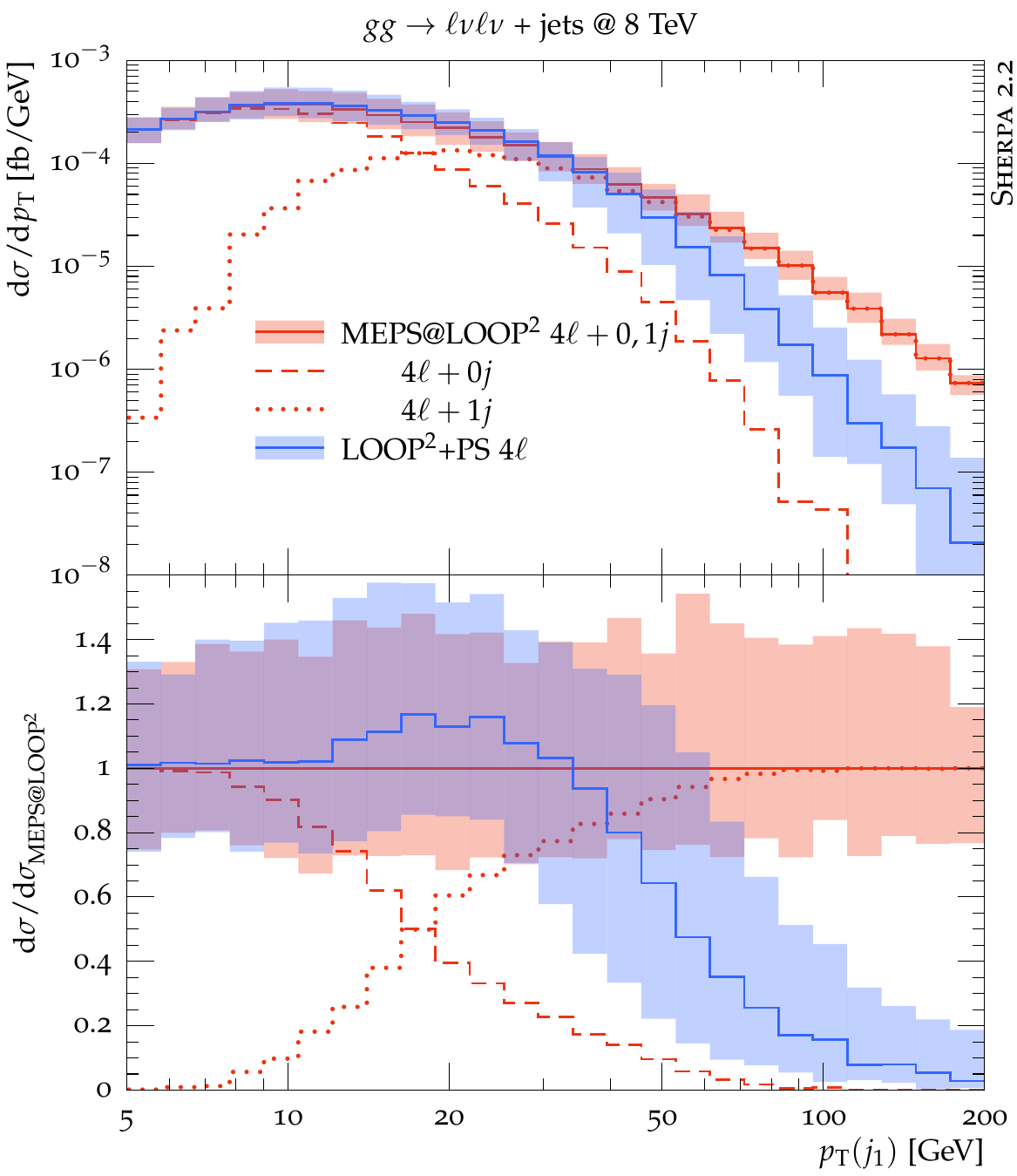}
    \caption{Predictions for the leading jet transverse-momentum distribution
      in loop-induced $pp\to \ell\nu\ell\nu$ production in association with jets
      at the \LHC.}%
    \label{fig:vvloop}}%
\end{figure}

We begin the discussion with one of the main Higgs-production backgrounds:
$pp\to WW^*$, \ie $pp\to \ell\nu\ell\nu$, with two charged leptons of different
flavour and the corresponding neutrinos. A dedicated analysis with \Sherpa\
has been presented in~\cite{Cascioli:2013gfa}. In~\FigRef{fig:vv} we present the
leading-jet $p_\mathrm{T}$ distribution for this off-shell diboson-production channel.
The upper panel shows the \Sherpa\ \MEPSatNLO\ prediction when merging the zero-
and one-jet contributions, with the QCD one-loop matrix elements provided by
\OpenLoops~\cite{Cascioli:2011va}. The uncertainty bands correspond to
the perturbative (red) and resummation (blue) scale variations, again added
in quadrature to yield an overall uncertainty estimate (yellow band).

In the first ratio plot the \MEPSatNLO\ prediction is compared to an inclusive
\MCatNLO (red dashed) calculation, based on the four-lepton ($4\ell$) NLO QCD
matrix element matched to the \Sherpa\ parton shower. Further, we present the
pure fixed-order result based on the $4\ell+1$jet NLO QCD matrix element (blue dashed).
Note that the inclusive \MCatNLO prediction describes
this observable only at LO precision, and is found not to be compatible with the
more precise \MEPSatNLO prediction over a wide range of the spectrum.
Details on the simulation setups and parameters used can be found in~\cite{Cascioli:2013gfa}.  

The lower panel displays the relative corrections and uncertainties of a
multijet-merged prediction of loop-induced $gg\to \ell\nu\ell\nu$
production in association with jets, dubbed \MEPSatLOOPsq, normalised to \MEPSatNLO at the central
scale. These squared quark-loop amplitudes constitute higher-order
corrections to the generic $4\ell$ and $4\ell+1$jet processes. However, their
relative contribution can be as large as $5\%$ around $p_\mathrm{T}(j_1)\approx 20\;\textrm{GeV}$. 

A more detailed view on these loop-induced corrections is provided in~\FigRef{fig:vvloop}.
Here the multijet-merged sample is compared to a simple \LOPS prediction of
$gg\to \ell\nu\ell\nu$ production, dubbed \LOOPsqPS here. Furthermore,
the contributions of the $4\ell+0j$ and $4\ell+1j$ matrix element to the full \MEPSatNLO sample
are indicated. It is evident that at high $p_\mathrm{T}$ the relevant contributions are those of the
one-jet process, which can not be fully accounted for by the pure parton shower in the
\LOOPsqPS sample.

Note, a very recent experimental measurement of this channel at $\sqrt{s}=13$ TeV, including
an extensive comparison of state-of-the-art theoretical predictions with data, among them those
from \Sherpa, has been presented by the ATLAS collaboration in~\cite{Aaboud:2019nkz}. A
similar study but for the final state of four charged leptons has been presented in
\cite{Aaboud:2019lxo}.

\subsection{\texorpdfstring{{$V\gamma$}}{Vgamma} production in association with jets}
\label{sec:res-mm:vgamma}
The second diboson channel we want to discuss here is the associated production of
a prompt photon and a lepton-pair, possibly accompanied by additional
QCD jets. The corresponding study has been presented in~\cite{Krause:2017nxq}, which
we refer to for details on the generator setup, parameter choices and object definitions.

\MEPSatNLO\ predictions for the transverse-momentum distribution of the photon based
on merging of $pp \rightarrow e^+ e^- \gamma + 0,1\textrm{ jets@NLO} + 2,3\textrm{ jets@LO} $
each matched to the parton shower is presented in \FigRef{fig:vgamma}. The prediction is compared
with an ATLAS measurement~\cite{Aad:2016sau} at $\sqrt{s}=8\;\textrm{TeV}$. Furthermore, results
based on \MEPSatLO and an inclusive \MCatNLO\ simulation are shown.

Most notably, the \MEPSatNLO calculation is in very good agreement with the data,
both in rate and shape over the whole range of the observable. It is interesting to note that,
similar to Sec.~\ref{sec:res-ttbar}, the \MEPSatLO\ prediction largely agrees in shape with
the NLO merged one as can be seen in the upper ratio panel. The effect of going from LO to NLO
accuracy in the simulation can be captured by a global $K$-factor which brings the central
prediction in good agreement with experimental data. More importantly,
NLO accurate predictions show
significantly reduced inherent uncertainties, which are estimated by
variations of the perturbative scales and PDFs, see the lower two ratio panels. 

\subsection{Diphoton production in association with jets}
\label{sec:res-mm:diphoton}

Predictions for prompt-photon production are notoriously difficult, especially
for low-energetic or not well isolated photons. Appropriate choices for the perturbative
scales need to be made that are valid for a wide range of kinematics and, potentially,
non-perturbative contributions need to be considered.
In particular, a fragmentation component has to be taken into account, where
soft or collinear photons are emitted from harder jets through QED $q\to
q\gamma$ splittings.
One option to do so is a
combined QCD $\otimes$ QED parton shower and related multijet merging, as
proposed in~\cite{Hoeche:2009xc}. As an implementation of such an algorithm
is not available at NLO accuracy yet, we use a QCD \MEPSatNLO setup here,
but take fragmentation-like configurations of a hard jet and a soft photon into
account through higher-multiplicity matrix elements. To make the fragmentation
component as inclusive as possible, we use a dynamic merging
cut~\cite{Carli:2009cg} with $\bar{Q}_{\textrm cut}=10$~GeV using the following
run-parameter settings:
\begin{lstlisting}[style=runcard,numbers=none]
  (run){
    % core scale m_yy
    CORE_SCALE VAR{Abs2(p[2]+p[3])};
    ...
  }(run)
  (processes){
    Process 93 93 -> 22 22 93{3};
    Order (*,2);
    % dynamical merging cut with Qcutbar=10.0 GeV and mu=m_yy
    CKKW sqr(10.0/E_CMS)/(1.0+sqr(10.0/0.6)/Abs2(p[2]+p[3]));
    ...
    End process;
  }(processes)
\end{lstlisting}
To mitigate the mismatch of the photon-isolation cuts between the generator
level and the experimental analysis, we choose a hybrid isolation approach as
described in more details in~\cite{Siegert:2016bre}.

\begin{figure}[]%
  \centering
  \parbox{0.47\textwidth}{%
    \includegraphics[width=0.48\textwidth]{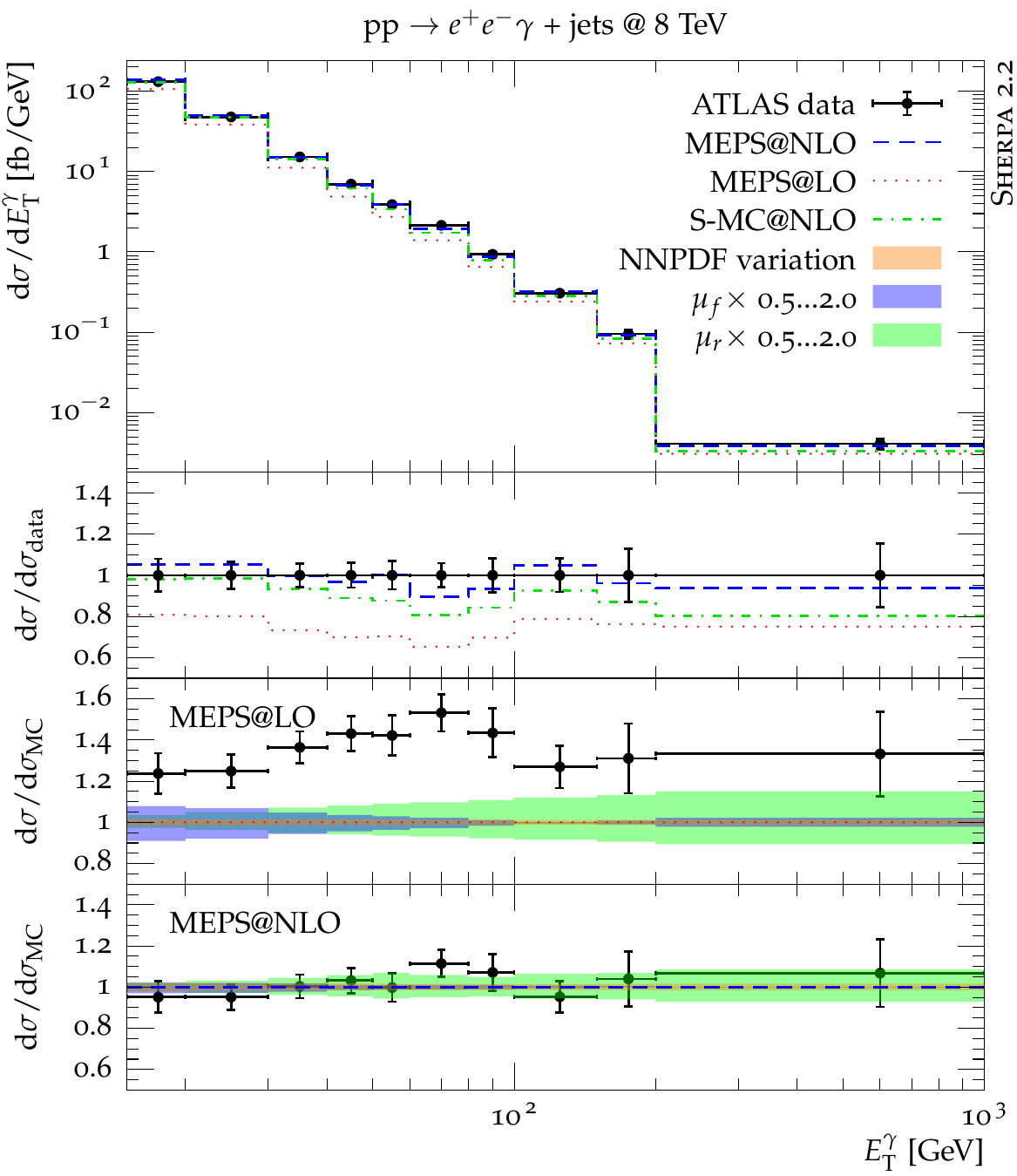}
    \caption{Results for the photon transverse-momentum distribution in $pp \to
      e^+ e^- \gamma$ production in association with jets at the \LHC,
      comparing to data from~\cite{Aad:2016sau}.}%
    \label{fig:vgamma}}%
  \hfill
  \parbox{0.47\textwidth}{%
    \includegraphics[width=0.48\textwidth]{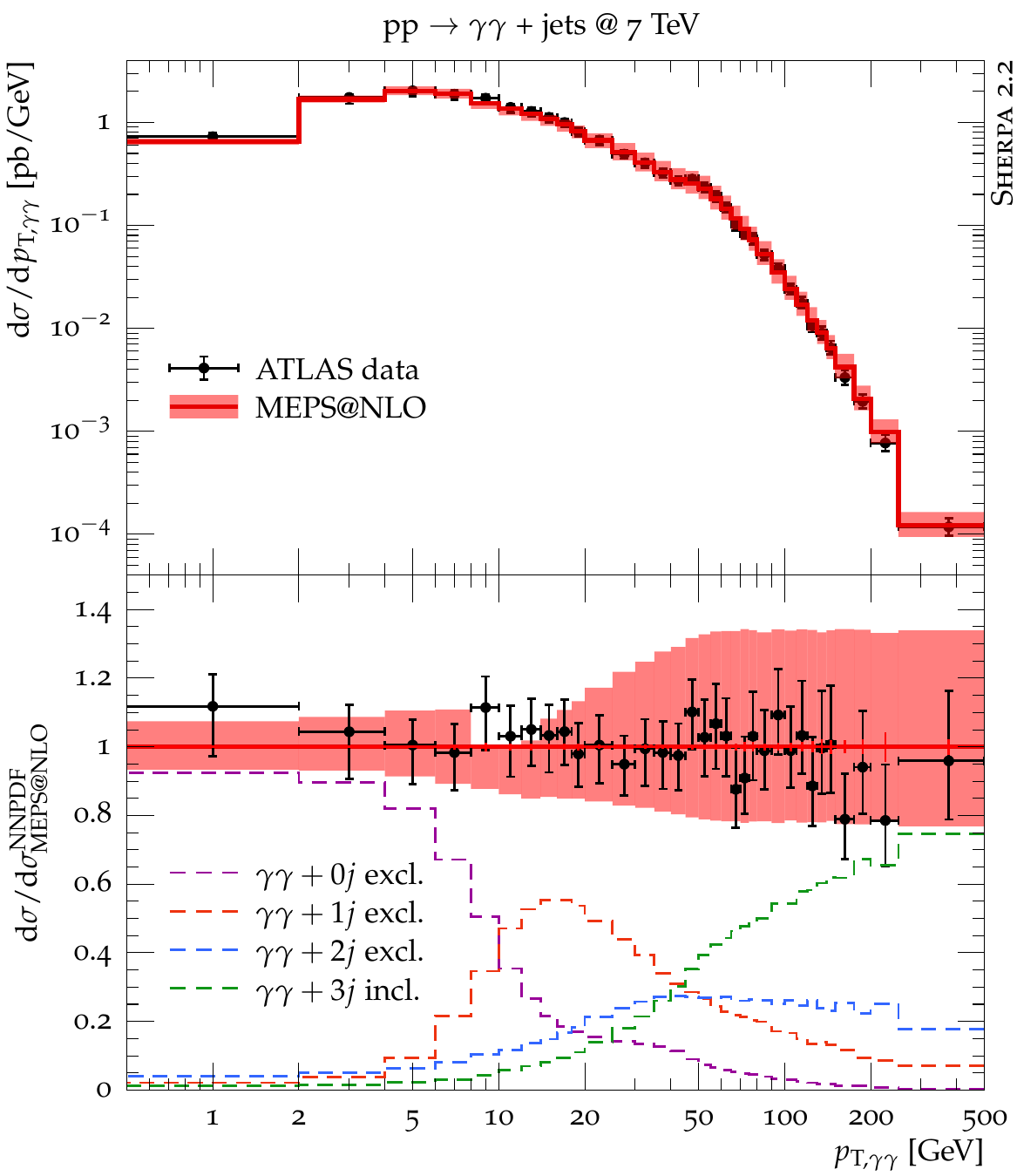}
    \caption{Result for the diphoton transverse-momentum distribution at the
      \LHC, in comparison to data taken
      from~\cite{Aad:2012tba}.\\\textcolor{white}{.}}%
    \label{fig:diphoton}}%
\end{figure}

Accordingly, \NLOPS matched simulations for $pp\to \gamma\gamma$ and
$pp\to \gamma\gamma$+jet production are merged into an inclusive sample and
additionally, matrix elements with up to three partons in the final state are
included at LO accuracy in the approach of~\cite{Hoeche:2010kg}.
The comparison with data from \ATLAS~\cite{Aad:2012tba} for the transverse-momentum
distribution for diphoton production in \FigRef{fig:diphoton} shows
good agreement in all regions of the spectrum. Note, the \CMS collaboration also
presented an analysis of diphoton production at $\sqrt{s}=7$ TeV~\cite{Chatrchyan:2014fsa},
where \MEPSatLO\ predictions from \Sherpa provided a very good description of the data. 

\FloatBarrier
\subsection{Physics beyond the Standard Model}
\label{sec:res-mm:bsm}
We now present two examples in which \Sherpa is used as a generator
for a New Physics signal. First, an analysis of dimension-six gluon operators in multijet
production at a Future Circular Hadron Collider (\FCC) with
$\sqrt{s}=100\;\textrm{TeV}$. Further details on this study can be found
in~\cite{Krauss:2016ely,Bothmann:2016loj}.
Second, a study on an anomalous triple gauge coupling in $Z$-boson pair production
at the \LHC, based on the corresponding \CMS measurement~\cite{CMS:2014xja}.

A study presented in~\cite{Krauss:2016ely} considers the impact of additional
dimension-six gluon interactions given by the effective operator
\begin{equation}
  \label{eq:extra-vertex}
c_G \mathcal{O}_G = \frac{g_s \, c_G}{\Lambda^2} \, f_{abc} 
         G_{a \nu}^\rho G_{b \lambda}^\nu G_{c \rho}^\lambda \quad
\text{with} \quad 
G_a^{\rho \nu} 
= \partial^\rho G_a^\nu - \partial^\nu G_a^\rho - i g_s f_{abc} G^{b
  \rho} G^{c \nu}
\end{equation} 
on multijet production at the \LHC. The corresponding model, which needs
to be invoked by \Sherpa, has been obtained through a \FeynRules implementation of the
interactions, subsequently interfaced to \Sherpa using the \UFO
standard, as described in Section~\ref{Sec::MEs}.
The matrix element generator \Comix has been used to evaluate all contributing Lorentz and
$SU(3)$ colour structures~\cite{Hoche:2014kca}. For SM backgrounds
as well as for the signal (which interferes with the SM amplitudes) corresponding leading-order
matrix elements for up to five jets are merged via the \MEPSatLO\
method described in Section~\ref{sec:matchmerge}.

\begin{figure}%
  \centering
  \parbox{0.47\textwidth}{%
    \includegraphics[width=0.48\textwidth]{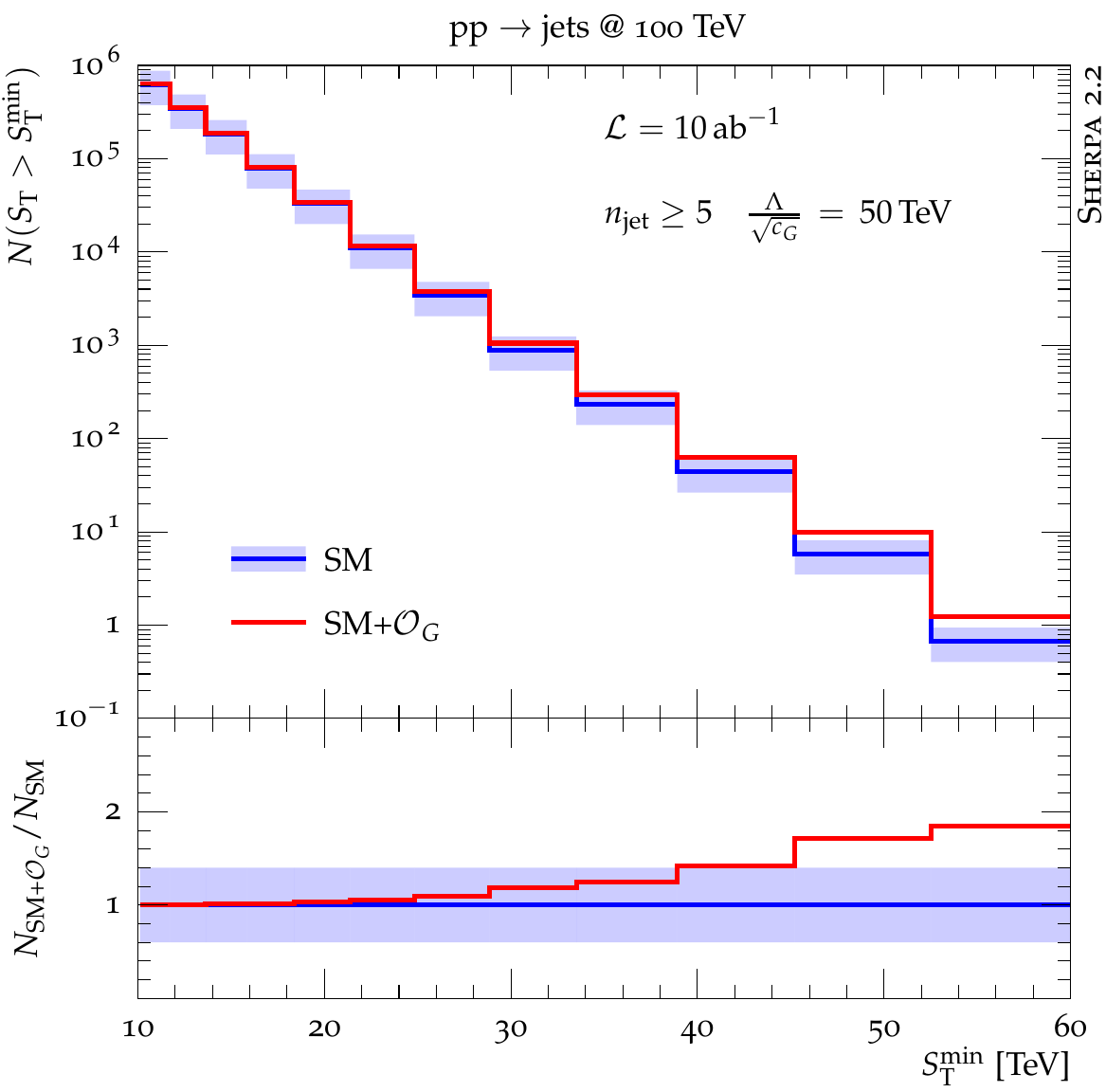}
    \caption{Prediction for the expected number of events with
      $S_\mathrm{T}>S_\mathrm{T}^\text{min}$ in inclusive five-jet production
      in proton-proton collisions at $\sqrt{s}=100\;\textrm{TeV}$.\\}
    \label{fig:d6jets}}%
  \hfill
  \parbox{0.47\textwidth}{%
    \includegraphics[width=0.48\textwidth]{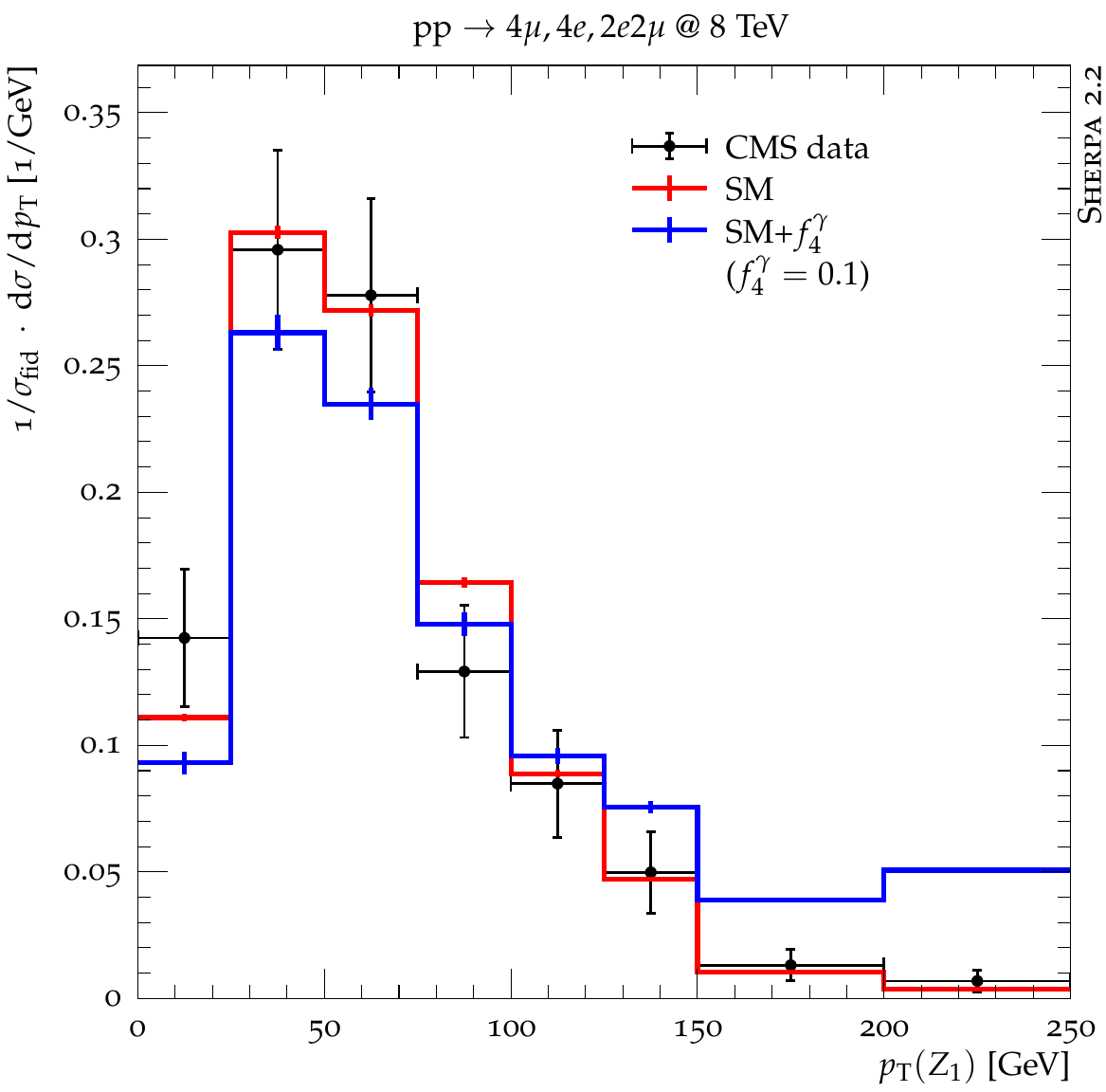}
    \caption{Results for the transverse momentum of the $Z$-boson candidate
      closest to the nominal $Z$-boson mass in four-lepton events at the \LHC,
      comparing to data taken from Ref.~\cite{CMS:2014xja}.}%
    \label{fig:zzagc}}%
\end{figure}

In \FigRef{fig:d6jets} we show the effect on the $S_\mathrm{T}$ distribution,
with or without the contributions from Eq.~\eqref{eq:extra-vertex} for a selection
of inclusive five-jet events at \FCC energies,
where $S_\mathrm{T}$ denotes the scalar sum of the transverse momentum of all reconstructed jets,
with $p_{\mathrm{T},j}>1\;\textrm{TeV}$ and $|\eta_{j}|<5.2$. Here the relevant ratio of the scale
$\Lambda$ and the Wilson coefficient $c_G$ is taken to be $\Lambda/\sqrt{c_G}=50\;\textrm{TeV}$.
For the considered luminosity of $10\;\textrm{ab}^{-1}$ the New Physics signal exceeds the given
uncertainty band of the SM prediction, based on variations of the
perturbative scales, at around
$S_\mathrm{T}\gtrsim 40\;\textrm{TeV}$.

The second example is related to anomalous triple gauge couplings in the electroweak sector of the
Standard Model. For this we prepared a \FeynRules\ implementation of the general $WWV$ and $ZZV$
Lagrangian considered in~\cite{Hagiwara:1986vm}, where $V$ denotes either a $Z$-boson or a photon.
This theory features for example a CP-violating $ZZ\gamma$ coupling, proportional to the form factor
$f^\gamma_4$, where it is assumed that the two $Z$ bosons are on-shell. The best testbed for this
type of interaction is $Z$-boson pair production. In Ref.~\cite{CMS:2014xja} the \CMS collaboration
reported on a corresponding search for anomalous $ZZZ$ and $ZZ\gamma$ interactions in
four-lepton production in $8\;\textrm{TeV}$ proton-proton collision events. The final states
$4e$, $4\mu$ and $2e2\mu$ are taken into account. The event-selection criteria used read
\begin{eqnarray}
  p_\mathrm{T}(\mu) >5\;\textrm{GeV},\;\; p_\mathrm{T}(e)>7\;\textrm{GeV},\;\;|\eta(\mu)|<2.4,\;|\eta(e)|<2.5\;\;\text{and}\;\; m_{e^+e^-/\mu^+\mu^-}\in [60,120]\;\text{GeV}\,.
\end{eqnarray}
In the experimental analysis the \Sherpa generator has been used for the signal predictions. 

In \FigRef{fig:zzagc} we compare leading-order plus parton-shower predictions from \Sherpa with
\CMS data published in~\cite{CMS:2014xja}. Besides the leading-order SM expectation we show as an
illustrative example the prediction when including a $ZZ\gamma$ vertex with coupling
$f^\gamma_4=0.1$, with all other New Physics couplings set to zero. Clearly, the latter hypothesis
is not compatible with the observed data. The \CMS collaboration extracted $95\%$ confidence
level limits on $f^\gamma_4\in[-0.005,0.005]$.
%


\clearpage
\subsection{Hadronisation, Underlying Events and Hadron Decays}\label{sec:full_results}
This section is devoted to highlight some aspects of the modelling of
non-perturbative phenomena in \Sherpa. In particular, we present results sensitive
to hadronisation, the underlying event and (soft) hadron decays, including spin
correlations in hadronic $\tau$-decays.

\paragraph{Hadronisation}
\Sherpa\ implements a cluster model for the fragmentation of partons into hadrons,
\cf Sec.~\ref{sec:hadronisation} and Ref.~\cite{Winter:2003tt}. Furthermore,
it offers an interface to the Lund string fragmentation model as
implemented in \Pythia 6.4~\cite{Sjostrand:2006za}. This allows for important
cross checks of the non-perturbative modelling. In particular it is possible to
extract theoretical uncertainties related to the parton-to-hadron transition,
keeping all perturbative aspects of the simulation identical. 

\begin{figure}[h!]%
  \centering
  \parbox{0.47\textwidth}{%
    \includegraphics[width=0.47\textwidth]{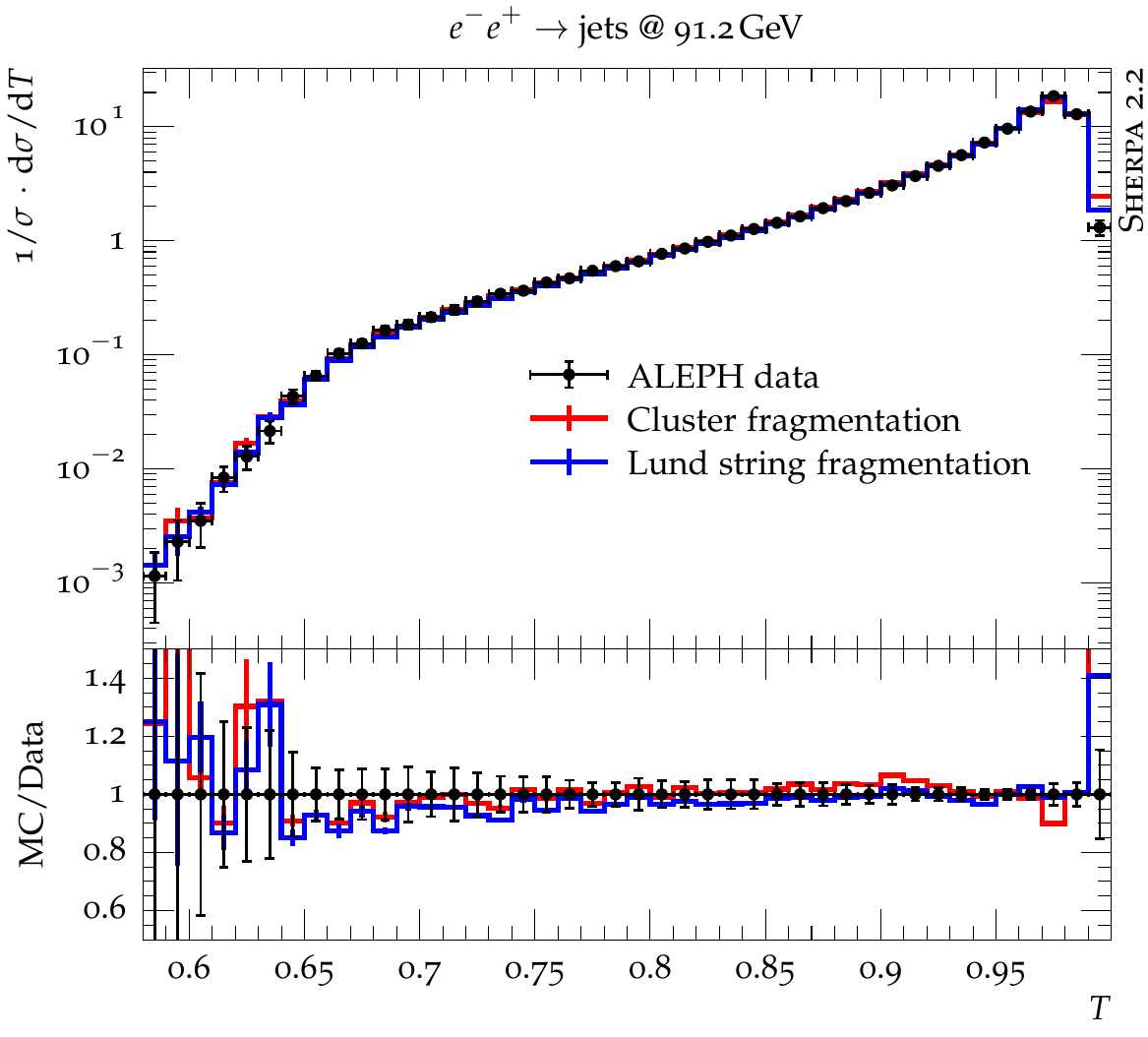}
    \caption{Results for the thrust distribution in jet production at \LEP for
        the two fragmentation models available in \Sherpa in comparison with
        \ALEPH data~\cite{Heister:2003aj}.}%
    \label{fig:thrust}}%
  \hfill
  \parbox{0.47\textwidth}{%
    \includegraphics[width=0.47\textwidth]{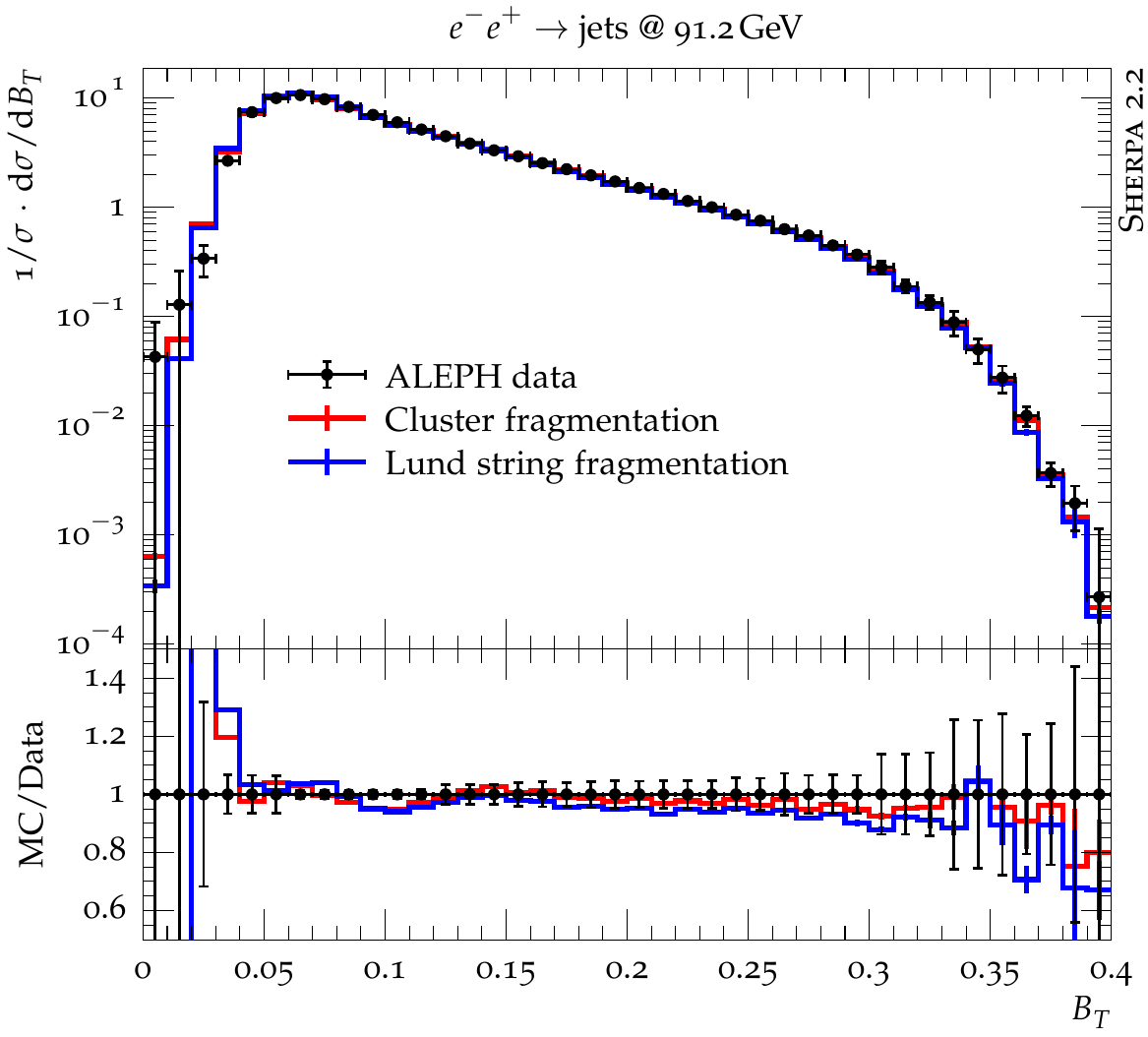}
    \caption{Results for the distribution of the total jet broadening at \LEP
        for the two fragmentation models available in \Sherpa in comparison
        with \ALEPH data~\cite{Heister:2003aj}.}%
    \label{fig:jetbroad}}%
\end{figure}

To illustrate this aspect, we show a comparison with \LEP data from
\ALEPH~\cite{Heister:2003aj} for the thrust and total jet broadening event-shape
variables in \FigRef{fig:thrust} and \FigRef{fig:jetbroad}, respectively. The
\Sherpa\ predictions presented there are based on an \MEPSatNLO sample, where
the $2\to2,3,4,5$-parton matrix elements are considered at NLO QCD. The merging
parameter is set to $y_{\textrm{cut}}=\left(Q_{\textrm{cut}}/E_{\textrm{CMS}}\right)^2=10^{-2.25}$.
We evolve the strong coupling at the two-loop order, assuming $\alphaS(m_Z)=0.117$.%
\footnote{ Note, the \Sherpa\ default value is $\alphaS(m_Z)=0.118$. However, we observed a
  marginally better description of \LEP observables, and in particular the thrust
  distribution, both for the cluster and the Lund string fragmentation
  using $\alphaS(m_Z)=0.117$.}
While for the cluster fragmentation model we have kept all relevant parameters at
their default values, we have set the main parameters of the Lund model to
\begin{equation}
  a = 0.3\;(\texttt{PARJ(41)}),\;\; b = 0.6\;\textrm{GeV}^{-2}\;(\texttt{PARJ(42)}),\;\; \sigma = 0.36\;\textrm{GeV}\;(\texttt{PARJ(21)})\,.
\end{equation}
For both hadronisation models a satisfactory agreement with data is observed. The
variations between the two predictions stay within the few percent range.

\FloatBarrier
\paragraph{Underlying Event}

\begin{figure}%
  \centering
  \parbox{0.47\textwidth}{%
    \includegraphics[width=0.47\textwidth]{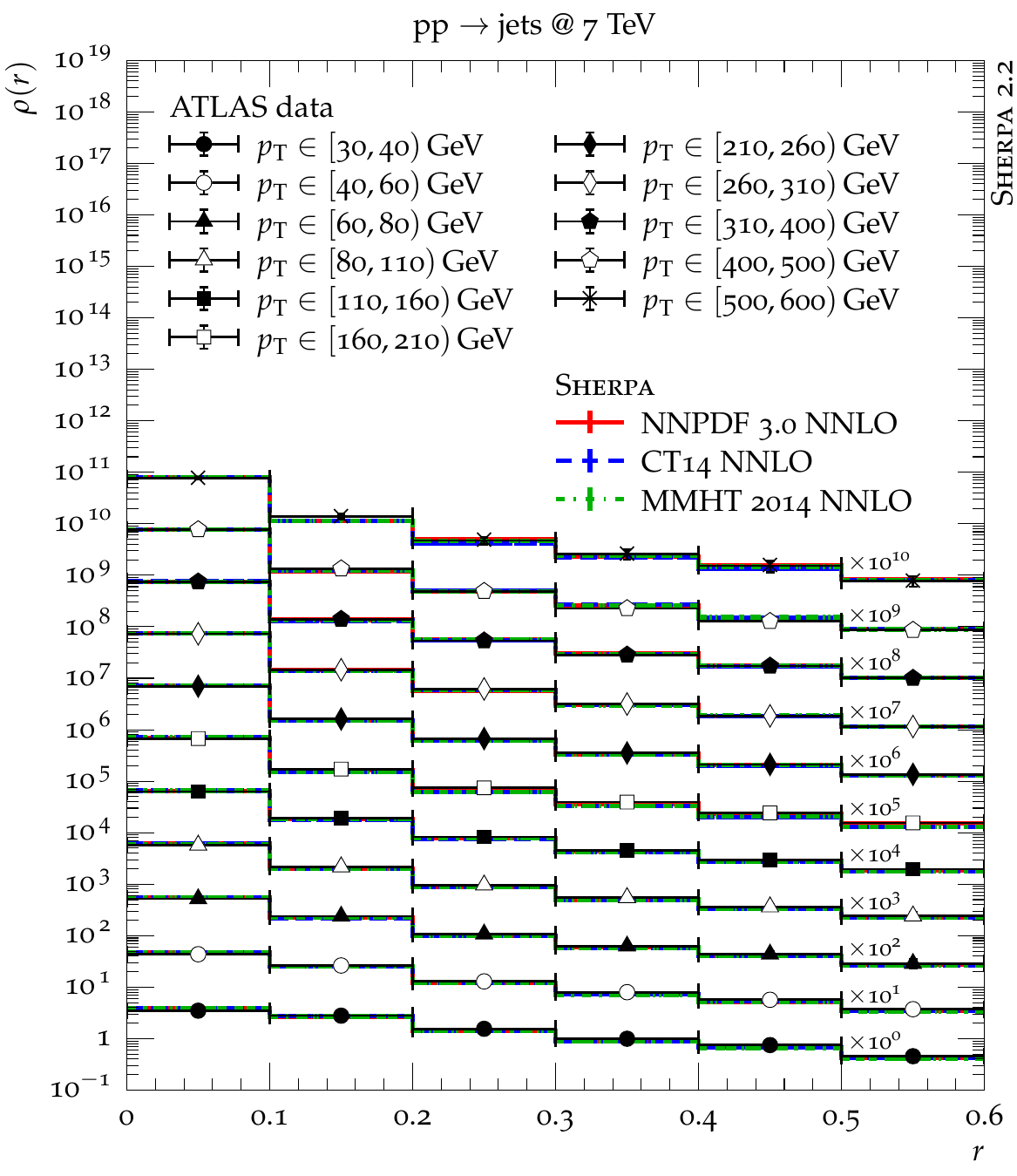}
    \caption{Results for the differential jet shape $\rho(r)$ in dependence on
        the jet transverse momentum within $|\eta|<2.8$ in comparison to \ATLAS
        data~\cite{Aad:2011kq}.\\\textcolor{white}{.}\\\textcolor{white}{.}}%
    \label{fig:jetshape}%
  }
  \hfill
  \parbox{0.47\textwidth}{%
    \includegraphics[width=0.47\textwidth]{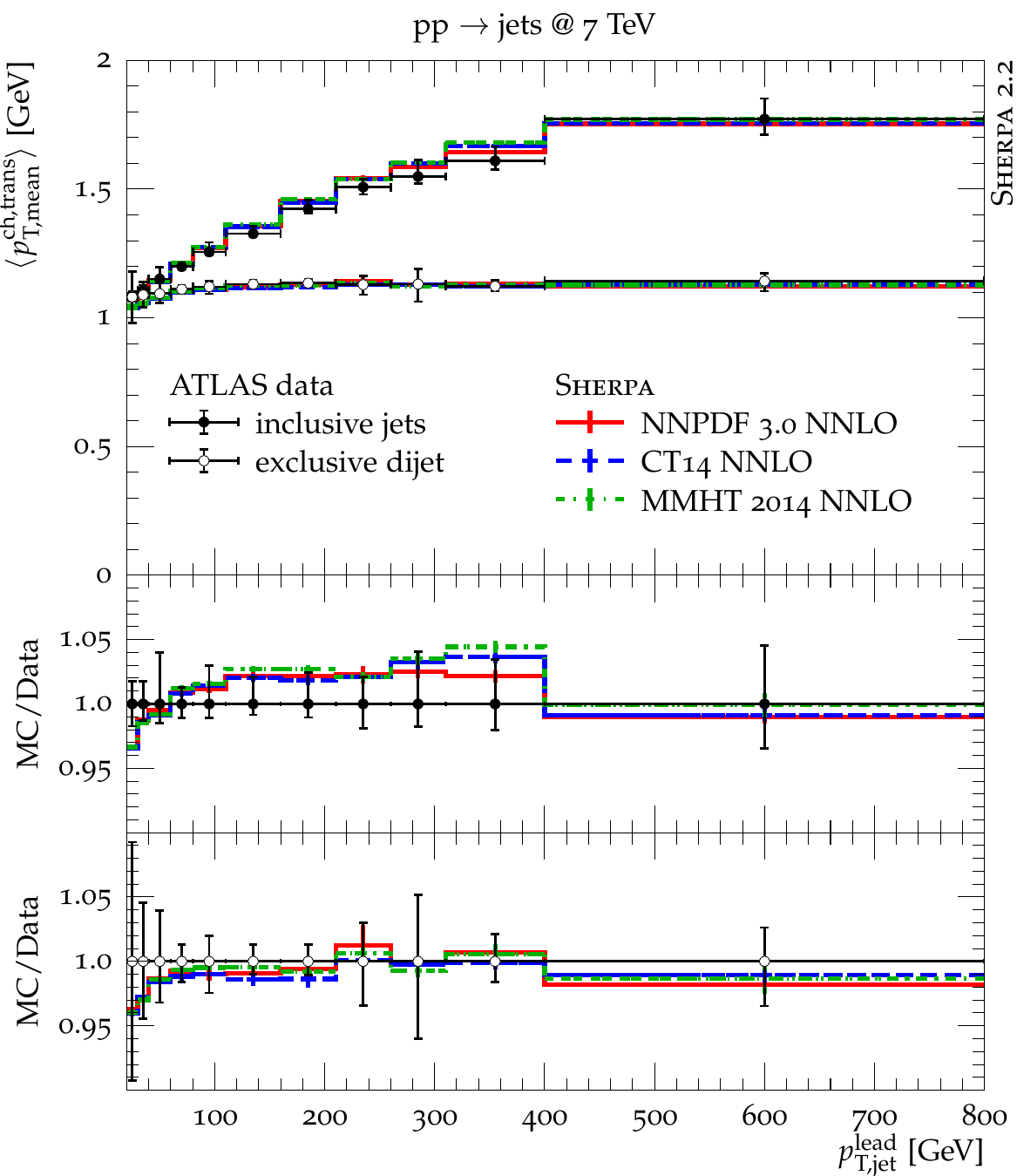}
    \caption{Results for the average mean charged-particle transverse momentum
        in the transverse region in dependence of the leading jet \pT in
        inclusive jet and exclusive dijet events in comparison to data from the
        \ATLAS experiment~\cite{Aad:2014hia}.}
    \label{fig:jetChmulti}%
  }
\end{figure}

As explained in \SecRef{sec:ue}, an accurate simulation of the
effects of secondary scattering and their subsequent evolution is
necessary to be able to describe observables at hadron colliders. As
an example for an observable that is impacted by non-perturbative
contributions from the underlying event and hadronisation we consider
the differential jet-shape variable $\rho(r)$. In  \FigRef{fig:jetshape}
we compare \Sherpa particle-level predictions based on the
two-jet-production matrix element, evolved by the \CSS, and including 
effects from multiple parton scatterings and hadronisation. 
We give predictions based on three
different PDF sets -- consistently used throughout the hard process, initial-state
parton showering and the underlying-event simulation -- namely
NNPDF~3.0~\NNLO~\cite{Ball:2014uwa}, MMHT 2014 \NNLO~\cite{Harland-Lang:2014zoa}
and CT14 \NNLO~\cite{Dulat:2015mca}. The predictions for various slices of
jet transverse momentum are compared with data from the ATLAS collaboration
taken in \LHC Run 1 at $\sqrt{s}=7$ TeV~\cite{Aad:2011kq}. Notably, the
predictions for all three PDF sets largely agree and yield a satisfactory
description of the measurements. 
Please note that \Sherpa's non-perturbative models have only been tuned 
using the NNPDF~3.0~\NNLO set, and thus this level of agreement is non-trivial. 
Furthermore,
jets at different transverse momenta receive different contributions from
hadronisation and the underlying event. Clearly, the softer the jet, the larger
the non-perturbative corrections the jet shape $\rho(r)$ receives. 

As a second example we consider in \FigRef{fig:jetChmulti} a more exclusive
observable, namely the average of the mean charged-particle transverse 
momentum per event in the region transverse to the leading jet, 
differential in the leading-jet transverse momentum. 
This transverse region, defined with respect to the azimuthal angle
of the leading jet as $60^\circ<|\Delta \phi| <120^\circ$, is expected to be very
sensitive to the underlying event. 
The fact that it measures charged particles only, renders it sensitive 
to the flavour structure of the hadronisation simultaneously. 
This observable has been
measured by the ATLAS collaboration in \cite{Aad:2014hia}, where results for
inclusive jet and exclusive dijet production have been presented. Jets thereby
have to fulfill $p_{\mathrm{T},j}>20$ GeV and $|y_j|<2.8$, for the charged particles in
the transverse region it is required that $p_\mathrm{T}>0.5$ GeV and $|\eta|<2.5$. The
\Sherpa predictions are in good agreement with the data, both for the inclusive
jet and the exclusive dijet selection. No significant dependence on the PDF
set employed in the simulation can be observed, despite \Sherpa only having been 
tuned using one of the PDF sets, as discussed above. 

\begin{figure}%
  \centering
  \parbox{0.47\textwidth}{%
    \includegraphics[width=0.47\textwidth]{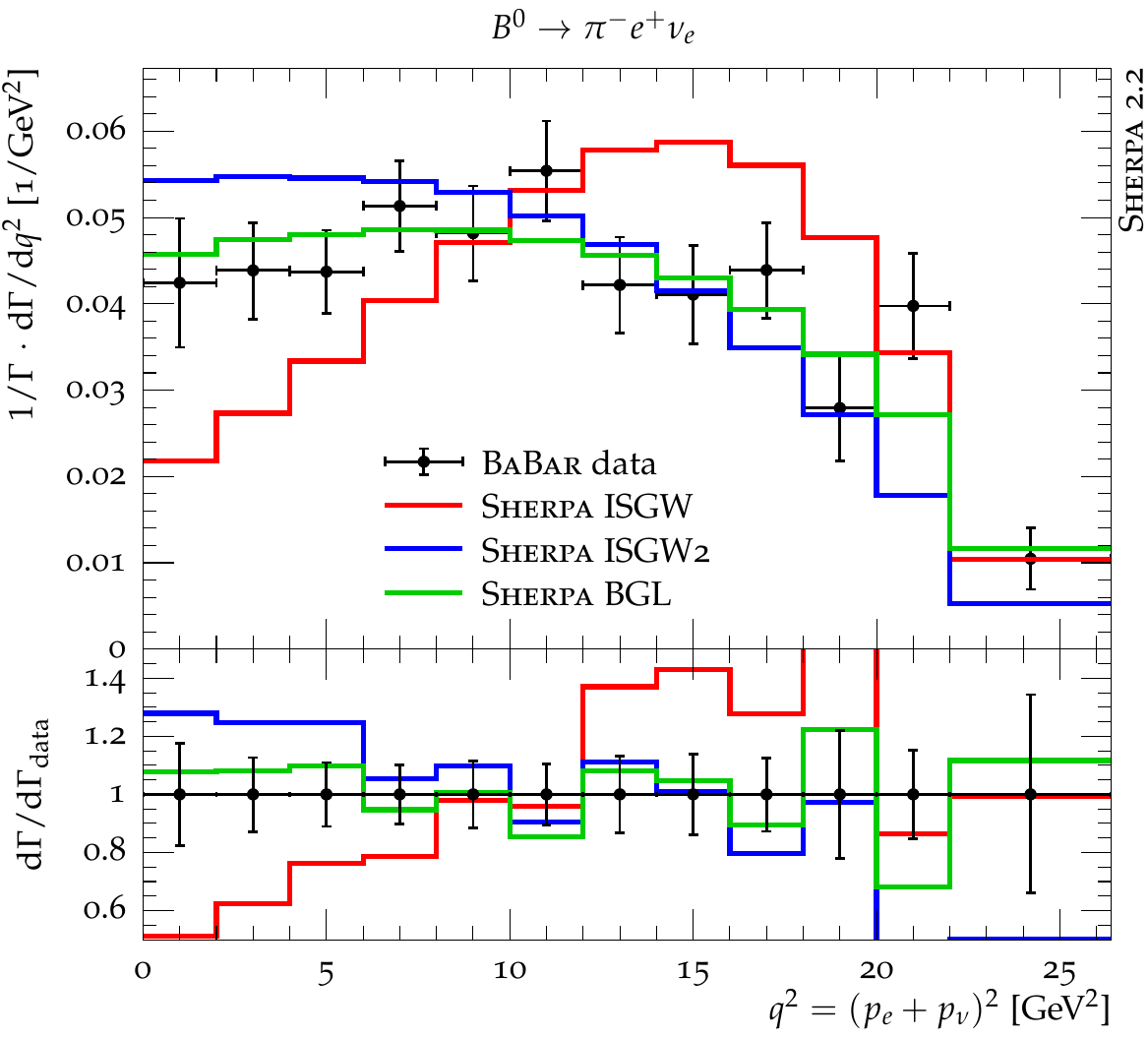}
    \caption{Results for semileptonic $B^0$ decays based on different
        form-factor models in comparison to data from the \babar
        experiment~\cite{Lees:2012vv}.}%
    \label{fig:hadrondecays}}%
  \qquad
  \parbox{0.47\textwidth}{%
    \includegraphics[width=0.47\textwidth]{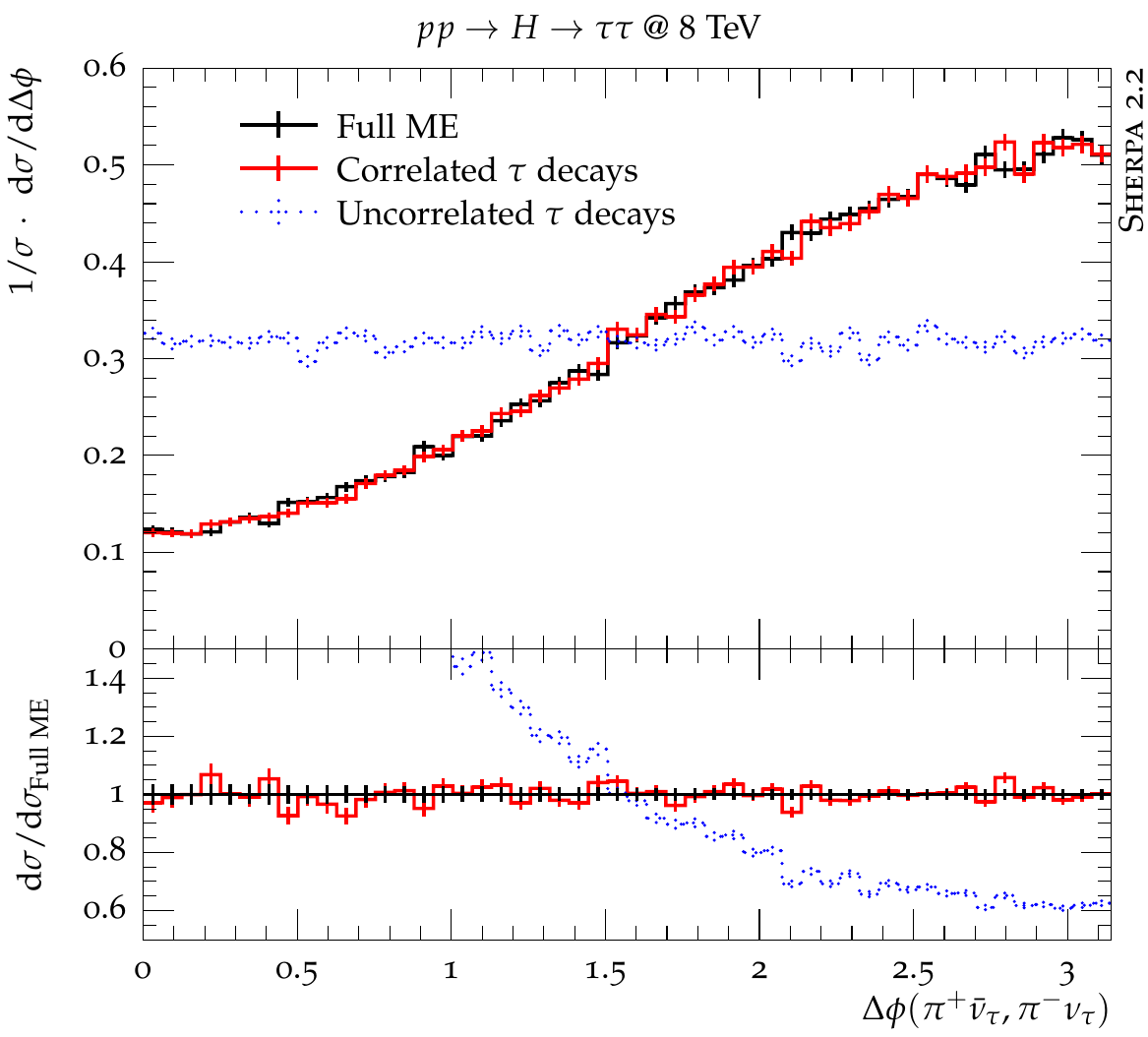}
    \caption{Different predictions for the decay-plane angle between
        hadronically decaying $\tau$-leptons from Higgs-boson decays.}%
    \label{fig:tau-spincorrs}}%
\end{figure}

\paragraph{Hadron and $\tau$ Decays}
As a last example, we show results which are sensitive to the modelling of
hadron and $\tau$-lepton decays. The distribution of momentum transfer in the
semileptonic $B^0\to \pi^- e^+ \nu_e$ decay for various form-factor models
implemented in \Sherpa is compared with data taken from the \babar
experiment~\cite{Lees:2012vv} in~\FigRef{fig:hadrondecays}.
The BGL parametrisation~\cite{Boyd:1995sq} with parameters extracted from that
measurement reproduces the data well, while the ISGW2~\cite{Scora:1995ty} and ISGW
models~\cite{Isgur:1988gb} with their original parameter sets do not agree well
with this measurement.

To illustrate the simulation of hadronic $\tau$ decays, we consider in
\FigRef{fig:tau-spincorrs} the production of Higgs bosons at the \LHC which
decay to a pair of $\tau$-leptons, which then are assumed to subsequently decay
via $\tau \to \pi\nu$. Clearly, the effect of including spin correlations in the
decay chain can have a dramatic effect when measuring correlations of the decay
products, as exemplified for the decay-plane angle in the $H\to \tau(\to \pi\nu)\tau(\to \pi\nu)$ final state.
In addition, one can see how the proper inclusion of spin-correlation effects leads to
an excellent agreement with the full result, obtained with exact matrix elements for
the decayed final state.

\FloatBarrier



\section{Conclusions}\label{sec:conclusions}

We have summarised essential features and improvements of the \Sherpa 2.2
event generator. The \Sherpa framework has been extensively used for event
generation during the \LHC Run 1 and Run 2, and represents a decade
of developments towards ever higher precision in the simulation.

Key building blocks of the \Sherpa\ generator are implementations of all the
physics aspects needed for a full event description, including automated
matrix-element generators, parton showers, a hadronisation model and a simulation
of multiple parton interactions. Supplemented by methods to deal with particle
decays, QED corrections and a large variety of interfaces, \eg to parton density
functions, New Physics models or event-output formats, this qualifies
\Sherpa\ as a full-fledged multi-purpose event generator for the modelling of
scattering events at past, current, and future collider experiments. Certainly
a highlight and unique feature of \Sherpa\ are its comprehensive methods to
combine higher-order perturbative matrix-element calculations with parton-shower
simulations and especially its automation of the \MEPSatNLO\ method. In
Sec.~\ref{sec:results} we have illustrated some applications of \Sherpa\ to
challenges posed in particular by the \LHC experiments. Through the inclusion of
exact NLO QCD matrix elements the jet activity accompanying signal processes gets
adequately modelled and at the same time theoretical uncertainties will be
systematically reduced. 

At this time development is moving towards \Sherpa 3.0.0, heralding a major new
development effort with exciting improvements across the board.  They will include
the description of higher-orders in the perturbative parts of the simulation,
for example incorporating advances in the resummation properties of parton
showers~\cite{Hoche:2017hno,Hoche:2017iem,Dulat:2018vuy}, the inclusion of
approximate high-energy EW effects as based on~\cite{Denner:2000jv}, a fully
massive five-flavour scheme for heavy quarks in the initial state~\cite{Krauss:2017wmx},
the improvements in efficient phase-space sampling~\cite{Kroeninger:2014bwa}, and extensions such as
the automation of QCD soft-gluon resummation at NLL accuracy following the \Caesar\
formalism~\cite{Gerwick:2014gya,Marzani:2019evv}. This will be supplemented with continuous
improvements in the non-perturbative modelling, such as an improved cluster
hadronisation or a new model for inclusive QCD scattering.

With the preparations for \LHC Run 3 in full swing, and many measurements with the full
Run 2 data to appear in the next years, this new version will play an important
role in the future analysis of \LHC data.


\section*{Acknowledgements}

We would like to thank all our former collaborators on the \Sherpa project, in particular the
various students that contributed with their theses works to further improve our physics
models and the corresponding computer codes. We are indebted to our users that have helped
us to find and fix short-comings in the program. 

This work has received funding from the European Union’s Horizon 2020 research and innovation
programme as part of the Marie Sk\l{}odowska-Curie Innovative Training Network MCnetITN3
(grant agreement no. 722104). It was supported by the U.S. Department of Energy under contract
DE-AC02-76SF00515. It used resources of the Fermi National Accelerator Laboratory
(Fermilab), a U.S. Department of Energy, Office of Science, HEP User Facility.
Fermilab is managed by Fermi Research Alliance, LLC (FRA), acting under
Contract No.\ DE--AC02--07CH11359.

SS acknowledges support through the Fulbright-Cottrell Award and from BMBF (contracts 05H15MGCAA and 05H18MGCA1). 
FS's research was supported by the German Research Foundation (DFG) under
grant No.\ SI 2009/1-1.
MS acknowledges the support of the Royal Society through the award 
of a University Research Fellowship.
The work of DN is supported by the French Agence Nationale de
la Recherche, under grant ANR-15-CE31-0016.


\appendix

\clearpage
\section{Default-tune non-perturbative model parameters}\label{app:tune}

The major parameters specifying the cluster-fragmentation model in \Sherpa and their default values
obtained by tuning in particular to LEPI data:

\begin{lstlisting}[style=runcard,numbers=none]
  (fragmentation){
    % strange quark fraction 
    STRANGE_FRACTION 0.6049;
    % baryon fraction
    BARYON_FRACTION 1.0;
    % various quark, diquark flavour selection weights
    P_{QS}/P_{QQ} 0.3;
    P_{SS}/P_{QQ} 0.01;
    P_{QQ_1}/P_{QQ_0} 1.0;
    % gluon splitting kinematics, \eta and p^2_{T,0}, cf. Eq. (2.4)
    G2QQ_EXPONENT 1.08;
    PT^2_0 1.56;
    % cluster splitting exponents, cf. Eq. (2.10)
    SPLIT_EXPONENT 0.1608;
    SPLIT_LEAD_EXPONENT 1.0;
    SPECT_EXPONENT 1.739;
    SPECT_LEAD_EXPONENT 8.0;
    % cluster to hadron decays
    DECAY_OFFSET 1.202; % \kappa
    DECAY_EXPONENT 2.132; % \chi
    % multiplet weights
    MULTI_WEIGHT_L0R0_VECTORS 0.75;   % $\rho$, $K^*$, $\omega$, $\phi$ etc
    MULTI_WEIGHT_L0R0_TENSORS2 0.30;  % $a_2(1320)$, $f_2(1270)$, $f'_2(1525)$, $K^*_2(1430)$ etc
    MULTI_WEIGHT_L0R0_DELTA_3/2 0.45; % $\Delta(1232)$, $\Sigma(1385)$, $\Xi(1530)$, $\Omega^-$ etc
    % flavour-specific enhancement factors
    HEAVY_CHARMBARYON_ENHANCEMENT 0.9;
    HEAVY_BEUATYBARYON_ENHANCEMENT 1.7;
    HEAVY_CHARMSTRANGE_ENHANCEMENT 1.0;
    HEAVY_BEAUTYSTRANGE_ENHANCEMENT 3.0;
 }(fragmentation)
\end{lstlisting}

\noindent
Parameters for the intrinsic transverse-momentum distribution of initial-state partons in
composite beam particles, cf. Eq.~(\ref{eq:ue_intrinsic_kt}). Per default initial-state
protons are assumed. In the case of proton--proton collisions the same values for the mean
and width of the corresponding Gaussian distributions are assumed, respectively, \ie
\begin{lstlisting}[style=runcard,numbers=none]
 (beam){
    % Gaussian distributed intrinsic k_T in initial-state protons
    K_PERP_MEAN_1 1.1; % <k_T> beam 1
    K_PERP_MEAN_2 1.1; % <k_T> beam 2      
    K_PERP_SIGMA_1 0.85; % \sigma beam 1
    K_PERP_SIGMA_2 0.85; % \sigma beam 2
 }(beam)
\end{lstlisting}
These values correspond to the reference collision energy of $E_{\text{ref}}=7$ TeV. For the width
parameter the value at other centre-of-mass energies is computed as
 \begin{equation}
    \sigma(E_{\text{cms}}) =  \sigma(E_{\text{ref}})\left(\frac{E_{\text{cms}}}{E_\text{ref}}\right)^{0.55}\,.
\end{equation}

\noindent
List of major parameters defining the multiple-parton interaction model in \Sherpa and their
default-tune values, cf. Eq.~(\ref{eq:ue_sudakov}):

\begin{lstlisting}[style=runcard,numbers=none]
 (mi){
    % MPI model parameters
    SIGMA_ND_FACTOR 0.3142; % factor multiplying non-diffractive cross section
    % cut-off parameter and regulator, given at reference scale E_{ref} = 1.8 TeV
    SCALE_MIN 2.895; % p_{T,min}(E_{ref})
    TURNOFF 0.7549;  % p_{T,0}(E_{ref})
    TURNOFF_EXPONENT 0.244; % \alpha
 }(mi)
\end{lstlisting}

\noindent
The parameters $p_{T,\text{min}}(E_{\text{ref}})$ and $p_{T,0}(E_{\text{ref}})$ are evolved from the reference
energy $E_{\text{ref}}$ (\texttt{REFERENCE\_SCALE}) to the actual collider energy $E$ according to 
 \begin{equation}
    p_{T,i}(E) =  p_{T,i}(E_\mathrm{ref})\left(\frac{E}{E_\mathrm{ref}}\right)^\alpha\,,
\end{equation}
with the power $\alpha$ specified by the parameter \texttt{TURNOFF\_EXPONENT}.

\bibliographystyle{bib/amsunsrt_modp}
\bibliography{bib/journal}

\end{document}